\documentclass{amsproc}
\usepackage[english]{babel}
\usepackage{amsfonts,amsmath,amssymb,amscd}
\usepackage{times}
\usepackage{array}
\usepackage{latexsym}
\usepackage{float,bbm}
\usepackage[arrow,matrix,tips]{xy}
\usepackage{animate}
\usepackage[T1]{fontenc}
\usepackage{flashmovie}
\usepackage{dsfont,mathrsfs,stmaryrd,wasysym,amsbsy}
\usepackage{xcolor}
\usepackage{upgreek}
\usepackage{mathtools}
\usepackage{kbordermatrix}
\usepackage{epic,eepic}
\usepackage{float}
\usepackage{color}
\usepackage{stackengine}
\usepackage{bbm}
\usepackage{hyperref}

\usepackage{color}
\usepackage{amsthm}
\usepackage{colortbl}
\usepackage{enumerate}
\usepackage{graphicx}
\usepackage{url}
\usepackage{subfigure}
\usepackage{xcolor}
\usepackage{dsfont}
\usepackage{fancybox}

\newtheorem{theorem}{Theorem}[section]
\newtheorem{lemma}[theorem]{Lemma}

\theoremstyle{definition}
\newtheorem{definition}[theorem]{Definition}

\theoremstyle{remark}
\newtheorem{remark}[theorem]{Remark}

\numberwithin{equation}{section}

\def\bc{\boldsymbol{c}}
\def\bp{\boldsymbol{p}}
\def\br{\boldsymbol{r}}

\def\balpha{\boldsymbol{\alpha}}
\def\bbeta{\boldsymbol{\beta}}

\def\btheta{\boldsymbol{\theta}}

\def\biota{\boldsymbol{\iota}}

\def\bmu{\boldsymbol{\mu}}
\def\btheta{\boldsymbol{\theta}}
\def\bnu{\boldsymbol{\nu}}

\def\bcM{\boldsymbol{\mathcal{M}}}

\newdimen\dropht \newbox\dropbox
\def\dropshadow#1#2{\setbox\dropbox=
\vbox{\hrule\hbox{\vrule\kern6pt
        \vbox{\kern6pt#1\kern6pt}\kern6pt\vrule}\hrule}
\dropht=\ht\dropbox\advance\dropht by -#2
       \vbox{\baselineskip0pt\lineskip0pt
             \hbox{\copy\dropbox\vrule width#2 height\dropht}
             \hbox{\kern#2\vrule height#2 width \wd\dropbox}}}

\definecolor{gray9}{rgb}{1,.894,.769}

\newcommand{\bone}{{\bf 1}}

\def\bL{{\mathbf L}}
\def\bM{{\mathbf M}}
\def\bW{{\mathbf W}}
\def\bX{{\mathbf X}}
\def\bY{{\mathbf Y}}
\def\bZ{{\mathbf Z}}

\def\AA{\mathbb{A} }
\newcommand{\EE}{\mathbb{E} }
\newcommand{\FF}{\mathbb{F} }
\newcommand{\HH}{\mathbb{H} }

\newcommand{\PP}{\mathbb{P} }

\newcommand{\RR}{\mathbb{R} }
\newcommand{\TT}{\mathbb{T} }

\newcommand{\cE}{\mathcal{E} }
\newcommand{\cF}{\mathcal{F} }

\newcommand{\cL}{\mathcal{L} }

\newcommand{\cP}{\mathcal{P} }

\newcommand{\cS}{\mathcal{S} }

\def\o{\overline}
\def\u{\underline}
\def\t{\tilde}

\graphicspath{{"/Users/rcarmona/Google Drive/Books/MFG/CARMONA_DELARUE_BOOK_FINAL/"}}
\graphicspath{{"/Users/rcarmona/Google Drive/Talks/ANR_Paris_Dec_2018/"}}
\graphicspath{{"/Users/rcarmona/Google Drive/Talks/Duke_3_16/"}}

\begin{document}

\title[FInancial Engineering \& Economics]{\textbf{Applications of Mean Field Games\\
in Financial Engineering and Economic Theory}}
\author{Ren\'e Carmona }
\address{Department of Operations Research \& Financial Engineering\\
Bendheim Center for Finance\\
Program in Applied \& Computational Mathematics\\
Princeton University}

\email{rcarmona@princeton.edu}
\thanks{The author was partially supported by NSF DMS-1716673 and ARO W911NF-17-1-0578}
\subjclass[2000]{Primary }

\begin{abstract}
This is an expanded version of the lecture given at the AMS Short Course on Mean Field Games, on January 13, 2020 in Denver CO. The assignment was to discuss applications of Mean Field Games in finance and economics.
I need to admit upfront that several of the examples reviewed in this chapter were already discussed in book form. Still, they are here accompanied with discussions of, and references to, works which appeared over the last three years. Moreover, several completely new sections are added to show how recent developments in financial engineering and economics can benefit from being viewed through the lens of the Mean Field Game paradigm. The new financial engineering applications deal with bitcoin mining and the energy markets, while the new economic applications concern models offering a smooth transition between macro-economics and finance, and contract theory.
\end{abstract}

\maketitle

\tableofcontents

%%%%%%%%%%%%%%%%%%%%%%%%%%%%%%%%%%%%%%%%%%%%%
\section[Introduction]{\textbf{Introduction}}
%%%%%%%%%%%%%%%%%%%%%%%%%%%%%%%%%%%%%%%%%%%%%
\label{se:intro}

The goal of this chapter is to review applications in financial engineering and economics which can be cast and tackled within the framework of the Mean Field Game (often abbreviated as MFG in the sequel) paradigm.

Sections \ref{se:fe}, \ref{se:energy} and \ref{se:macro} revisit some of the applications discussed in \cite[Chapter 1]{CarmonaDelarue_book_I}. Here, we try to give some context, and examine developments which occurred since the publication of the book. While we do not delve into proofs, we provide detailed references where the interested reader will be able to find complements on the subject matters. This is clearly the case for Section \ref{se:fe} reviewing applications to financial engineering.

Economic models often involve optimization over sets of actions, behaviors and strategies. Engineers and biologists facing similar optimization problems would be likely to restrict themselves to finite action sets, but many economists prefer continuous action sets in order to be able to write first order conditions of optimality in differential forms, in hope of resolving them with explicit solutions. Furthermore, they have no qualms about using models with a continuum of agents. They explain that this guarantees an environment with perfect
competition and tractable equilibrium behavior. However, these formulations often raise eyebrows from colleagues, and especially mathematicians. Our discussion of macro-economic models is motivated in part by the desire to reconcile the approach to competitive equilibria with a continuum of players with the paradigm of mean field games. Additionally, while not all economists like to work with continuous time, it has been argued over the last two decades that models in macro-economic, contract theory, finance, $\ldots$, greatly benefit from the switch to continuous time from the more traditional discrete time models. Yuli Sannikov is certainly one of the most visible crusaders in this respect, and we shall follow his lead and consider only continuous time models in this review chapter.

\vskip 4pt
Like it is often the case in the published literature in economics, economies are modeled as populations with a continuum of players, and to capture the fact that the influence of any individual member on the aggregate should be infinitesimally small, these individuals are modeled as elements of a measure space with a continuous measure, say the unit interval $[0,1]$ with its Lebesgue's measure. This starting point is different from the typical set-up of mean field games which usually starts from $N$ player games with $N$ finite, and considers the limit $N\to \infty$. We shall use both modeling paradigms, and refere the reader to \cite[Section 3.7]{CarmonaDelarue_book_I} for a discussion of the links between the two.

None of the macro-economic papers which we review in this chapter mention the name \emph{Nash}, or use the terminology \emph{Nash equilibrium}. The authors of these papers are concerned with \emph{general equilibria}, not \emph{Nash equilibria}. They first assume that all the aggregate quantities in the economy are given, and they let the participants in the economy optimize their utilities independently of each other. Next, they identify the constraints to be satisfied according to economic theory, compile them in a set which they call set of clearing conditions, and check that the results of the participants' optimizations are compatible with the clearing conditions. As long as the clearing conditions are satisfied while the participants in the economy optimize their expected utilities, a general equilibrium is said to take place. This procedure is very similar to the search for a Nash equilibrium. Identifying the best response function amounts to having the participants optimize their expected utilities given the strategies of the other players, and given the constraints of economic theory, the search for a fixed point of the best response function appears as an analog of the check that the clearing conditions are satisfied. The parallel is even more striking when the clearing conditions can be written in terms of aggregate quantities. Indeed these aggregates quantify the interactions between the economy participants, and since these aggregates are most often nothing more than empirical means of state variables, they spell out \emph{mean field} interactions in the model. It is thus reasonable to cast these general equilibria as Nash equilibria for Mean Field Game models, and take full advantage of the technology already developed for the analysis of these game models to gain a better understanding of these general equilibria.

\vskip 4pt\noindent
\emph{Acknowledgements: } I would like to thank Markus Brunnermeier for relentless attempts at educating me on some of the subtleties of the economic models discussed in Section \ref{se:macro_to_finance}. Also, I would like to acknowledge the useful comments and suggestions from an anonymous referee.
 
%%%%%%%%%%%%%%%%%%%%%%%%%%%%%%%%%%%%%%%%%%%%%
\section[Financial Engineering Applications]{\textbf{Financial Engineering Applications}}
%%%%%%%%%%%%%%%%%%%%%%%%%%%%%%%%%%%%%%%%%%%%%
\label{se:fe}

%%%%%%%%%%%%%%%%%%%%%%%%%%%%%%%%%%%%%%%%%%%%%
\subsection[Systemic Risk]{\textbf{Systemic Risk}}
%%%%%%%%%%%%%%%%%%%%%%%%%%%%%%%%%%%%%%%%%%%%%
The study of systemic risk is concerned with the identification and analysis of events or a sequences of events which could trigger severe instability, or even collapse of the financial system and the entire economy as a result. In finance, systemic risk is approached differently than the risk associated with any one individual institution or portfolio.
In the US, it was first brought to the forefront by the Federal Reserve Bank of New York after the September 11, 2001 attack. See \cite{KambhuWeidmanKrishnan}. The Central Bank spearheaded several collaborative research initiatives involving economists and academic researchers including engineers working on the safety of the electric grid, mathematicians experts in graph theory and network analysis, population biologists, .... .  But unfortunately, what brought this line of research in the limelight is the financial crisis of 2008 for which systemic risk was a major contributor. While the mean field game models we discuss below give identical importance to the various institutions involved in the model, it is clear that a realistic model of systemic risk in finance should differentiate the roles of those companies considered to be \emph{too big to fail}, and which have been identified officially under the name of  SIFI for Systemically Important Financial Institutions. They are the banks, insurance companies, or other financial institutions that U.S. federal regulators think would pose a serious risk to the economy if they were to collapse. See \cite{FouqueLangsam} for an account of the state of the art after the financial crisis.
We shall briefly discuss extensions of the models reviewed below which could possibly account for the presence of such important players in the models.

%%%%%%%%%%%%%%%%%%%%%%%%%%%
\subsubsection{\textbf{A Toy Model of Systemic Risk }}

Our first example is borrowed from  \cite{CarmonaFouqueSun}. We view its main merit as being pedagogical. It gives us the opportunity to review some of the main features of Linear Quadratic (LQ) finite player game models, and in so doing, explain how one can solve them. Moreover, despite its very simple structure, it exhibits several very useful characteristics: for each integer $N\ge 1$, the $N$-player version of the game has a unique closed loop equilibrium, a unique open loop equilibrium, this open loop equilibrium is in closed loop form, but it is different from the closed loop equilibrium. However, when $N\to\infty$, both equilibria converge toward the unique equilibrium of the mean field game. 

\vskip 2pt
To describe the model, let us assume that the log-monetary reserves of $N$ banks, say $X^i_t, i=1,\dots, N$, satisfy
$$
dX^i_t=\left[a(\overline{X}_t-X^i_t)+\alpha^i_t\right]dt +\sigma \bigg(\sqrt{1-\rho^2} dW^i_t+\rho dW^0_t\bigg), \quad i=1,\ldots,N
$$
where $W^i_t, i=0,1,\dots, N$  are independent  Brownian motions and $\sigma>0$ is a constant. 
The notation $\bar X_t$ stands for the empirical mean of the $X^i_t$ for $i=1,\ldots,N$, $a$ is a constant regulating the \emph{mean reversion} of $X^i_t$ toward the mean, and $\rho$ correlates the idiosyncratic shocks $dW^i_t$ and the common shock $dW^0_t$. So in this model,  borrowing and lending is done through the drifts. In fact,
\begin{itemize}
\item If $X^i_t$ is small (relative to the empirical mean $\overline{X}_t$) then bank $i$  wants to borrow ($\alpha^i_t>0$) 
\item If $X^i_t$ is large  then bank $i$ will want to lend ($\alpha^i_t<0$) 
\end{itemize}
The adapted stochastic process $\balpha^i=(\alpha^i_t)_{t\ge 0}$ is the control strategy of bank $i$ which tries to minimize the quantity:
\begin{equation}
\begin{split}
&J^i(\balpha^1,\ldots,\balpha^N)\\
&\hskip 15pt
=\EE\left\{\int_0^T\left[\frac{1}{2}(\alpha^i_t)^2-q\alpha^i_t(\overline{X}_t-X^i_t)+\frac{\epsilon}{2}(\overline{X}_t-X^i_t)^2\right]dt+\frac{\epsilon}{2}(\overline{X}_T-X^i_T)^2\right\}
\end{split}
\end{equation}
We could imagine that the quantity $q>0$ is chosen by the regulator to control the cost of borrowing and lending. 

\vskip 4pt
This model is a simple example of an $N$ - player stochastic differential game with \emph{Mean Field Interactions}  since the interactions are through the empirical means of the $N$ states, and a \emph{common noise}.

%%%%%%%%%%%%%%%%%%%%%%%%%%%%%%%
\vskip 12pt\noindent
\subsubsection*{\textbf{Explicit Solutions for Finite $N$}}
While it is usually very difficult to identify and compute Nash equilibria for finite player games, especially when the games are stochastic and dynamic, the very special Linear-Quadratic (LQ) nature of the model allows for explicit solutions.

\vskip 2pt
Searching for an \emph{open loop} Nash equilibrium $\balpha=(\alpha_t)_{0\le t\le T}$ for the game with $\alpha_t=(\alpha^1_t,\ldots,\alpha^N_t)$ being adapted to the filtration generated by the Brownian motions can be done using the Pontryagin stochastic maximum principle. In this particular model, one finds that the strategy profile $\balpha$ defined by  
\begin{equation}
\label{fo:open_loop_control}
\alpha^i_t=\bigl[q+(1-\frac1N)\eta_t\bigr](\o X_t-X^i_t)
\end{equation}
where the deterministic function $t\mapsto\eta_t$ solves the Riccati equation
\begin{equation}
\label{fo:open_loop_Riccati}
\dot\eta_t=2\bigl[a+q -\frac1Nq]\eta_t +\bigl(1-\frac1N\bigr)\eta_t^2+q^2-\epsilon
\end{equation}
with terminal condition $\eta_T=c$. This equation is uniquely solvable if we assume $\epsilon\ge q^2$, is the unique open loop Nash equilibrium. Notice that $\alpha_t$ is in \emph{closed loop / feedback form} since it is of the form $\balpha_t=\phi_t(\bX_t)$ for the deterministic function 
\begin{equation}
\label{fo:phi_t(x)}
(t,x)\mapsto \phi^i_t(x)=\bigl[q+(1-\frac1N)\eta_t\bigr](\o x-x^i), \qquad \text{with}\quad \bar x=\frac1N\sum_{i=1}^N x^i,
\end{equation}
for $i=1,\ldots,N$. Note also that in equilibrium, the states of the banks satisfy:
\begin{equation}
\label{fo:open_loop_state}
dX^i_t= \bigl[a+q+(1-\frac1N)\eta_t \bigr] \bigl(\o X_t-X^i_t\bigr)dt +\sigma\rho dW^0_t+\sigma\sqrt{1-\rho^2}dW^i_t,
\end{equation}
for $i=1,\ldots,N$. So the states evolve according to an Ornstein-Ulhenbeck like process which is Gaussian if the initial conditions are Gaussian (or deterministic).

\vskip 6pt
Searching for a \emph{closed loop} Nash equilibrium $\bbeta=(\beta_t)_{0\le t\le T}$ with $\beta_t=\psi_t(\bX_t)$
is usually done by deriving the Hamilton-Jacobi-Bellman equation for the system of value functions of the players, but in this particular instance, it can also be done using the Pontryagin stochastic maximum principle.
In any case, one finds that there exists a unique Nash equilibrium, and like in the open loop case, the equilibrium strategy profile is given by feedback functions $\psi=(\psi^1,\ldots,\psi^N)$ given by the same formula \eqref{fo:phi_t(x)} except for the fact that the deterministic function $t\mapsto \eta_t$ now solves a slightly different Riccati equation, namely:
\begin{equation}
\label{fo:closed_loop_Riccati}
\dot\eta_t=2(a+q)\eta_t  +\bigl(1-\frac1{N^2}\bigr)\eta_t^2+q^2-\epsilon, 
\end{equation}
with the same terminal condition $\eta_T=c$.
The same remarks apply to the Ornstein-Ulhenbeck nature of the state evolutions in equilibrium.
However, what is remarkable is the fact that the open loop Nash equilibrium happens to be in closed loop form, but still, it is not a closed loop Nash equilibrium. This is very different from the situation for plain optimization. In fact, it reinforces the message that looking for a Nash equilibrium is a far cry from solving an optimization problem.

\vskip 2pt
The second point we want to emphasize is that both Riccati equations, and hence both solutions, \textbf{coincide} in the limit $N\to\infty$. More on that below.

\vskip 4pt
The interested reader can find complements and detailed proofs in \cite[Section 2.5]{CarmonaDelarue_book_I}.

\vskip 12pt\noindent
\subsubsection*{\textbf{Relevance to Mean Field Games (MFGs)}}
Mean Field Games have been touted as the appropriate limits of $N$ player games when $N\to\infty$. However, beyond appealing intuitive arguments and rigorous proofs that MFG solutions can be used to construct strategy profiles forming approximate Nash equilibria for $N$ player games, the larger $N$, the better the approximation, proving actual convergence is a difficult problem. See for example D.Lacker chapter in this volume.

\vskip 2pt
In the present situation, because of the explicit nature of the solutions of the finite player games,
we can take the limit $N\to\infty$
in the dynamics of the states,
in the Nash equilibrium controls (open and closed loop),
and in the expected costs optimized by the players,
all that despite the presence of the common noise.
In fact, we can read off the impact of the common noise in the limit $N\to\infty$ where the open and closed loop models coincide. 
This limit can formally be identified to the so-called Mean Field Game model, and we can even identify the \emph{Master Equation} from the large $N$ behavior of the system of HJB equations. 

\vskip 12pt\noindent
\subsubsection*{\textbf{MFGs as Models for Systemic Risk}}
Being set up in continuous time, the above model is multi-period in nature. This is in sharp contrast with most of the existing mathematical models of systemic risk which are most often cast as static one-period models. Still, one of the most valuable feature of the model presented above is being \emph{explicitly solvable}. 

On the shortcoming side, I need to admit that this is a very naive model of bank lending and borrowing. Among its undesirable features, the model does not have any provision for borrowers to repay their debts !
Moreover, despite explicit solutions, it gives only a small jab at the stability properties of the system.

Still, the model raises interesting challenges and it seems reasonable that it can be made more realistic and more useful with some mathematically tractable add-ons. We cite three of them for the sake of illustration.

\begin{itemize}
\item The introduction of major and minor players in this model will allow to capture the crucial role played by the SIFIs discussed in the introduction to this section. Game models with major and minor players experienced a renewal of interest recently, and systemic risk seems to be a perfect testbed for their implementation.

\item In order to palliate some of the unrealistic features of the original model, Carmona, Fouque, Moussavi and Sun suggested the introduction of time delays in the controls in \cite{CarmonaFouque_delay}. While increasing the technicalities of the proofs and precluding explicit solutions,  this extension of the model includes provisions for borrowers to repay their debts in a fixed amount of time.

\item As another example of the fact that systemic risk is a fertile ground for the introduction and analysis of new MFG models, we mention the recent paper by Benazzoli, Ciampi and Di Persio \cite{BenazzoliCiampiDIPersio} in which the authors study a simple illiquid inter-bank market model, where the banks can change their reserves only at the jump times of some exogenous Poisson processes with a common constant intensity.

\item Also noteworthy is a recent paper \cite{NadtochyiShkolnikov} of Nadtochyi and Shkolnikov who study the mean field limit of systems of particles interacting through hitting times. Their model was motivated by the contagion of the times of default of financial institutions in periods of economic stress. It would be interesting to add the optimization component to their model and study the endogenously made decisions of the participants.

\item Finally, the introduction of graphical constraints should be a good way to quantify the various levels of exchanges between the financial institutions. Introducing a weighted graph of interactions between the players of the game changes dramatically the mean field nature of the model, and new solution approaches will have to be worked out for any significant progress to be made in this direction. Still, despite the obvious challenges of this research program, it is under active investigation by many financial engineers. 
\end{itemize}

%%%%%%%%%%%%%%%%%%%%%%%%%%%%%%%%%%%%%%%%%%%%%
\subsection{\textbf{Price Impact and Optimal Execution}}
%%%%%%%%%%%%%%%%%%%%%%%%%%%%%%%%%%%%%%%%%%%%%
High frequency markets offer another fertile ground for applications of financial engineering.
Among other things, they highlighted the importance of \emph{price impact} and \emph{optimal execution}.
The search for the best possible way to execute a given trade is an old problem and the presence of price impact did not have to wait for the popularity of the high frequency markets.

\vskip 12pt\noindent
\subsubsection*{\textbf{A Model for Price Impact. }}
 We briefly review an MFG model of price impact introduced by Carmona and Lacker in \cite{CarmonaLacker1}.
 There, it was solved in the weak formulation, but for the purpose of the present discussion the specific approach used to get to a solution will not really matter.
\vskip 4pt
We start with a model for $N$ traders.
We denote by $X^i_t$ the inventory (i.e. the number of shares owned) at time $t$ by trader $i$, and we assume that this inventory evolves as an It\^o process according to 
\begin{equation}
\label{fo:dX^i_t}
dX^i_t=\alpha^i_t\,dt+\sigma^i dW^i_t
\end{equation}
where $\alpha^i_t$ represents the rate of trading of trader $i$. This will be their control.
$\bW^i=(W^i_t)_{t\ge 0}$ are independent Wiener processes for $i=1,\ldots,N$, and
$\sigma^i$ represents an idiosyncratic volatility. We assume that it is independent of $i$ for simplicity. Note that in essentially all the papers on the subject this volatility is assumed to be  $0$. In other words, inventories are assumed to be differentiable in time. Our decision to work with $\sigma^i=\sigma>0$ is backed by empirical evidence, at least in the high frequency markets, as demonstrated by Carmona and Webster in \cite{CarmonaWebster_fs}.
Next, we denote by $K^i_t$ the amount of cash held by trader $i$ at time $t$. We have:
$$
dK^i_t=-[\alpha^i_t S_t +c(\alpha^i_t)]\, dt,
$$
where $S_t$ is the transaction price of one share at time $t$, and the function $\alpha \to c(\alpha)\ge 0$  models the cost for trading at rate $\alpha$. As explained in  \cite{CarmonaWebster_fs}, this function $c$ should be thought of as the Legendre transform of the shape of the order book. So  for a flat order book we should have:
$$
c(\alpha)=c\alpha^2.
$$
We model the time evolution of the price $S_t$ using the natural extension to the case of $N$ traders of the price impact formula of Almgren and Chriss \cite{AlmgrenChriss}:
$$
dS_t=\frac1N\sum_{i=1}^N h(\alpha^i_t)\, dt +\sigma_0 dW^0_t
$$
for some non-negative increasing function $\alpha\mapsto h(\alpha)$ and a Wiener process $\bW^0=(W^0_t)_{t\ge}$ independent of the other ones.
In this model, the wealth $V_t^i$ of trader $i$ at time $t$ is given by the sum of his holdings in cash and the value of his holdings in the stock, as marked at the current value of the stock:
$$
V^i_t=K^i_t+X^i_tS_t.
$$
Using the standard self-financing condition and It\^o's formula we see that:
\begin{equation}
\label{fo:dVit}
\begin{split}
 dV_t^i &= dK^i_t +X^i_t\,dS_t + S_t \,dX^i_t\\
 &=\bigg[-c(\alpha_t^i) +  X_t^i\frac{1}{N} \sum_{j=1}^N  h(\alpha_t^j)\bigg] dt +  \sigma S_tdW_t^i+ \sigma_0 X_t^i dW^0_t,
\end{split}
\end{equation}
so if player $i$ minimizes their expected trading costs 
$$
J^i(\balpha^1, ..., \balpha^N )= \EE \bigg[  \int_0^T c_X(X_t^i) dt + g(X_T^i) - V_T^i\bigg]
$$
where $x\mapsto c_X(x)$ represents the cost of holding an inventory of sixe $x$ and $g(x)$ a form of terminal inventory cost.
Using \eqref{fo:dVit}, we can rewrite these expected costs as;
$$
J^i(\balpha^1, ..., \balpha^N )= \EE \bigg[  \int_0^T f(t,X^i_t,\bar\nu^N_t,\alpha^i_t)dt + g(X^i_T)\bigg]
$$
if $\bar\nu^N_t$ denotes the empirical distribution of $\alpha^1_t$, $\ldots$, $\alpha^N_t$, and the function $f$ is defined by  $f(t,x,\nu,\alpha)=c(\alpha)+c_X(x)-x\int h \,d\nu$.

\begin{remark}
Several important remarks are in order at this stage.
\begin{enumerate}
\item The above model is the epitome of a $N$-player stochastic differential game.
\item The state equations are given by \eqref{fo:dX^i_t}. They are driven by the $N$ independent idiosyncratic noise term $dW^i_t$. The common noise $\bW^0$ disappeared from the optimization problem only because of the \emph{risk neutrality} of the traders, namely the fact that they minimize the expectations of their costs. Should they choose to minimize a nonlinear utility function, the common noise would not disappear !
\item Another remarkable property of this model, and one of the reasons its analysis was of great interest, is that it is one of the earliest MFG models for which the mean field interaction occurs naturally through the controls.
An early analysis of these MFGs within the probabilistic approach was given in \cite[Section 4.6]{CarmonaDelarue_book_I}.
\end{enumerate}
\end{remark}
While the model formulated above is for $N$ traders, the mean field formulation $N\to\infty$ is clear.
A complete solution of this limit MFG in the weak formulation can be found in 
\cite{CarmonaLacker1} and in \cite[Section 4.7]{CarmonaDelarue_book_I} for the strong formulation. 
Later on, it was revisited by Cardaliaguet and Lehalle in \cite{CardaliaguetLehalle} where the authors consider agents' possible heterogeneities, and the introduction of fictitious plays which gives a learning twist to the model. It was also extended by Cartea and Jaimungal in \cite{CarteaJaimungal},
to formulate an optimal execution problem for which the authors could still provide solution formulas in quasi explicit forms. Game models for optimal execution in the presence of price impact did not wait for the theory of MFGs to catch the interest of financial economists and financial engineers. The interested reader may want to check the papers 
\cite{BrunnermeierPedersen}, \cite{CarlinLoboViswanathan}, and \cite{CarmonaYang} for games models shedding light on predatory trading.

%%%%%%%%%%%%%%%%%%%%%%%%%%%%%%%%%%%%%%%%%%%%%
\subsection[Models for Bank Runs \& Mean Field Games of Timing]{\textbf{Models for Bank Runs \& Mean Field Games of Timing}}
%%%%%%%%%%%%%%%%%%%%%%%%%%%%%%%%%%%%%%%%%%%%%

I came across the potential application of Mean Field Games to the important problem of bank runs by attending Jean Charles Rochet's lectures at  the Vancouver Systemic Risk Summer School in  July 2014, and the talk given by Olivier Gossner during the conference following the summer school. Both works \cite{RochetVives, FongGossnerHornerSannikov} are reported in detail in \cite{CarmonaDelarue_book_I,CarmonaDelarue_book_II}. Here, we only review the second one because it fits better in the class of continuous time models on which we concentrate in this chapter.

In the spirit of the discussion to follow, it is worth mentioning the works of Morris and Shin \cite{MorrisShin1998} and He and Xiong \cite{HeXiong}.  Like most economists, they use a model of the economy with a continuum of players based on an atomless measure space. Our goal is to recover their models starting with \emph{finitely many players} and then, analyze the mean field limit.

%%%%%%%%%%%%%%%%%%%%%%%%%%%%%
\vskip 12pt\noindent
\subsubsection*{\textbf{A Continuous Time Model for Bank Runs}}

Following Gossner's talk mentioned earlier, we consider a group of $N$ depositors
with individual initial deposits in the amount  $D^i_0=1/N$ for $i=1,\ldots, N$.
They are promised a rate of return $\o r>r$ where $r$ is the current prevailing interest rate.
We assume that the value  $Y_t$ of the assets of the bank at time $t$ follows 
an It\^o process and that $Y_0\ge 1$. We also assume the existence of a deterministic function
$y\mapsto L(y)$ giving the liquidation value of the bank assets when $Y_t=y$.
One can imagine that the bank has a rate $\o r$  credit line of size $L(Y_t)$ at time $t$,
and that the bank uses this credit line each time a depositor runs (withdraws their deposit).

Next, we assume that the assets mature at time $T$, and that no transaction takes place after that.
If at that time $Y_T\ge 1$, every one is paid in full, but if $Y_T<1$  we treat this case as an \emph{exogenous default}. We talk about an \emph{endogenous default} at time $t$ if depositors try to withdraw more than $L(Y_t)$ at that time.

\vskip 2pt
As time passes, each depositor has access to a \emph{private signal} $X^i_t$ satisfying:
\begin{equation*}
dX^i_t = dY_t + \sigma dW^i_t,\qquad i=1,\ldots,N,
\end{equation*}
and at a time $\tau^i$ of their choosing, they can attempt to withdraw their deposit,
de facto collecting the return $\o r$ until time $\tau^i$,
and trying to maximize:
$$
J^i(\tau^{1},\ldots,\tau^{N})=\EE\bigg[g(\tau^i,Y_{\tau^i},\tau^{-i})\bigg]
$$
where we use the standard notation $\tau^{-i}=(\tau^1,\ldots,\tau^{i-1},\tau^{i+1},\ldots,\tau^N)$, and for example 
$$
g(t,Y_t,\tau^1,\ldots,\tau^N)=e^{(\o r-r)t\wedge\tau} + e^{-rt\wedge\tau}(L(Y_t)-N_t/N)^+\wedge\frac1N,
$$ 
$N_t=\#\{i;\; \tau^i\le t\}$ is number of withdrawals before $t$,
and 
$$
\tau=\inf\{t;\;L(Y_t)<N_t/N\}
$$
is the first time the bank cannot withstand the withdrawal requests. 

\vskip 6pt
First, let us try to derive some conclusions if the depositors had \emph{full information}, in which case
$Y_t$ would be public knowledge, and $\sigma$ would be $0$, i.e.$\sigma=0$.
If we also assume that the function $y\hookrightarrow L(y)$ is known to the depositors, then it is easy to check that
in any equilibrium:
$$
\tau^i=\inf\{t;\; L(Y_t)\le 1\}.
$$
So all the depositors withdraw at the \emph{same time} (they all run on the bank simultaneously)
and each depositor gets their deposit back: \emph{no one gets hurt!}.
Clearly this scenario is unfortunately, highly unrealistic. We should expect that depositors wait longer before running on the bank, presumably because they only have imperfect information (i.e. noisy private signals) on the health of the bank.

%%%%%%%%%%%%%%%%%%%%%%%%
\vskip 12pt\noindent
\subsubsection*{\textbf{Games of Timing}}
Let us consider a population of $N$ players with individual states $X^{N,i}_t$ at time $t$ satisfying equations of the form
$$
dX^{N,i}_t=b(t,X^{N,i}_t,\o\nu^N_t)dt+\sigma(t,X^{N,i}_t) dW^{i}_t,\qquad i=1,\ldots,N
$$
coupled through their empirical distribution
$$
\o\nu^N_t=\frac1N\sum_{i=1}^N\delta_{X^{N,i}_t}.
$$
Each player chooses a $\cF^{X^i}$ - stopping time $\tau^i$ and tries to maximize
$$
J^i(\tau^{1},\ldots,\tau^{N})=\EE\bigg[g(\tau^i,X_{\tau^i},\o\mu^N([0,\tau^i])\bigg]
$$
where $\o\mu=\frac1N\sum_{i=1}^N\delta_{\tau^i}$ is the empirical distribution of the $\tau^i$'s,
 $g(t,x,p)$ is the reward to a player for exercising their timing decision at time $t$ when
their private signal is $X^i_t=x$, 
and the proportion of players who already exercised their right is $p$. 
Taking formally the limit $N\to\infty$ in this set-up, we obtain the following MFG formulation of a mean field game of timing.

Assuming that the drift is independent of the empirical distribution of the states for the sake of simplicity,
i.e. $b(t,x,\nu)=b(t,x)$ the dynamics of the state of a generic player are given by an It\^o equation of the form:
\begin{equation*}
dX_t = b(t,X_t)dt + \sigma(t,X_t) dW_t.
\end{equation*}
We denote by $\FF^X=(\cF^X_t)_{0\le t\le T}$ the information available to the agent at time $t$,
and by $\cS^X$ the set of $\FF^X$-stopping times.
The MFG of timing paradigm can then be formulated as follows:
\begin{enumerate}
\item \emph{Best Response Optimization}: for each fixed environment $\bmu\in\cP([0,T])$ solve
$$
\hat\theta\in\text{arg}\sup_{\theta\in\cS^X, \theta\le T}\EE[g(\theta,X_\theta, \mu([0,\theta]))]
$$
\item \emph{Fixed Point Step}: find $\bmu$ so that
$$
\mu([0,t])=\PP[\hat\theta\le t]
$$
\end{enumerate}
Here and throughout, we denote by $\cP(A)$ the space of probability measures on $A$.
\vskip 12pt\noindent
\subsubsection*{\textbf{Existence with randomized stopping times}}
In an unpublished PhD thesis, Geoffrey Zhu proposed an existence proof for randomized stopping times, providing an analog of Nash's original existence theorem for the existence of equilibria in mixed strategies.
\vskip 4pt
Before we go any further, recall the sobering shortcoming of convergence in distribution which says that even if
$$
\begin{cases}
&\lim_{n\to\infty} (X,Y_n) = (X,Y) \quad\text{in law}\\
& Y_n\;\text{ is a function of }\; X
\end{cases}
$$
then $Y$ is not necessarily a function of $X$, in other words, $Y\in\sigma\{X\}$ may not hold.
\vskip 6pt
For the purpose of this existence proof, let us assume that the reward function $g:[0,T]\times\RR\times \cP([0,T]) \ni (t,x,\mu)\mapsto g(t,x,\mu)\in \RR$ is bounded,
continuous in $(t,x)$ for $\mu$ fixed, 
and Lipschitz continuous in $\mu$ for $(t,x)$ fixed. Note that, unfortunately, this last assumption is not satisfied for functions $t\hookrightarrow \mu([0,t])$, unless $t\in\TT\subset[0,T]$ with $\TT$ finite! This will prevent us from using this existence result in the model of bank run discussed earlier.
In any case, under the present assumptions
$$
\Pi:\cP([0,T])\times \cP(C([0,T]\times [0,T])) \mapsto \RR
$$
$$
(\mu,\xi) \hookrightarrow  \Pi(\nu,\xi)=\int_{C([0,T])\times [0,T]} g(t,x_t,\mu)\xi(dx,dt)
$$
is continuous, and since the space $\t{\cS}$ of randomized stopping times is compact because of an old result of Baxter and Chacon, Berge's maximum theorem implies that the multivalued function
$$
\cP([0,T])\ni\nu\hookrightarrow \text{arg}\sup_{\xi\in\t{\cS}}\Pi(\nu,\xi)
$$
is upper hemi-continuous and compact-valued. 
Followed by the projection on the first marginal, it is still upper hemi-continuous and compact-valued,  
and Kakutani's fixed point theorem implies the desired existence
result.

%%%%%%%%%%%%%%%%%%%%%%
\vskip 12pt\noindent
\subsubsection*{\textbf{Existence with usual stopping times}}
Existence for standard stopping times can be shown to hold under a different set of assumptions, using for example the order structure of the space of stopping times instead of topological properties of this space. For example, if we assume that the time increments of $g$ are monotone in $\nu$, then we can use the fact that the
space of stopping times is a complete lattice, check that
$$
\tau \hookrightarrow \text{arg}\sup_{\tau'\in\cS}\EE[g(\tau',X_{\tau'},F_\tau(\tau'))]
$$
is monotone, and use Tarski's fixed point theorem.
Here $F_\tau(t)=\PP[\tau\le t]$ is the cumulative distribution function of $\tau$.
\vskip 2pt
Unfortunately, once more, this existence result does not apply to the model of bank run discussed earlier.

%%%%%%%%%%%%%%%%%%%%
\vskip 12pt\noindent
\subsubsection*{\textbf{Solution in the general case (including common noise)}}
Beyond a simple example presented by M. Nutz in \cite{Nutz}, the solution in the general bank run set-up introduced earlier is much more difficult and technical than originally thought. A complete solution was given by Carmona, Delarue and Lacker in \cite{CarmonaDelarueLacker_timing}. See also  \cite{Bertucci} by C.  Bertucci for an approach relying purely on partial differential equations and quasi-variational inequalities, and \cite{BouveretDumitrescuTankov} by Bouveret, Dumitrescu and Tankov for more on the use of relaxed stopping times.

%%%%%%%%%%%%%%%%%%%%%%%%%%%
\vskip 12pt\noindent
\subsection{\textbf{Cryptocurrencies and Bitcoin Mining}}

Given the fact that the Bitcoin mania hits the financial markets on a recurrent basis, and because of the competitive nature of the mining process involving a large number or \emph{miners}, it is no surprise that mean field game models have been proposed to analyze some of the features of the cryptocurrency space. Here we briefly review two very recent papers by Li, Reppen and Sircar \cite{LRS} and Bertucci, Bertucci, Lasry and Lions \cite{BBLL} which use continuous time mean field games, though in very different ways, to analyze some of the features of cryptocurrency generation. Both papers envision a continuum of miners interacting through
the aggregate computational power they allocate to mining the blockchain.

\vskip 2pt
Even though it is not the only cryptocurrency, we shall only talk about Bitcoin because it is the one getting the most press. There are many reasons for that, wild price moves being definitely one of them.
After briefly exceeding \$12,000, it crashed down to the \$10,000 range, before quickly moving up again to high levels. 

The generation of bitcoins is based on blockchain technology. The latter was introduced for the purpose of record keeping in a decentralized ledger. Still, it is at the core of bitcoin generation.
In bitcoin production, independent miners compete for the right to record the next transaction block on the blockchain. They follow a Proof-of-Work (PoW) protocol. Their goal is to solve mathematical puzzles designed to be solved by brute force only. Computations to solve puzzles (create a block and earn a reward) are otherwise totally useless as they are not applicable anywhere else. Once a miner obtains a solution, the corresponding block is added on top of the blockchain and the miner obtains their reward. This reward is paid out in cryptocurrency (a fixed number of bitcoins, currently $12.50$ bitcoins for adding a block) while electricity and mining hardware need to be paid with traditional fiat currency (like the US Dollar).

The supply of bitcoins is constantly growing. However, it is limited to $21$ million, of which more than $17$ million are already in circulation.
The security of the network is a serious issue. A major fear is a \emph{majority attack} also called $51\%$ attack when a group of users controls the majority of mining power. These instances are rare, mostly because they are very difficult to realize due to their enormous costs. They are not  considered in \cite{LRS,BBLL}.

The computing power devoted to mining is called the hash rate. It captures the
number of trials per second trying to solve the mining puzzle. 
In order to maintain stability in the blockchain, the mining puzzle difficulty
is dynamically adjusted so that, on average, the time between the creation of two
consecutive blocks is constant, currently approximately $10$ minutes. 
Therefore as the hash rate increases, the difficulty
increases so that it is required to compute more hashes for a given block. 

Miners control their hash rate to increases their share of the blockchain reward, all other things being
equal, while reducing the share of the other miners. On the other hand,
intensive computations consume a lot of energy and each miner faces significant electricity costs. 
In a nutshell, this dilemma is the core of the individual miner optimization problem.
The aggregate hash rate in the system represents
the total computational power devoted to block creation. In both papers this aggregate hash rate will
be the source of the mean field interaction between the miners.

\vskip 4pt
In \cite{LRS}, Li, Reppen, and Sircar focus
on the risk borne by risk-averse miners and study mining concentration. 
They use a jump process to represent the acquisition of the reward, the jump intensity being the control of the typical miner. In their model, the jump intensity reflects the computer power, or hash rate, devoted to the effort, and the mean of the controls of the individual miners is what creates the mean field interaction in the model.
Given our simplistic description of how bitcoins are generated, it is natural to assume that the miner’s probability of receiving the next mining reward is proportional to the ratio of their hash rate to that of the population. The number of rewards each miner receives is modeled by a counting process $N_t$ with jump intensity $\lambda_t>0$. If the number of miners is $M+1$, this intensity is given by
$$
\lambda_t=\frac{\alpha_t}{D(\alpha_t+M\bar\alpha_t)}
$$
where $M\bar\alpha_t$ approximates the total hash rate of the other miners.
The wealth of the miner is used as state variable. It evolves as an It\^o process of the form
$$
dX_t=-c\alpha_t dt+PdN_t
$$
where $P$ is the bitcoin price, so the value of each reward is the product of its quantity by $P$.
As we mentioned earlier, $12.50$ bitcoins are granted to a miner for adding a block successfully. The total number of rewards in the system as a whole is a Poisson process with a constant intensity $D > 0$. This will play the role of a common noise in the model.

Given an adapted process $\bar{\balpha}=(\bar\alpha_t)_{t\ge 0}$ representing the conditional mean of the controls given the common noise, the miner optimization problem is to maximize the expected utility of wealth at a fixed terminal time $T$:
$$
v^{\bar\alpha}(t,x) = \sup_{\alpha\in[0,A(x)]}\EE\bigl[U(X_T)\,|\, X_t=x\bigr]
$$
where the controls $\alpha$ are restricted to the interval $[0,A(x)]$ when the state is $x$. The authors tackle the optimization problem by solving the HJB equation:
$$
\partial_tv^{\bar\alpha} + \sup_{\alpha\in[0,A(x)]}\Bigl(-c\alpha \partial_x v^{\bar\alpha} +\frac{\alpha}{D(\alpha+M\bar\alpha_t)}\Delta v^{\bar\alpha}\Bigr)=0
$$
and the solution is completed by solving the fixed point equation for the average effort rate $\bar{\balpha}$. The authors provide an explicit solution for exponential utility and no liquidity constraint. They go on to the analysis of the effect of liquidity constraints and more general utility functions. They provide robust numerical procedures to compute the equilibrium. In the case of the CRRA power utility function, they study the concentration of wealth among the miners. Their conclusion is that the richer the miner, the wealthier they will get.

\vskip 2pt
The authors also introduce a model in which one special miner is singled out for having a significant cost advantage (e.g. benefitting from cheaper electricity prices) over the remaining field of miners. Naturally this special miner is shown to contribute more to the hash rate.

%%%%%%%%%%%%%%%%%
\vskip 4pt
Bertucci, Bertucci, Lasry and Lions use a different modeling approach. Very much in the spirit of the work \cite{LachapelleLasryLehalleLions} proposing a mean field approach to the dynamics of an order book in the high frequency markets, they directly introduce the master equation, arguing that this is the best way to study mean field games. Notice that this is in sharp contrast with the usual approach starting from the introduction of the agents maximization problems.

Using their notation system, $P$ is now the \emph{nominal} hashrate, (number of hashes per second). They define the \emph{real} hashrate $K_t=e^{-\delta t}P_t$ arguing that the rate  $\delta$ quantifies technological progress. The evolution of $K_t$ is modeled in continuous time as miners continuously acquire computing power to compute hashes.
As we explained earlier, the blockchain outputs a fixed number of coins per unit of time, so the miners compete
against each other to earn a share of this fixed output. From the above description, it is reasonable to assume that the share they get is proportional to their relative share of the total computational power.
The authors posit dynamics of the real hash rate in the form:
$$
d K_t=-\delta K_t dt +\lambda U_t(K_t) dt,\qquad K_0=K
$$
where $U_t$ represents the flow of entry of computing devices, or in other words, the value of a unit of real hash rate. They introduce the relationship:
$$
0=-(r+\delta)U+(-\delta K +\lambda)\partial_K U +(K+\epsilon)^{-1}-c.
$$
This Partial Differential Equation (PDE) should be viewed as the master equation of the competitive mean field game with finitely many states. The players are the miners, and we should think of $K$ as a measure of an aggregate of the population of miners responsible for the mean field interaction.

\vskip 2pt
If we now assume that the reward is of the form $g(P_t)/(K_t+\epsilon)$
where $g$ is a smooth positive function of $P_t$ which evolves according to
$$
\begin{cases}
dP_t&=\alpha(P_t)dt +\sqrt{2\nu}dW_t, \qquad P_0=P\\
d K_t&=-\delta K_t dt +\lambda U_t(K_t,P_t) dt,\qquad K_0=K
\end{cases}
$$
where the function $\alpha$ is Lipschitz, $\nu>0$ and $\bW=(W_t)_{t\ge 0}$ is a standard Wiener process, then
$$
U(K,P)=\int_0^\infty e^{-(r+\delta)t}\Bigl(\frac{g(P_t)}{\epsilon+K_t}-c\Bigr) dt
$$ 
is the value function of one unit of real hashrate. In this model, $P_t$ captures the exchange rate between the value of the cryptocurrency and fiat money and
$\bW=(W_t)_{t\ge 0}$ is a form of common noise.
In this case the master equation on $[0,\infty)\times \RR$  reads:
$$
0=-(r+\delta)U+(-\delta K +\lambda U)\partial_K U +\alpha\partial_PU +\nu\partial^2_{PP}U+\frac{g(P)}{K+\epsilon}-c
$$
This is the master equation of a MFG with finite state space and common noise.
Again, the monotone structure of the MFG plays a key role in the well-posedness of these
models. Indeed, existence and uniqueness follow from monotonicity, and the existence of a stationary state is also proven and analyzed. This is proven in \cite{BBLL} when $g$ is bounded from above and below.

\vskip 2pt
\cite{BBLL} also discusses model security against attacks, proposes extensions to several competing populations of miners facing different mining costs, and a market where \emph{mining equipment} can be bought and sold.

%%%%%%%%%%%%%%%%%%%%%%%%%%%%%%%%%%%%%%%%%%%%%
\section[Game Models for Energy and the Environment]{\textbf{Games Models for Energy and the Environment}}
%%%%%%%%%%%%%%%%%%%%%%%%%%%%%%%%%%%%%%%%%%%%%
\label{se:energy}

In this section we review some of the MFG models which have been touted and used to revisit and extend earlier economic analyses of energy and environment markets. We first summarize the discussion given in \cite[Section 1.4.4]{CarmonaDelarue_book_I} of a first model proposed by Gu\'eant, Lasry and Lions in \cite{GueantLasryLions.pplnm}.

\vskip 12pt\noindent
\subsection{\textbf{MFG Models for Oil Production}}

If we denote by $x^1_0,\ldots, x^N_0$ the initial reserves of  $N$ oil producers who control their own rates of production, and if we denote by $X^i_t$ the oil reserves of producer $i$ at time $t$, the changes in reserves
should be given by equations of the form
\begin{equation}
\label{fo:dX^i_t}
dX^i_t=-\alpha^i_tdt+\sigma X^i_t dW^i_t
\end{equation}
where $\sigma>0$ is a volatility level common to all the producers,  the non-negative adapted and square integrable processes $\balpha^i=(\alpha^i_t)_{t\ge 0}$ being the controls exerted by the producers, and the  $\bW^i=(W^i_t)_{t\ge 0}$  independent standard Wiener processes. If we denote by $P_t$ the price of one barrel of oil at time $t$, and if we denote 
by $C(\alpha)=\frac b2 \alpha^2 + a\alpha$ the cost of producing $\alpha$ barrels of oil, then producer $i$ tries to maximize:
\begin{equation}
\label{fo:exhaustibleJi}
J^i(\balpha^1,\ldots,\balpha^N)=\sup_{(\alpha_t)_{t\ge 0}, \alpha_t\ge 0}\EE\bigg[\int_0^\infty e^{-rt}[\alpha^i_t P_t - C(\alpha^i_t)] dt\bigg],
\end{equation}
where $r>0$ is a discount factor. As we are about to explain, the price $P_t$ is the source of coupling between the producers' strategies. The notion of general equilibrium is intended to characterize situations in which all the producers manage to maximize their profits simultaneously, and \emph{the market clears} in the sense that demand matches supply. Let us denote by $D(t,p)$ the demand at time $t$ when the price is $p$. The function $D(t,p)=we^{\rho t}p^{-\gamma}$ was used in  \cite{GueantLasryLions.pplnm}. We use the obvious notation $D^{-1}$  for the inverse demand function, i.e. $q=D(t,p)\Longleftrightarrow p=D^{-1}(t,q)$.

%%%%%%%%%%%%%%%%%
\vskip 12pt\noindent
\subsubsection*{\textbf{Mean Field Formulation. }}
In the present context, the MFG paradigm can be articulated easily.
For each fixed deterministic flow $(\mu_t)_{t\ge 0}$ of probability measures, we compute the price $P_t$ from the formula:
$$
P_t=D^{-1}\bigg(t,-\frac{d}{dt}\int x\mu_t(dx)\bigg),
$$
and the best response to this flow of distributions is 
given by the solution of the discounted infinite horizon optimal control problem for the instantaneous cost function
$$
f(t,x,\mu,\alpha)=[\alpha p - C(\alpha)]e^{-rt}
$$
under the dynamic constraint \eqref{fo:dX^i_t}. The MFG will be solved if one can find a measure flow $(\mu_t)_{t\ge 0}$ such that the marginal distribution $\cL(X_t)$ of the solution of the control problem matches the flow we started from, namely $\mu_t=\cL(X_t)$ for all $t\ge 0$. The analytic approach to MFGs based on the solution of a system of coupled HJB and Fokker-Planck-Kolmogorov equations is used in \cite{GueantLasryLions.pplnm} to give a solution to this problem. Numerical illustrations provide comparative statics of the solutions are also given.

%%%%%%%%%%%%%%%%%%
\vskip 12pt\noindent
\subsubsection*{\textbf{Variations and Extensions}}

In \cite{ChanSircar}, Chan and Sircar propose to look at dynamics
\begin{equation}
\label{fo:csdynamics}
dX_s=-\alpha_s ds +\sigma dW_s,\qquad X_t=x>0.
\end{equation}
with a Dirichlet boundary condition at $X_t=0$ to guarantee that the reserves of a generic oil producer do not become negative.  As before, $\alpha_t$ represents the rate of production of a generic producer and $X_t$ represents the remaining reserves. As in most models for Cournot games, 
 the price experienced by the producer, call it $P_t$ for the sake of definiteness,  is given by a linear inverse demand function of the rates of production, and it is chosen to be of the form
$$
P_t=1-\alpha_t -\epsilon\bar\alpha_t
$$
where $\bar\alpha_t$ is the mean production rate for all the exhaustible resources. so that the cost function becomes
$$
f(t,x,\mu,\alpha)=\alpha[1-\alpha p - \epsilon \o\alpha]
$$
where $\mu$ denotes the distribution of the control $\alpha$ giving the rate of production, $\bar\alpha$ the mean of this distribution, and $p$ the price given by the above inverse demand function. This is a typical extended MFG (because the mean field interaction is through the controls) with a boundary condition to guarantee that the remaining reserves do not become negative.
In parallel, the authors propose a slightly modified model for producers of renewable energy and analyze the oil market in the presence of both populations of producers.
In this paper, they also propose several variations on the above model. Their goal was to include several realistic features like strategic blockading the entrance of renewable producers, and exploration and discovery of new reserves. While not always worrying about all the subtleties of mathematical existence theorems, they provide enlightening numerical illustrations of their conclusions. This prompted more mathematically inclined 
authors like Bensoussan and Graber to pursue in \cite{GraberBensoussan} a complete existence analysis based on partial differential equations techniques of the models proposed by Sircar and Chan.

\vskip 4pt
For the sake of completeness, we mention that plain models for a macro perspective on the behavior of a large population of oil producers  were proposed by Giraud, Gu\'eant, Lasry, and Lions. See for example \cite{GiraudGueantLasryLions,FDD09,GueantLasryLions.pplnm}. More recently, Achdou, Giraud, Lasry and Lions revisited some of these models including the presence of major and minor players. See \cite{AchdouGiraudLasryLions}. Also, note that game theoretical approaches, though not involving mean field games per se, were used by Ludkovski and Sircar in \cite{LudkovskiSircar} to analyze oil production.

%%%%%%%%%%%%%%%%%%%
\vskip 12pt\noindent
\subsection{\textbf{MFG Models for the Electricity Markets and the Grid}}
General equilibrium models have a long history in the engineering literature on electricity pricing. See for example \cite{HinzWarsaw} and the references therein.  More recently, Djehiche, Barreiro-Gomez and Tembine proposed a mean field game model for pricing electricity in a smart grid. See \cite{DjehicheBarreiro-GomezTembine}. Still, to model individual decision in a smart 
grid,  Alasseur, Ben Tahar and Matoussi use in \cite{AlasseurBenTaharMatoussi} a game with mean field interactions through the controls as a framework to manage storage. 
In \cite{AidDumitrescuTankov}, A\"{i}d, Dumitrescu, and Tankov use one of the mean field game of timing models reviewed earlier to capture the time at which renewable producers choose to enter a new market, and when conventional producers using fossil fuels should exit the market. In a different context, A\"{i}d, Basei and Pham investigate in \cite{AidBaseiPham} a Stackelberg game model where the leader (an electricity producing firm) and the follower (consumer) choose strategies possibly depending upon their distributions. So they solve optimal control problems of the McKean-Vlasov type. The main emphasis of the paper is to show that the Stackelberg equilibrium is not Pareto optimal, and to explain the economic consequences of this disparity.

Investigating the valuation of demand response contracts in a model with
a continuum of consumers with mean field interactions and the presence of a common noise impacting their consumptions, Elie, Hubert, Mastrolia, and Possama\"{i}  formulate in \cite{Elie_et_al} the problem as a contract theory problem with moral hazard as those we shall discuss in more detail in Section \ref{se:contract} below. In their model, the Principal is an electricity
producer who observes continuously the consumption of a continuum of risk averse consumers, and designs contracts
in order to reduce the production costs. To be more specific, the producer incentivizes the consumers to reduce the average consumptions as well as their volatilities in different regimes, without observing the efforts they potentially make. This is exactly the type of models we shall investigate in Section \ref{se:contract}.

The recent paper by Shrivats, Firoozi, and Jaimungal \cite{ShrivatsFirooziJaimungal}, still in the context of the electricity markets, offers a smooth transition with the next discussion of the environment markets. Indeed, it proposes MFG models to derive the optimal behavior of electricity producers and an equilibrium price for Solar Renewable Energy Certificate (SRECs) in market-based systems designed to incentivize solar energy generation.

\vskip 12pt\noindent
\subsection{\textbf{Environment Economics}}
Early general equilibrium models aimed at understanding the effects and the control of externalities and taxes (in the spirit of Tobin taxes) were proposed by Golosov, Hassler, Krusell, and Tsyvinski in  \cite{GolosovHasslerKrusellTsyvinski}. General equilibrium models were also used in early works on the emissions markets by Bueler \cite{bueler} and Haurie \cite{Haurie} and more significantly in the analysis of the European Union Emission Trading System (EU ETS) by Carmona, Fehr, Hinz and Porchet in \cite{CFHP}. See also the references therein.
We argue later on in Section \ref{se:macro} that many general equilibrium models can be recast as Mean Field Game models. More recently, ideas which first appeared in the treatment of MFGs were used by Bahn, Haurie and Malham\'e to model negotiations related to environment policies. See  \cite{BahnHaurieMalhame}.

\vskip 2pt
Finally, we mention the recent work of Carmona, Dayanikli and Lauri\`ere \cite{CarmonaDayanikliLauriere} who use MFG models with major and minor players, very much in the spirit of the contract theory models we review in Section \ref{se:contract}, to derive equilibrium analyses of externalities and regulation on one end, and investments in renewables on the other, when dealing with  electricity production.

%%%%%%%%%%%%%%%%%%%%%%%%%%%%%%%%%%%%%%%%%%%%%
\section[Macro-Economic Growth Models]{\textbf{Macro-Economic Growth Models}}
%%%%%%%%%%%%%%%%%%%%%%%%%%%%%%%%%%%%%%%%%%%%%
\label{se:macro}

In this section, we review several  general equilibrium economic growth models.  
We borrowed the first one from a paper by Gu\'eant, Lasry and Lions on mean field games \cite{GueantLasryLions.pplnm}. We chose to present it here because, by cleverly adapting ideas from a model of Aghion and Howitt, the authors present a model with common noise which can be solved explicitly, all the way to the master equation. 

\vskip 2pt
Our choice of the second model was driven by a remarkable property: even though the original contribution \cite{KrusellSmith} of Krusell and Smith appeared long before the mean field game paradigm was articulated, the numerical algorithm proposed by the authors to approximate numerically the equilibrium characteristics, reads as if it had been designed for the computation of an MFG equilibrium. Indeed, it is eerie to see how closely the description of their numerical algorithm mimics, step by step, the mean field game strategy based on the alternate iteration of steps to approach the solution of the HJB equation and the Fokker-Planck Kolmogorov equation.

\vskip 2pt
We learned of the third example presented in this section, from a private conversation with Benjamin Moll. We include it in this review because it can be solved completely, both analytically and numerically. Unfortunately, those examples are few and far between.

\vskip 2pt
The interested reader may also want to consult \cite{AchdouHanLLMoll} by Achdou, Han, Lasry, Lions and Moll for another discussion of continuous time macro-economic models recast as MFGs.

%%%%%%%%%%%%%%%%%%%%%%%%%%%%%%%%%
\subsection[A First Example based on Calculus with Pareto Distributions]{\textbf{A First Example based on Calculus with Pareto Distributions}}
\label{sub:macro1}
We introduce directly the mean field formulation of the game without starting from the definition of the finite player game because the interaction between the players is \emph{local} in the sense that it is a function of the density of the statistical distribution of the states of the players. In the case of finitely many players, this distribution is the empirical distribution of the finitely many states and as such, it does not have a density per se. So in order to avoid the introduction of smoothing of the empirical measures to define the costs to the players, we jump directly to the mean field game formulation which can be done directly with densities without any need for mollification arguments.

\vskip 2pt
In this model, there are no idiosyncratic shocks, just aggregate shocks common to all the players. They are given by the increments of a one dimensional Wiener process $\bW^0=(W^0_t)_{ t\ge 0}$. 
We denote by $\FF^0=(\cF^0_t)_{t\ge 0}$ its filtration. We also assume that the volatility of the state of a generic player is linear, that is $\sigma^0(x)=\sigma x$ for some positive constant $\sigma$, and that each player controls the drift of their state so that the dynamics of their state read:
\begin{equation}
\label{SDE:pareto}
dX_t=\alpha_t dt+\sigma X_t dW_t^0.
\end{equation}
We shall restrict ourselves to Markovian controls of the form $\alpha_t=\alpha(t,X_t)$ for a deterministic function 
$(t,x)\mapsto \alpha(t,x)$, which will be assumed to be non-negative and Lipschitz in the variable $x$. Under these conditions, $X_{t} \geq 0$ at all times $t>0$ if $X_{0} \geq 0$.
Note that if $X_t$ and $\tilde X_t$ are solutions of \eqref{SDE:pareto} for the same linear control 
$
\alpha(t,x)=\gamma_{t} x
$ 
for some continuous path $[0,T] \ni t \mapsto \gamma_{t} \in [0,+\infty)$, with initial conditions $X_0\le \tilde X_0$,
then
\begin{equation}
\label{fo:2pts}
\tilde X_t=X_t+(\tilde X_0-X_0)e^{\int_{0}^t \gamma_{s} ds - (\sigma^2/2) t+\sigma W_t^0}.
\end{equation}
We assume that $k>0$ is a fixed parameter, and we introduce a special notation for the family of one-sided scaled Pareto distributions with decay parameter $k$.
For any real number $q>0$, we denote by $\mu^{(q)}$ the one-sided Pareto distribution on the interval $[q,\infty)$:
\begin{equation}
\label{fo:Pareto}
\mu^{(q)}(dx)=k\frac{q^k}{x^{k+1}}\bone_{[q,\infty)}(x)dx.
\end{equation}
Notice that for any random variable $X$, $X\sim \mu^{(1)}$ is equivalent to $qX\sim\mu^{(q)}$. 

\vskip 4pt
For each $t\ge 0$ we define $\mu_t(dx)=\PP[X_t\in dx|\cF_t^0]$. The flow $(\mu_t)_{t\ge 0}$ of probability measures is adapted to the filtration $\FF^0$ of the common noise. Recall that the MFG paradigm in the presence of a common noise is to solve, for each fixed $\FF^0$-adapted flow of probability measures $(\mu_t)_{t\ge 0}$, the optimization problem of a generic player, and then solve the fixed point problem to guarantee that the flow $(\mu_t)_{t\ge 0}$ we started from is in fact the flow of conditional marginal laws of the solution of the optimization problem.

\vskip 2pt
For this particular family of distributions, if $\mu_0=\mu^{(1)}$, then $\mu_t=\mu^{(q_t)}$ 
where $q_t=e^{\int_{0}^t \gamma_{s} ds - (\sigma^2/2) t+\sigma W_t^0}$. In other words, conditioned on the history of the common noise, the distribution of the states of the players remains Pareto with parameter $k$ if it starts that way, and the left-hand point of the distribution $q_t$ can be understood as a sufficient statistic 
characterizing the distribution $\mu_t$. So if $X_0\sim\mu^{(1)}$, then $\mu_t\sim\mu^{(q_t)}$. This simple remark provides an explicit formula for the time evolution of the (conditional) marginal distributions of the states given the common noise. In general MFGs with a common noise, this time evolution is difficult to determine as it requires the solution of a forward Stochastic Partial Differential Equation  (SPDE for short).

\vskip 2pt
Using the same notation as in \cite{GueantLasryLions.pplnm}, we define the running cost function $f$ by
$$
f(x,\mu,\alpha)=  c \frac{x^a}{[(d\mu/dx)(x)]^b}-\frac{E}{p}\frac{\alpha^p}{[\mu([x,\infty))]^b},
$$
for positive constants $a$, $b$, $c$, $E$ and $p >1$. The economic rationale for the form of this cost function and the meanings of the parameters are discussed in \cite{GueantLasryLions.pplnm}. By convention, the density appearing in this formula is the density of the absolutely continuous part of the Lebesgue's decomposition of the measure $\mu$, and it is set to $0$ when the measure is singular. The argument of the optimization of the Hamiltonian is given by
$$
\hat\alpha(x,\mu,y)=\Bigl(\frac{y}{E}\bigl[\mu([x,\infty))\bigr]^b\Bigr)^{1/(p-1)}.
$$
This formula can be used to write the master equation which, when restricted to one-sided Pareto distributions, can be reduced to a finite dimensional PDE because of the above remark. Accordingly, Nash equilibria can be identified in this family of Pareto distributions. The details, far too technical for this review, can be found in Section 4.5.2 of \cite{CarmonaDelarue_book_II}.

%%%%%%%%%%%%%%%%%%%%%%%%%%%%
\subsection[The Krusell - Smith's Growth Model]{\textbf{The Krusell - Smith's Growth Model}}
\label{sub:ks}

\vskip 2pt
One major difference with the growth model discussed in the previous subsection is the fact that, on the top of the common noise affecting all the states, we also have idiosyncratic random shocks specific to each individual agent in the economy. In \cite{KrusellSmith} the shocks take only finitely many values. We suspect that this restrictive assumption was made for the purpose of numerical implementation.
In the next subsection, we change the nature of the random shocks by introducing Wiener processes to recast the model in the framework of stochastic differential games.

\vskip 12pt\noindent
\subsubsection*{\textbf{Description of the Economy}}
While economists usually work with models comprising a continuum of players (this is indeed the case in \cite{KrusellSmith}), in order to avoid the discussion of measurability issues related to continuum families of independent random variables, we first discuss the model of an economy comprising $N$ consumers. The random shocks are given by a set of $N$ continuous time Markov chains $(z_t,\eta^i_t)_{t\ge 0}$ for $i=1,\ldots, N$. The common component $z$ captures the health of the overall economy, like an aggregate productivity measure, so for some constant $\Delta_z\ge 0$, $z_t=1+\Delta_z$ in good times, and $z_t=1-\Delta_z$ in bad times. The idiosyncratic component $\eta$ is specific to the consumer, $\eta^i_t=1$ when consumer $i$ is employed, and $\eta^i_t=0$ whenever they are unemployed. 
$\Delta_z=0$ corresponds to the absence of common noise.

The production technology is modeled by a Cobb - Douglas production function 
in the sense that the per-capita output is given by
\begin{equation}
\label{fo:cobb_douglas}
Y_t=z_t K^\alpha_t(\o\ell L_t)^{1-\alpha}
\end{equation}
where $K_t$ and $L_t$ stand for per-capita capital and employment rates respectively.  The constant $\o\ell$ can be interpreted as the number of units of labor produced by an employed individual. The power $\alpha\in(0,1)$ is a constant of the model. In such a model, two quantities play an important role: the capital rent $r_t$ and the wage rate $w_t$.  Economic theory says that in equilibrium, these marginal rates are defined as the partial derivatives of the per-capita output $Y_t$ with respect to capital and employment rate respectively. So
\begin{equation}
\label{fo:r_t}
r_t=r(K_t,L_t,z_t)=\alpha z_t\big(\frac{K_t}{L_t}\big)^{\alpha -1}
\end{equation}
and 
\begin{equation}
\label{fo:w_t}
w_t=w(K_t,L_t,z_t)=(1-\alpha) z_t\big(\frac{K_t}{L_t}\big)^\alpha.
\end{equation}

\vskip 12pt\noindent
\subsubsection*{\textbf{Consumer's Optimization Problem}}
Consumers control their capital consumption rate $c^i_t$ at time $t$, and maximize their expected utilities of  consumption
$$
\EE\bigg[\int_0^\infty e^{-\rho t} u(c^i_t)dt\bigg],
$$
for some discount factor $\rho >0$. We use the power utility function  
\begin{equation}
\label{fo:CRRA}
u(c)=\frac{c^{1-\gamma}-1}{1-\gamma}
\end{equation}
for some $\gamma\in(0,1)$, also known as CRRA (short for Constant Relative Risk Aversion) utility function. 

\vskip 4pt
Consumers must choose their consumptions while making sure that their individual capitals $k^i_t$ remain non-negative at all times. The individual capitals evolve according to the equation
$$
d k^i_t=\big[ (r_t-\delta)k^i_t + [(1-\tau_t)\o\ell \eta^i_t + \mu(1-\eta^i_t)]w_t\big] dt - c^i_t dt.
$$
Here, the constant $\delta>0$ represents a depreciation rate. The second term in the above right hand side represents the wages earned by the consumer.
It is equal to $\mu w_t$ when the consumer is unemployed, quantity which should be understood as an unemployment benefit rate. On the other hand, it is equal  to $(1-\tau_t)\o\ell w_t$  after adjustment for taxes, when they are employed. Here 
$$
\tau_t=\frac{\mu u_t}{\o\ell L_t}
$$
where $u_t=1-L_t$ is the unemployment rate.

\vskip 12pt\noindent
\subsubsection*{\textbf{MFG Formulation}}
By de Finetti's law of large numbers, we expect that the empirical measures $\o\mu^{k,N}_t$ of capital and $\o\mu^{\eta,N}_t$ of labor:
$$
\o\mu^{k,N}_t=\frac1N\sum_{i=1}^N\delta_{k^i_t}, 
\qquad\text{and}\qquad
\o\mu^{\eta,N}_t=\frac1N\sum_{i=1}^N\delta_{\eta^i_t}, 
$$
converge as $N\to\infty$ toward a limit which we denote by $\mu^{k,z}_t$ and $\mu^{\eta,z}_t$.
These limits give the conditional distributions of capital and labor
$k_t$ and $\eta_t$ given the state $z_t=z$ of the economy. Since $z_t$ can only take two values $1-\Delta_z$ and $1+\Delta_z$, we only need the knowledge of deterministic flows of measures, $(\mu^{k,d}_t)_{t\ge 0}$, $(\mu^{k,u}_t)_{t\ge 0}$, $(\mu^{\eta,d}_t)_{t\ge 0}$, and $(\mu^{\eta,u}_t)_{t\ge )}$ corresponding to the two values of $z$, say down and up, namely  $d=1-\Delta_z$ and $u=1+\Delta_z$.

Once the flows of conditional measures are known, the computation of the best response  of a representative agent reduces to the solution of the optimal control problem
 $$
\max_c\EE\bigg[\int_0^\infty e^{-\rho t} u(c_t)dt\bigg]
$$
under the constraints $k_t\ge 0$ and 
$$
d k_t=\big[ (r(K_t,L_t,z_t)-\delta)k_t + [(1-\tau_t)\o\ell \eta_t + \mu(1-\eta_t)]w(K_t,L_t,z_t)\big] dt - c_t dt.
$$
Here, $(z_t,\eta_t)_{t\ge 0}$ is a continuous time Markov chain with the same law as any of the $(z_t,\eta^i_t)_{t\ge 0}$ introduced earlier, the rental rate function $r$ and the wage level function $w$ are as in \eqref{fo:r_t} and \eqref{fo:w_t}, and $K_t=\o k_t^{z_t}$ is the mean of the conditional measure $\mu^{z_t}_t$, namely
$$
K_t=\int_{[0,\infty)} k\mu^{k,u}_t(dk)\;\;\text{if } z_t=1+\Delta_z,\quad\text{and}\quad K_t=\int_{[0,\infty)} k\mu^{k,d}_t(dk)\;\;\text{if } z_t=1-\Delta_z,
$$
and where the aggregate labor $L_t$ is defined similarly as the conditional mean of $\mu_t^{\eta,z}$. The aggregates $K_t$ and $L_t$ are the conditional means of the capital and labor given the common noise: they carry the mean field interactions in the model.

\vskip 4pt
As we recalled during our discussion of the previous example,  the MFG paradigm in the presence of a common noise is to solve, for each fixed flow of probability measures adapted to the filtration of the common noise, the optimization problem of a generic player, and then solve the fixed point problem to guarantee that the flow we started from is in fact the flow of conditional marginal laws of the solution of the optimization problem.
Note that in the Krusell-Smith's model, the common noise and the idiosyncratic noise are correlated and that the labor state variable $\eta_t$ (whose aggregate is $L_t$) is nothing but the idiosyncratic noise. So clearly, the MFG paradigm reduces to the solution of the individual optimization problem given the aggregate $K_t$ and then solving for the fixed point. This is exactly what the numerical algorithm proposed in \cite{KrusellSmith} does. Time discretization is not needed in \cite{KrusellSmith} since the model is introduced in discrete time there. Then a form of dynamic programming is used to solve the optimization problem given the aggregate $K_t$, and then an update of $K_t$ is done by Monte Carlo simulation, before going back to the solution of the optimization problem given the update of $K_t$, and so on and so forth. While the authors realize that the entire distribution $\mu_t^{k,z}$ should be updated, they argue that updating the mean is sufficient to get reasonable numerical results for a problem whose complexity should have been prohibitive. As explained in the introduction, even though they never used the term Nash equilibrium, their numerical search for a \emph{recursive competitive equilibrium} is exactly the algorithm based on the iteration of the numerical approximation of the solution of the HJB equation followed by the Fokker-Planck-Kolmogorov equation, algorithm (re)introduced and used over 15 years later for the numerical solution of Mean Field Games.

%%%%%%%%%%%%%%%%%%%%%%%%%
\subsection[A Diffusion Form of Aiyagari's  Growth Model]{\textbf{A Diffusion Form of Aiyagari's  Growth Model}}
\label{sub:Aiyagari}

As explained in the introduction to this section, we learned about the model presented in this subsection from a private conversation with Benjamin Moll.
It is one of the models discussed in the review \cite{AchdouBueraLLMoll}
by Achdou, Buera, Lasry, Lions and Moll devoted to partial differential equation models in macroeconomics.
As far as we know, the first, and most likely the only, complete mathematical solution as a mean field game of this model can be found in Chapter 3 of \cite{CarmonaDelarue_book_I}.

\vskip 2pt
We first describe the finite player form of the model. We shall solve it as a mean field game model later on.
The $N$ agents $i\in\{1,\ldots,N\}$ are the workers comprising the economy. The private state at time $t$ of agent $i$ is a two-dimensional vector $X^i_t=(Z^i_t,A^i_t)$. For the purpose of this model,  $A^i_t$ is the wealth at time $t$ of worker $i$, and $Z^i_t$ their labor productivity. The time evolutions of the states are given by stochastic differential equations
\begin{equation}
\label{fo:a_dynamics}
\begin{cases}
dZ^i_t&=\mu_Z(Z^i_t)dt\, + \, \sigma_Z(Z^i_t) dW^i_t\\
dA^i_t&=[w^i_tZ^i_t \, + \, r_t A^i_t \,-\,c^i_t] dt.
\end{cases}
\end{equation}
The functions $\mu_{Z},\sigma_{Z}: \RR \rightarrow \RR$ are known. We shall specify them later on in the examples we treat theoretically and numerically. The random shocks are given by $N$ independent Wiener processes $\bW^i=(W^i_t)_{t\ge 0}$, for $i=1,\ldots, N$. $r_t$ is the interest rate at time $t$, $w^i_t$ represents the wages of worker $i$ at time $t$ and the consumption process ${\boldsymbol c}^i=(c^i_t)_{t\ge 0}$ is the control of player $i$.

\begin{remark}
\label{re:a_1}
In many economic applications, a borrowing limit is imposed. Mathematically, this means that the wealths must satisfy the constraints $A^i_t\ge \u a\;$ for some nonpositive constant $\u a\le 0$. Moreover, the labor productivity processes $\bZ^i=(Z^i_t)_{t\ge 0}$ are also restricted by requiring that they are ergodic, or even restricted to an interval $[\u z,\o z]$ for some finite constants $0\le \u z <\o z<\infty$. We do not know of an economic rationale for these constraints and we suspect that these assumptions are made for the sole benefits of the technical proofs.
\end{remark}

In this model, given adapted processes ${\boldsymbol r}=(r_t)_{t\ge 0}$ and ${\boldsymbol w}^i=(w^i_t)_{t\ge 0}$ for $i=1,\ldots,N$, the workers choose their consumptions $\bc^1,\ldots,\bc^N$ in order to maximize their expected discounted utilities:
\begin{equation}
\label{fo:Ji}
J^i({\boldsymbol c}^1,\ldots,{\boldsymbol c}^N)=\EE\int_0^\infty e^{-\rho t}u(c^i_t)dt.
\end{equation}
As usual in economic applications, the model is set up in infinite horizon, and $u$ is an increasing concave utility function, the same for all the workers. So far, it seems like the workers do not interact.
Also, we need to explain how the interest rate and the wage processes appear in equilibrium. As in the Krusell-Smith model discussed earlier, we assume that the aggregate production in the economy is given by a production function $Y=F(K,L)$, the total capital supplied in the economy at time $t$, say $K_t$ being given by the aggregate wealth
\begin{equation}
\label{fo:K_t}
K_t=\int a \;d\o\mu^N_{X_t}(dz,da)=\frac1N\sum_{i=1}^N A^i_t
\end{equation}
while the total amount of labor $L_t$ supplied in the economy at time $t$ is normalized to $1$. Here, we denote by $\o\mu^N_{X_t}$ the empirical measure of the sample $X^1_t,\ldots,X^N_t$. Note that only the $A$-marginal enters the definition of $K_t$.

\begin{remark}
As explained in the introduction, the fact that the economic agents interact through aggregate quantities is the reason why mean field models and mean field game formulations are so natural for these macro-economic models.
\end{remark}

Economic theory says that in a competitive equilibrium, the interest rate and the wages are given by the partial derivatives of the production function
$$
\begin{cases}
r_t &= [\partial_KF](K_t,L_t)|_{L_t=1} -\delta\\
w_t &=  [\partial_LF](K_t,L_t)|_{L_t=1}
\end{cases}
$$
where $\delta\ge 0$ is the rate of capital depreciation. So in equilibrium, the interaction between the agents in the economy is through the mean $K_t=\int a\o\mu^N_X(a,z)$ of the empirical distribution of the workers' wealths $A^i_t$. 

\vskip 12pt\noindent
\subsubsection*{\textbf{Practical Solution}} 
We now specify the model further to solve it as a mean field game.
We use the CRRA isoelastic utility function with constant relative risk aversion introduced above in \eqref{fo:CRRA}.
Note that
\begin{equation}
\label{fo:u'}
u'(c)=c^{-\gamma}
\qquad\text{and}\qquad
(u')^{-1}(y)=y^{-1/\gamma}.
\end{equation}
Next, we  use the Cobb - Douglas production function
\begin{equation}
\label{fo:cd}
F(K,L)=A\,K^\alpha\, L^{1-\alpha}
\end{equation} 
for some constants $A>0$ and $\alpha\in (0,1)$. With this choice, in equilibrium,
$$
r_t=\alpha A K_t^{\alpha-1}L_t^{1-\alpha} - \delta
\qquad\text{and}\qquad
w_t=(1-\alpha) A K_t^\alpha L_t^{-\alpha}
$$
and since we normalized the aggregate supply of labor to $1$,
\begin{equation}
\label{fo:cd_equilibrium}
r_t=\frac{\alpha A}{ K_t^{1-\alpha}} - \delta
\qquad\text{and}\qquad
w_t=(1-\alpha) A K_t^\alpha,
\end{equation}
where $K_t$ is given by \eqref{fo:K_t} and provides the mean field interaction.
Finally, we use an Ornstein-Uhlenbeck process for the mean reverting labor productivity process $\bZ=(Z_t)_{t\ge 0}$ 
by choosing $\mu_{Z}(z) = 1-z$ and $\sigma_{Z} \equiv 1$ for the sake of definiteness. Moving to the mean field game formulation of the model, the state $X_t=(A_t,Z_t)$ evolves according to
\begin{equation*}
\begin{cases}
&dZ_{t} = -(Z_{t}-1)\, dt + dW_{t},
\\
&dA_{t} = \bigl[ (1-\alpha) \bar{\mu}_{t}^\alpha Z_{t}
+ \bigl( \alpha \bar{\mu}_{t}^{\alpha-1}
-\delta \bigr)
A_{t} - c_{t}\bigr] dt, \quad t \in [0,T],
\end{cases}
\end{equation*}
where $(\bar{\mu}_{t})_{0 \leq t \leq T}$ denotes the
flow of average wealths in the population in 
equilibrium. It is assumed to take (strictly) positive values. The set $\AA$ of admissible controls is the set $\HH^{2,1}_+$ of real valued square-integrable  $\FF$-adapted processes ${\boldsymbol c}=(c_{t})_{0 \leq t \leq T}$
with non-negative values, and the cost functional is 
defined by: 
\begin{equation*}
J({\boldsymbol c}) = \EE \biggl[ 
\int_{0}^T (-u)(c_{t}) dt - \tilde{u}(A_{T}) \biggr],
\end{equation*}
for the CRRA utility function $u$ given by \eqref{fo:CRRA} and $\tilde{u}(a)=a$. Notice the additional minus signs due to the fact that we want to treat the optimization problem as a minimization problem. Here we chose to take $0$ for the discount rate since we are working on a finite horizon.
Throughout the analysis, we shall assume that $A_{0}>0$ and $Z_{0}=1$, so that 
$\EE[Z_{t}]=1$ for any $t \geq 0$. 

\vskip 4pt
In order to solve this MFG using the Pontryagin Maximum Principle, we introduce the Hamiltonian:
$$
H(t, z,a,\mu,y_z,y_a,c)=(1-z)y_z+\big(-c +(1-\alpha)\o\mu_t^\alpha z+ (\alpha \o\mu^{\alpha-1} -\delta) a \big) y_a - u(c),
$$
where $\o\mu=\int a\;d\mu(z,a)$ denotes the mean of the second marginal of the measure $\mu$. 
The first adjoint equation reads
$$
dY_{z,t}=-\partial_z H(t,Z_t,A_t,Y_{z,t},Y_{a,t},c_t)dt +\t Z_{z,t} dW_t=Y_{z,t}dt+\t Z_{z,t} dW_t.
$$
Its solution is $Y_{z,t}=0$ because its terminal condition is $Y_{z,T}=0$. Since the variables $z$ and $y_z$ do not play any role in the minimization of the Hamiltonian with respect to the control variable $c$, we use the reduced Hamiltonian:
$$
H(t, a,\mu,y,c)=\big(-c + (\alpha \o\mu^{\alpha-1} -\delta) a \big) y - u(c),
$$
which is convex in $(a,c)$  and strictly convex in $c$. The form \eqref{fo:u'} of the derivative of the utility function implies that the value of the control minimizing the Hamiltonian is $\hat c=(-u')^{-1}(y) = (-y)^{-1/\gamma}$. 
Therefore, the FBSDE derived from the Pontryagin stochastic maximum principle
reads
\begin{equation}
\label{eq:ch:master:KS:stoc}
\begin{cases}
&dA_t= \bigl[
(1-\alpha) \o\mu^\alpha_t Z_t
+ [\alpha  \o\mu_t^{\alpha-1}-\delta] A_t 
- (-Y_{t})^{-1/\gamma}
\big] dt 
 \\
&dY_t =-Y_t[ \alpha  \o\mu_t^{\alpha-1}-\delta] dt + Z_t' dW_{t}, \quad
t \in [0,T] \ ; \quad Y_{T}=-1, 
\end{cases}
\end{equation}
where we used the notation $(Z_{t}')_{0 \leq t \leq T}$
to denote the integrand in the quadratic variation part of the backward equation in order 
to distinguish it from the process $(Z_{t})_{0 \leq t \leq T}$
used in the model as the first component of the state. Despite the fact that the utility function has a singularity at $0$, it is not difficult to check that the proof of the sufficient part of the Pontryagin principle goes through
provided that  the adjoint process $(Y_{t})_{0 \leq t \leq T}$ lives, with probability 
$1$, in a compact subset of $(-\infty,0)$. 

We shall refrain from going through the gory details of the rest of the proof. We refer the interested reader to \cite[Section 3.6.3]{CarmonaDelarue_book_I}.
The major insight is to notice that the backward equation may be decoupled from the forward equation and that
its solution is deterministic and is obtained by solving 
the backward ordinary differential equation:
\begin{equation*}
dY_{t} = -Y_t[ \alpha  \o\mu_t^{\alpha-1}-\delta] dt, \quad
t \in [0,T] \ ; \quad Y_{T}=-1. 
\end{equation*}
The remaining of the proof follows easily.

%%%%%%%%%%%%%%%%%%%%%%%%%%%%%%%%%%%%%%%%%%%%%
\section[From Macro to Finance]{\textbf{From Macro to Finance}}
%%%%%%%%%%%%%%%%%%%%%%%%%%%%%%%%%%%%%%%%%%%%%
\label{se:macro_to_finance}

In this section, we review two recent works of M. Brunnermeier and Y. Sannikov \cite{BrunnermeierSannikov1,BrunnermeierSannikov2} in which the authors compare the historical evolutions of macro-economic and finance models, arguing that properly framed, the analysis of continuous time stochastic models should provide a unifying thread for these sub-fields of economics which so far, developed in parallel. To make their point, the authors introduce models of the economy comprising households maximizing consumption like in classical macro-economic growth models, as well as experts trading in financial markets. 

As explained in the introduction, those models lead to MFGs with a common noise. The importance of common shocks in macro-economics points to the need of a better mathematical understanding of MFGs with a common noise. The first model reviewed in this section fits in the class of MFGs with one population of individuals facing idiosyncratic noise terms as well as random shocks common to all.
It was first introduced in  \cite{BrunnermeierSannikov1} in discrete time. While the second model does not have idiosyncratic noise terms, it involves two populations of agents. This gives us an opportunity to quickly review some of the features of MFGs with several populations, which are not discussed often enough in the mathematical literature on mean field games.

%%%%%%%%%%%%%%%%%%%%%%%%%%%%%%%%%%%%%%%%%%%%%
\subsection[Economy with One Type of Agents]{\textbf{Economy with One Type of Agents}}
%%%%%%%%%%%%%%%%%%%%%%%%%%%%%%%%%%%%%%%%%%%%%

We consider a one-sector economy with a continuum of households with identical preferences (we shall use the logarithmic utility function $u(x)=\log x$) and different levels of wealth. 
We denote by $I$ the set of households. We choose $I=[0,1]$ for the sake of definiteness. In this model, because each agent's influence on the economy is infinitesimal, we use a continuous probability measure  $\lambda$ on $I$ to sample households. For practical purposes, we can think of $\lambda$ as the Lebesgue measure on $[0,1]$.

\begin{remark}
\label{re:Fubini_extension}
This suggestion to think of the space of households as the unit interval $[0,1]$ equipped with its Lebesgue measure is a flagrant expediency. Indeed, mathematically speaking, \emph{it does not pass the smell test}, as in order to manipulate a continuum of idiosyncratic shocks without having to face severe measurability issues, we would have to jump through several hoops, for example using \emph{rich Fubini extensions} instead of the Lebesgue unit interval. See for example \cite[Section 3.7]{CarmonaDelarue_book_I} for a discussion of such a rigorous approach.
\end{remark}

\vskip 2pt
In this model, each household operates a firm and holds money. The capital stock of a generic household $h$ evolves according to the equation:
\begin{equation}
\label{fo:dk^h_t}
\frac{dk^h_t}{k^h_t}= (\phi(\iota^h_t) -\delta) dt  +\sigma^0 dW^0_t + \sigma dW^h_t 
\end{equation}
where $\delta>0$ is a depreciation rate, and the function $\phi$ reflects adjustment costs in capital stock. This function is assumed to satisfy $\phi(0)=0$, $\phi'(0)=1$, $\phi'(\cdot)>0$ and $\phi''(\cdot)<0$. Its concavity captures technological illiquidity. 
$\iota^h_t$ represents the investment rate of household $h$ in physical capital at time $t$. Essentially, it gives how many units of physical capital are used in order to produce new physical capital.
$\bW^0=(W^0_t)_{t\ge 0}$ and $\bW^h=(W^h_t)_{t\ge 0}$ are independent Wiener processes modeling random shocks. $dW^h_t$ represents an idiosyncratic shock specific to the household $h$, while $dW^0_t$ represents a shock common to all the households. We shall often call it the \emph{common noise}. The volatilities $\sigma^0$ and $\sigma$ are positive constants.

\vskip 2pt
Households hold money. We denote by $m^h_t$ the amount of money held at time $t$ by household $h$. They consume in the amount $c^h_t$. We denote by $\theta^h_t$ the fraction of the household wealth in money at time $t$. So the control of a household is the triple $(\iota^h_t,\theta^h_t, c^h_t)$. The goal of a household is  to maximize its long-run discounted expected utility of consumption:
\begin{equation}
\label{fo:J^h}
J^h(\biota,\btheta,\bc)=\EE\Bigl[\int_0^\infty e^{-\rho t}u(c_t)dt\Bigr]
\end{equation}
over the control strategies $(\biota,\btheta,\bc)=(\iota_t,\theta_t,c_t)_{t\ge 0}$. The constant $\rho>0$ provides actualization. We now derive the dynamic constraint under which this optimization is performed by each household.
It is expressed in terms of the wealth $n^h_t$ of the household at time $t$.
\vskip 4pt
We use capital letters to denote the aggregates (i.e. the empirical means) of each of the variables $k^h_t$, $m^h_t$ and $n^h_t$. In other words:
\begin{equation}
\label{fo:aggregates}
K_t=\int_{I^h}k^h_t\;\lambda(dh),\qquad
M_t=\int_{I^h}m^h_t\;\lambda(dh),\qquad
N_t=\int_{I^h}n^h_t\;\lambda(dh).
\end{equation}
Anticipating on the fact that we shall discover that in equilibrium, $(\iota^h_t)_{t\ge 0}$ is independent of the household and adapted to the filtration of the common noise (which is the case if $\iota^h_t$ depends only upon aggregate quantities at time $t$), which implies that all the households use the same investment in physical capital strategy, we can integrate \eqref{fo:dk^h_t} over $h$ and find that 
\begin{equation}
\label{fo:dK_t}
\frac{dK_t}{K_t}= (\phi(\bar \iota_t) -\delta) dt  +\sigma^0 dW^0_t 
\end{equation}
which is a stochastic differential equation only driven by the common noise. The idiosyncratic shocks disappear because of a continuous form of the exact law of large numbers. See for example \cite[Section 3.7]{CarmonaDelarue_book_I}. We used the notation $\bar \iota_t$ to distinguish this aggregate return on capital from the individual households' $\iota^h_t$.

\vskip 2pt
We introduce two more constants: $q$ for the price of one unit of physical capital (so the real value of aggregate physical capital is $qK_t$), and $p$ for the real value of money normalized by the size of the economy as measured by $K_t$ (so $pK_t$ is the real value of total money supply). These could be It\^o processes, say $(q_t)_{t\ge 0}$ and $(p_t)_{t\ge 0}$ driven by the common noise $\bW^0$, but for the sake of simplicity, we shall assume them to be deterministic constants for the purpose of this presentation. Given the definition of the constants $q$ and $p$, the total wealth in the economy is 
\begin{equation}
\label{fo:N_t}
N_t=(p+q)K_t,
\end{equation}
$qK_t$ representing the value of the physical capital and $pK_t$ the value of the nominal capital.
We denote by $\vartheta$ the fraction of nominal wealth:
\begin{equation}
\label{fo:vartheta}
\vartheta=
\frac{p}{p+q}.
\end{equation}
The quantity of money in the economy is controlled exogenously
by a central bank. We assume that money supply follows
the following stochastic differential equation
\begin{equation}
\label{fo:dM_t}
\frac{dM_t}{M_t} = \mu^M dt +\sigma^M dW^0_t
\end{equation}
driven by the common noise.

%%%%%%%%%%%%%%%%%%%%%%%
\vskip 12pt\noindent
\subsubsection*{\textbf{Individual household optimization problem}}
We first derive the stochastic differential equation driving the dynamics of the wealth of a generic household, and then  tackle the optimization of the expected utility of consumption by the Pontryagin stochastic maximum principle.

\vskip 4pt
Changes in the wealth $n^h_t$ of household $h$ at time $t$ are the sums of three contributions. We have:
\begin{equation}
\label{fo:dn^h_t}
dn^h_t= \theta^h_tn^h_t dr^{M}_t +(1-\theta^h_t)n^h_t\; dr^{h,K}_t(\iota^h_t) -c^h_tdt 
\end{equation}
where $r^{M}_t$ denotes the rate of return on money, and $r^{h,K}_t(\iota^h_t)$ the rate of return on capital.
If $ \theta^h_tn^h_t $ is the amount the household holds in money, and if we denote by $p^m_t$ the value of one unit of money, namely
\begin{equation}
\label{fo:p^m_t}
p^m_t=\frac{pK_t}{M_t},
\end{equation}
then the return on this investment is 
$$
dr^M_t=\frac{d p^m_t}{p^m_t}=\frac{d(K_t/M_t)}{K_t/M_t}
$$
since we assume that $p$ is a constant.
Using It\^o's formula with \eqref{fo:dK_t} and \eqref{fo:dM_t} we get:
\begin{equation}
\label{fo:dr^M_t}
dr^{M}_t= \Bigl[\phi(\bar\iota_t)-\delta-[\mu^M-\sigma^M(\sigma^M-\sigma^0)]\Bigr]dt+(\sigma^0-\sigma^M)dW^0_t.
\end{equation}
We now identify the time evolution of the rate of return on capital $r^{h,K}_t(\iota^h_t)$. It has three components: the return of investment in physical capital, the return on the household capital $qk^h_t$, and the seigniorage.
Seigniorage is the amount of money which is transferred to money holders proportionally to their capital.
Given the definition \eqref{fo:p^m_t} of the value of one unit of money, we can easily understand the change in the seigniorage over a period $[t, t+\Delta t]$. It is given by:
$$
T^h_{t+\Delta t}-T^h_t=p^m_{t+\Delta t}(M_{t+\Delta t}-M_t)=p^m_t(M_{t+\Delta t}-M_t)+(p^m_{t+\Delta t}-p^m_t).(M_{t+\Delta t}-M_t).
$$
So in continuous time, i.e. after taking the limit $\Delta t\searrow 0$:
\begin{equation}
\label{fo:dT_t}
\begin{split}
dT_t&=p^m_t\; dM_t + d[p^m,M]_t\\
&= pK_t\Bigl[ [\mu^M + (\sigma^0-\sigma^M)\sigma^M]dt +\sigma^M dW^0_t\Bigr].
\end{split}
\end{equation}
Consequently:
\begin{equation}
\label{fo:dr^h,K_t}
\begin{split}
dr^{h,K}_t(\iota^h_t)&=\frac{a-\iota^h_t}{q}dt + \frac{d\bigl(qk^h_t\bigr)}{qk^h_t}+\frac{dT_t}{qK_t}\\
&= \Bigl[\frac{a-\iota^h_t}{q} + \phi(\iota^h_t)-\delta + \frac{p}{q}[\mu^M+(\sigma^0-\sigma^M)\sigma^M]\Bigr]dt+(\sigma^0+\frac{p}{q}\sigma^M)dW^0_t + \sigma dW^h_t.
\end{split}
\end{equation}
Putting together \eqref{fo:dn^h_t}, \eqref{fo:dr^M_t} and \eqref{fo:dr^h,K_t} we get:
\begin{equation}
\label{fo:dn^h_t_final}
\begin{split}
dn^h_t&= \theta^h_tn^h_t dr^{M}_t +(1-\theta^h_t)n^h_t\,dr^{h,K}_t(\iota^h_t) -c^h_tdt\\
&=\theta^h_tn^h_t \Bigl[\phi(\bar\iota_t)-\delta-[\mu^M+(\sigma^0-\sigma^M)\sigma^M]\Bigr]dt+\theta^h_tn^h_t (\sigma^0-\sigma^M)dW^0_t\\ 
&\hskip 35pt
+(1-\theta^h_t)n^h_t\Bigl[\frac{a-\iota^h_t}{q} + \phi(\iota^h_t)-\delta + \frac{p}{q}[\mu^M+(\sigma^0-\sigma^M)\sigma^M]\Bigr]dt\\
&\hskip 35pt
+(1-\theta^h_t)n^h_t(\sigma^0+\frac{p}{q}\sigma^M)dW^0_t + (1-\theta^h_t)n^h_t\sigma dW^h_t
-c^h_tdt\\
&=\Bigl(n^h_t \Bigl[\theta^h_t\bigl(\phi(\bar\iota_t)-\delta\bigr)+(1-\theta^h_t)\bigl[\frac{a-\iota^h_t}{q}+ \phi(\iota^h_t)-\delta\bigr]\\
&\hskip 35pt
+[\mu^M+(\sigma^0-\sigma^M)\sigma^M]\bigl(\frac{p}{q}-\theta^h_t\frac{p+q}{q}\bigr)\Bigr]
-c^h_t\Bigr)dt\\
&\hskip 35pt
+n^h_t\bigl[\sigma^0+\bigl(\frac{p}{q}-\theta^h_t\frac{p+q}{q}\bigr)\sigma^M\bigr]dW^0_t + (1-\theta^h_t)n^h_t\sigma dW^h_t
\end{split}
\end{equation}
The Hamiltonian of the optimization problem of a generic household reads:
\begin{equation}
\label{fo:H}
\begin{split}
&H(t,n,y,z^0,z,\iota,\theta,c)\\
&\hskip 25pt=\Bigl(n \Bigl[\theta\bigl(\phi(\bar\iota_t)-\delta\bigr)+(1-\theta)\bigl[\frac{a-\iota}{q}+ \phi(\iota)-\delta\bigr]\\
&\hskip 75pt
+[\mu^M+(\sigma^0-\sigma^M)\sigma^M]\bigl(\frac{p}{q}-\theta\frac{p+q}{q}\bigr)\Bigr]
-c\Bigr)y\\
&\hskip 75pt
+n\bigl[\sigma^0+\bigl(\frac{p}{q}-\theta\frac{p+q}{q}\bigr)\sigma^M\bigr]z^0 + (1-\theta)n\sigma z
-e^{-\rho t} u(c)
\end{split}
\end{equation}
if we use the notations $y$, $z^0$ and $z$ for the adjoint variables (sometimes called co-states). The necessary part of the Pontryagin maximum principle suggests to minimize the Hamiltonian with respect to the control variables $\iota$, $\theta$ and $c$.

Moreover, since $(1-\theta)\ge 0$, we can isolate $\iota$ and minimize $(a-\iota)/q+\phi(\iota)$ which gives Tobin's $q$ equation:
$$
 -\frac{1}{q}+\phi'(\iota)=0 \quad\Leftrightarrow\quad \iota=(\phi')^{-1}\bigl(\frac{1}{q}\bigr),
$$
and in the case of the function $\phi(\iota)=(1/\kappa)\log(1+\kappa\iota)$ used in \cite{BrunnermeierSannikov2}, we get:
\begin{equation}
\label{fo:hat_iota_t}
\kappa\hat \iota_t=q-1.
\end{equation}
Notice that the optimal $\hat\iota_t$ is a constant independent of $t$. In general, if $q$ is an It\^o process adapted to the filtration of the common noise, so is $\hat\iota$. But the fact to remember at this stage is that the optimal $\hat\iota_t$ is the same for all the households. So from now on $\bar\iota_t=\hat\iota_t=\kappa^{-1}(q-1)$.

\vskip 2pt
Expecting that $\theta\in[0,1]$, minimizing the Hamiltonian over $\theta$ \emph{could} lead to $\partial_\theta H=0$, i.e.
$$
ny\Bigl[\frac{a-\iota}{q} + \frac{p+q}{q}[\mu^M+(\sigma^0-\sigma^M)\sigma^M]\Bigr]
+nz^0\frac{p+q}{q}\sigma^M+n\sigma z=0.
$$
For obvious reasons we write the adjoint variables $z$ and $z^0$ in the form $z=-y\zeta$ and 
$z^0=-y\zeta^0$ so we can rewrite the first order condition $\partial_\theta H=0$ in equilibrium as:
\begin{equation}
\label{fo:pricing}
\frac{a-\iota}{q} +\frac{p+q}{q}[\mu^M+(\sigma^0-\sigma^M)\sigma^M]=
\frac{p+q}{q}\sigma^M\zeta^0+\sigma \zeta.
\end{equation}
This equation does not determine directly the optimal value of $\theta_t$. It is sometimes called the pricing equation because it can also be derived from the HJB equation of the optimization problem offering a pricing interpretation.

\vskip 4pt\noindent
Finally, since we restrict ourselves to $y<0$, we can write the third First Order Condition (FOC) as:
$$
\partial_cH=0\quad\Leftrightarrow\quad -y-e^{-\rho t}u'(c)=0 \quad\Leftrightarrow\quad c=(u')^{-1}\bigl(-e^{\rho t}y\bigr)
$$
so that in the case of logarithmic utility, the optimal consumption rate should be the process $(\hat c_t)_{t\ge 0}$  given by
$$
\hat c^h_t=-e^{-\rho t}\frac1{y^h_t}
$$
where the adjoint process $(y_t)_{t\ge 0}$ is the first component of the solution of the adjoint equation, namely the Backward Stochastic Differential Equation (BSDE) equation:
\begin{equation}
\label{fo:dy_t}
dy^h_t=-\partial_nH(t,n^h_t,y^h_t,z^{0,h}_t,z^h_t,\hat\iota_t,\hat\theta_t,\hat c_t) dt+z^{0,h}_t dW^0_t+z^h_t dW^h_t.
\end{equation}
Computing $\partial_nH$ from \eqref{fo:H} and using \eqref{fo:pricing} we get:
\begin{equation}
\begin{split}
\partial_nH&=y\bigl[\phi(\iota)-\delta - [ \mu^M+\sigma^M(\sigma^0-\sigma^M)] -(\sigma^0-\sigma^M)\zeta^0\bigr] \\
&=yr^h_t
\end{split}
\end{equation}
if we define the individual household effective interest by:
\begin{equation}
\label{fo:r^h_t}
r^h_t=\phi(\iota)-\delta - [ \mu^M+\sigma^M(\sigma^0-\sigma^M)] -(\sigma^0-\sigma^M)\zeta^{0,h}_t.
\end{equation}
So the adjoint Backward Stochastic Differential Equation (BSDE) reads
\begin{equation}
\label{fo:adjoint}
\frac{dy^h_t}{y^h_t}=-r^h_tdt -\zeta^{0,h} dW^0_t -\zeta^h_t dW^h_t,
\end{equation}
hence the interpretation of $r^h_t$ as an individual household short interest rate and $y^h_t$ as an individual stochastic discount factor.
Using It\^o's formula with \eqref{fo:dn^h_t} and \eqref{fo:adjoint}, and the definition \eqref{fo:H} of the Hamiltonian we get:
$$
\frac{d (y^h_t n^h_t)}{y^h_t n^h_t}= -\frac{c^h_t}{n^h_t} dt
+\Bigl(\bigl[\sigma^0+\bigl(\frac{p}{q}-\theta^h_t\frac{p+q}{q}\bigr)\sigma^M\bigr]-\zeta^{0,h}_t\Bigr) + 
\Bigl((1-\theta^h_t)\sigma -\zeta^h_t\Bigr),
$$
so if we choose: 
\begin{equation}
\label{fo:zetas}
\zeta^{0,h}_t=\sigma^0+\bigl(\frac{p}{q}-\theta^h_t\frac{p+q}{q}\bigr)\sigma^M
\quad\text{and}\quad
\zeta^h_t=\sigma (1-\theta_t)\;,
\end{equation}
we have that $y^h_t n^h_t=-e^{-\rho t}/\rho$ and consequently:
\begin{equation}
\label{fo:expert_c}
\hat c^h_t=\rho n^h_t.
\end{equation}
\emph{NB: } The fact that the optimal rate of consumption is proportional to the wealth is to be expected when using logarithmic utility.
\vskip 2pt
Plugging the expressions \eqref{fo:zetas} for $\zeta^{0,h}_t$ and $\zeta^h_t$ into the pricing equation \eqref{fo:pricing}, we find:
$$
\frac{a-\iota}{q} +\frac{p+q}{q}[\mu^M+(\sigma^0-\sigma^M)\sigma^M]=
\frac{p+q}{q}\sigma^M\Bigl(\sigma^0+\bigl(\frac{p}{q}-\theta^h_t\frac{p+q}{q}\bigr)\sigma^M\Bigr)+\sigma^2 (1-\theta_t)$$
from which we derive
\begin{equation}
\label{fo:final_theta}
1-\theta^h_t=\frac{\frac{a-\iota}{q} + \frac{p+q}{q}\mu^M}{\bigl(\frac{p+q}{q}\bigr)^2(\sigma^M)^2+\sigma^2}.
\end{equation}
Using the fraction of nominal wealth defined in \eqref{fo:vartheta} this gives:
\begin{equation}
\label{fo:another_theta}
1-\theta^h_t=\frac{\frac{a-\iota}{q}(1-\vartheta) +\mu^M}{(\sigma^M)^2+\sigma^2(1-\vartheta)^2}(1-\vartheta).
\end{equation}
So not only is the optimal portfolio the same for all the households, but we also learn that it is a constant. 
Moreover:
$$
\zeta^{0,h}_t=\sigma^0-\sigma^M +\sigma^M\frac{1-\theta^h_t}{1-\vartheta}=\sigma^0-\sigma^M +\sigma^M\frac{\frac{a-\iota}{q}(1-\vartheta) +\mu^M}{(\sigma^M)^2+\sigma^2(1-\vartheta)^2},
$$
and inserting this value of $\zeta^{0,h}_t$ into the formula \eqref{fo:r^h_t} we get:
\begin{equation}
\label{fo:new_r^h_t}
r^h_t=\phi(\iota)-\delta - [ \mu^M+\sigma^M(\sigma^0-\sigma^M)] -(\sigma^0-\sigma^M)\Bigl(\sigma^0-\sigma^M +\sigma^M\frac{\frac{a-\iota}{q}(1-\vartheta) +\mu^M}{(\sigma^M)^2+\sigma^2(1-\vartheta)^2}\Bigr),
\end{equation}
which shows that the individual interest rate is in fact the same constant for all the households.

%%%%%%%%%%%%%%%%%%%%%%%%%%%%%%%
\vskip 12pt\noindent
\subsubsection*{\textbf{Clearing Conditions}}
The goods market clears if total output $aK_t$ equals the sum of investment $\iota_t K_t$ and consumption $C_t$.
So the overall consumption $C_t=\int c^h_t \lambda(dh)$ should be equal to $(a-\iota_t)K_t$ since $aK_t$ represents the overall production and $\iota_tK_t$ represents the overall reinvestment in capital. If we recall that we are using logarithmic utility, we saw that the optimal consumption was proportional to the wealth so:
$$
C_t=\int c^h_t \lambda(dh)=\rho\int n^h_t\lambda(dh)=\rho N_t=\rho(p+q)K_t
$$
so that the clearing condition amounts to
$$
\rho(p+q)=a-\iota_t
$$
which gives 
\begin{equation}
\label{fo:iota_t}
\frac{a-\hat\iota}{q}= \frac{\rho}{1-\vartheta}.
\end{equation}

\vskip 6pt
The capital market clears if aggregate capital demand equals capital supply $K_t$, in other words if:
$$
1-\theta_t\frac{N_t}{q}=K_t
$$
and using the fact that $N_t=(p+q)K_t$ we get:
\begin{equation}
\label{fo:thetas}
1-\hat\theta_t=\frac{q}{p+q}=1-\vartheta.
\end{equation}
The money market clears by Walras law.

\vskip 2pt
Using the clearing condition \eqref{fo:iota_t} and the optimal value of $\hat\iota_t$ \eqref{fo:hat_iota_t} we get:
$$
q=(1-\vartheta)\frac{1+\kappa a}{1-\vartheta +\kappa\rho},
$$
from which we derive:
$$
\hat\iota=\frac{(1-\vartheta) a -\rho}{1-\vartheta +\kappa\rho},
\qquad\text{and}\qquad
p=\vartheta\frac{1-\kappa a }{1-\vartheta +\kappa\rho}.
$$
Finally, injecting \eqref{fo:thetas} and \eqref{fo:iota_t} into the pricing equation \eqref{fo:pricing} we get:
$$
1-\vartheta=\sqrt{\frac{\rho+\mu^M-(\sigma^M)^2}{\sigma^2}}.
$$
which shows that a stationary (meaning the processes $(p_t)_{t\ge 0}$ and $(q_t)_{t\ge 0}$ are deterministic and constant given by the real numbers $p$ and $q$) general equilibrium is possible if 
\begin{equation}
\label{fo:existence_conditions}
\rho+\mu^M-(\sigma^M)^2 >0
\qquad\text{and}\qquad
\sigma>\sqrt{\rho+\mu^M-(\sigma^M)^2}.
\end{equation}

%%%%%%%%%%%%%%%%%%%%%%%%%%%%
\vskip 12pt\noindent
\subsubsection*{\textbf{Does all this have anything to do with Mean Field Games?}}
Since the state variable of an individual household is  its wealth $n^h_t$, the typical interaction one should expect if this general equilibrium can be recast as a mean field game should be the aggregate wealth $N_t$. So in the presence of the common noise $W^0_t$ one should fix the flow of conditional distributions of the wealth $n^h_t$ given the filtration of the common noise, and search for the best response of this household. In other words, given the knowledge of $(N_t)_{t\ge 0}$ which is a stochastic process adapted to the filtration $\FF^0$ of $\bW^0$, the individual household should find optimal investment rate in physical capital $(\hat\iota_t)_{t\ge 0}$, optimal investment portfolio $(\hat\theta_t)_{t\ge 0}$, and optimal consumption rate $(\hat c_t)_{t\ge 0}$, to maximize its long-run discounted expected utility of consumption \eqref{fo:J^h}. This is exactly what was done in the section dealing with the individual household optimization problem. The next step of the MFG paradigm is the fixed point step according to which one tries to identify a flow of conditional distributions which ends up being the flow of conditional distributions of the solution of the optimization problem underpinning the search for the best response.

\vskip 2pt
In typical macro-economic general equilibrium problems, individual optimizations are performed assuming that the aggregates are known. If aggregates can be interpreted as means of some state variables, fixing the aggregates amounts to fixing the distributions of these state variables. In this example, assuming the knowledge of  $(N_t)_{t\ge 0}$ is the same thing as assuming the knowledge of $(K_t)_{t\ge 0}$ since as we saw, $N_t=\rho(p+q)K_t$, which in turn, is assuming the knowledge of the process $(\bar\iota_t)_{t\ge 0}$ representing the aggregate investment rate in physical capital. This is the \emph{mean field interaction} appearing explicitly in the dynamics \eqref{fo:dn^h_t_final} of the state of the individual household. Since the individual household optimal investment rate in capital is constant as given by Tobin's $q$ equation \eqref{fo:hat_iota_t}, a necessary condition for the fixed point step is that $\bar\iota_t=\hat\iota_t$. Added to the necessary conditions of optimality (which we derived from the Pontryagin stochastic maximum principle) and the capital market clearing condition, this fixed point step leads to the equilibrium solution under the conditions \eqref{fo:existence_conditions}.

\vskip 2pt
Because we chose to restrict ourselves to the search for a stationary general equilibrium in which the processes $(p_t)_{t\ge 0}$ and $(q_t)_{t\ge 0}$ are deterministic and constant given by the real numbers $p$ and $q$, the deterministic nature of most of the characteristics of the equilibria are rather anti-climatic, and the reformulation of the solution as the search for Nash equilibria in a mean field game is rather contrived. We chose to present this model because of the presence of both idiosyncratic and common shocks. We refer the interested readert to \cite{BrunnermeierSannikov1,BrunnermeierSannikov2} for extensions with deeper financial meaning.
The next example will be more illustrative of the deep connection with the paradigm of mean field games. While it does not involve idiosymcratic shocks, it involves two populations and this will give us a chance to highlight the possible benefits of a mean field game reformulation of the model.

%%%%%%%%%%%%%%%%%%%%%%%%%%%%%%%%%%%%%%%%%%%%%
\subsection[Economy  with Two Types of Agents]{\textbf{Economy with Two Types of Agents}}
%%%%%%%%%%%%%%%%%%%%%%%%%%%%%%%%%%%%%%%%%%%%%
\label{sub:BS2}
We present the analysis of the model discussed in \cite{BrunnermeierSannikov2} mutatis mutandis.
We consider an economy with a continuum of households and experts. 
We denote by $I^h$ (resp. $I^e$) the space of households (resp. experts). Typically, we choose $I^h=I^e=[0,1]$ which we assume to be equipped with its Borel $\sigma$-field. We shall alo use  continuous probability measures  $\lambda^h$ and $\lambda^e$ on $I^h$ and $I^e$ respectively. Again, for practical purposes, and modulo the contents of Remark \ref{re:Fubini_extension} at the beginning of the discussion of the previous model, we can think of them both as equal to the Lebesgue measure on $[0,1]$. 

\vskip 2pt
In this economy, households consume and lend money to experts. On the other end, experts borrow money from households, invest in the production of a single good, and consume. The goal of each agent is to maximize their long run expected utility. In this model, all agents use the logarithmic utility function $u(c)=\log c$. So if we denote by $c^e_t$ and $c^h_t$ the consumptions at time $t$ of expert $e$ and household $h$ respectively, the optimization problem is:
$$
\sup_{(c^i_t)_{t\ge 0}} \EE\Bigl[\int_0^\infty e^{-\rho t}\log c^i_t\;dt\Bigr],\qquad i=e,h,
$$
where $\rho>0$ is a discount factor common to the two classes of agents. To be consistent with the computation done throughout this chapter, we shall in fact minimize the negative of the above expected utility of consumption.

%%%%%%%%%%%%%%%%%%%%%%%
\vskip 12pt\noindent
\subsubsection*{\textbf{Individual household optimization problem}}

If we denote by $n^h_t$ the wealth of household $h$ at time $t$, we have:
\begin{equation}
\label{fo:new_dn^h_t}
dn^h_t= r_tn^h_t dt  -c^h_tdt 
\end{equation}
Here, the process $(r_t)_{t\ge 0}$ represents the interest rate common to all agents. It is one of the stochastic processes to be determined endogenously.

The Hamiltonian of the optimization problem of a generic household reads:
$$
H(t,n,\xi,c)=(r_t n-c)\xi - e^{-\rho t}u(c)
$$
if we use the notation $\xi$ for the adjoint variable (sometimes called the co-state) which we shall restrict to be negative. The necessary part of the Pontryagin maximum principle suggests to minimize the Hamiltonian with respect to the control variable $c$.
This gives the First Order Condition (FOC):
$$
\partial_cH=0\quad\Leftrightarrow\quad -\xi-e^{-\rho t}u'(c)=0 \quad\Leftrightarrow\quad c=(u')^{-1}\bigl(-e^{\rho t}\xi\bigr)
$$
so that in the case of logarithmic utility, the optimal consumption rate is given by
$$
\hat c_t=-e^{-\rho t}\frac1{\xi_t}
$$
where the adjoint function $t\mapsto\xi_t$ solves the adjoint equation:
$$
d\xi_t=-\partial_nH dt= -r_t \xi_t dt.
$$
The differentiation product rule gives:
$$
d (\xi_t n^h_t)= c^h_t\xi_t dt=-e^{-\rho t} dt
$$
implying that $\xi_t n^h_t=-e^{-\rho t}/\rho$ and consequently:
\begin{equation}
\label{fo:household_c}
\hat c^h_t=\rho n^h_t.
\end{equation}
As noted in the previous example, the fact that the optimal rate of consumption is proportional to the wealth (and is independent of the interest rate) is a well known property of the logarithmic utility function.

%%%%%%%%%%%%%%%%%%%%%%%
\vskip 12pt\noindent
\subsubsection*{\textbf{Individual expert optimization problem. }}

If  at time $t$ we denote by $n^e_t$ the wealth of expert $e$, by $\theta^e_t$ the proportion of self worth invested in bonds (i.e. borrowed from the households, so $\theta^e_t\le 0$) and by $\iota^e_t$ the investment in physical capital,
we have:
\begin{equation}
\label{fo:dn^e_t_def}
dn^e_t= \theta^e_tn^e_t\;r_t dt  +(1-\theta^e_t)n^e_t \; dr^k_t(\iota^e_t)-c^e_tdt 
\end{equation}
where $r^k_t(\iota^e_t)$ denotes the return from the investment $\iota^e_t$ in physical capital.
The capital stock of a generic expert $e$ evolves according to the equation:
\begin{equation}
\label{fo:dk^e_t}
\frac{dk^e_t}{k^e_t}= (\phi(\iota^e_t) -\delta) dt  +\sigma dW^0_t 
\end{equation}
where $\delta>0$ is a depreciation rate, and the function $\phi$ reflects adjustment costs in capital stock. It is assumed to satisfy $\phi(0)=0$, $\phi'(0)=1$, $\phi'(\cdot)>0$ and $\phi''(\cdot)<0$. Its concavity captures technological illiquidity. The volatility $\sigma>0$ is a positive constant and $\bW^0=(W^0_t)_{t\ge 0}$ is a Wiener process modeling random shocks. Note that this is the same process for all the experts. This is an instance of what we call a \emph{common noise}. There is no source of idiosyncratic noise in this model.

\vskip 2pt
Let the price $q_t$ at time $t$ of one unit of capital be an It\^o process satisfying
\begin{equation}
\label{fo:dq_t}
\frac{dq_t}{q_t}= \mu^q_tdt +\sigma^q_t dW^0_t 
\end{equation}
for two processes $(\mu^q_t)_{t\ge 0}$ and $(\sigma^q_t)_{t\ge 0}$ adapted to the filtration $\FF^0$ of the common noise, which will be specified later on.

The return on capital $r^k_t(\iota^e_t)$ is defined as:
\begin{equation}
\label{fo:dr^k_t_def}
dr^k_t(\iota^e_t)= \frac{a-\iota^e_t}{q_t}dt + \frac{d(q_tk^e_t)}{q_tk^e_t}.  
\end{equation}
The first term in the right hand side represents the dividend yield while the second one gives the capital gain.
Using the definitions \eqref{fo:dq_t} and \eqref{fo:dk^e_t} and It\^o's formula for the differential of a product we get:
\begin{equation}
\label{fo:dr^k_t}
dr^k_t(\iota^e_t)=\Bigl[ \frac{a-\iota^e_t}{q_t} +\phi(\iota^e_t)-\delta+\mu^q_t+\sigma\sigma^q_t\Bigr]dt + (\sigma+\sigma^q_t)dW^0_t.  
\end{equation}
Plugging this formula in the dynamics \eqref{fo:dn^e_t_def} of the wealth of a generic expert we get:
\begin{equation}
\label{fo:dn^e_t}
dn^e_t= \Bigl[\theta^e_tn^e_t\;r_t +(1-\theta^e_t)n^e_t \; \Bigl( \frac{a-\iota^e_t}{q_t} +\phi(\iota^e_t)-\delta+\mu^q_t+\sigma\sigma^q_t \Bigr)-c^e_t\Bigr]dt + (1-\theta_t)n^e_t(\sigma+\sigma^q_t)dW^0_t.
\end{equation}
This equation should be viewed as giving the dynamics of the state variable $n^e_t$ as controlled by $(c^e_t,\iota^e_t,\theta^e_t)$. 
As before, we use Pontryagin stochastic maximum principle to solve the optimization of the expected utility of consumption. For the sake of simplicity of notation, we skip the superscript ${}^e$ throughout the remaining of this subsection. No confusion is possible since we are only dealing with the expert optimization problem. The Hamiltonian of this optimization problem reads:
\begin{equation}
\label{fo:H^e}
\begin{split}
&H(t,n,\xi,\zeta,c,\iota,\theta)
= \Bigl[\theta n\;r_t   +(1-\theta)n \; \Bigl( \frac{a-\iota}{q_t} +\phi(\iota)-\delta+\mu^q_t+\sigma\sigma^q_t \Bigr)-c\Bigr]\xi\\
&\hskip 125pt
- (1-\theta) n(\sigma+\sigma^q_t)\zeta\xi- e^{-\rho t}u(c)
\end{split}
\end{equation}
where for reasons which will become clear soon, we used the notations $\xi$ (which is assumed to be negative) and $-\xi\zeta$ for the adjoint variables.
We now have three First Order Conditions. 

\vskip 2pt
Since $\xi\le 0$ and $(1-\theta)\ge 1$ we can isolate the contribution of the control $\iota$. This leads to the maximization of the quantity $(a-\iota)/q_t+\phi(\iota)$ which leads to"
$$
 -\frac1{q_t}+\phi'(\iota)=0 \quad\Leftrightarrow\quad \iota=(\phi')^{-1}\bigl(-\frac1{q_t}\bigr).
$$
In the case of the function $\phi(\iota)=(1/\kappa)\log(1+\kappa\iota)$ used in \cite{BrunnermeierSannikov2}, we get:
$$
\hat \iota_t=\frac1\kappa(q_t-1).
$$
In any case, this value is the same for all the experts $e$, and as a control process, it is adapted to the filtration of the common noise.

\vskip 2pt
As before:
$$
\partial_cH=0\quad\Leftrightarrow\quad -\xi-e^{-\rho t}u'(c)=0 \quad\Leftrightarrow\quad c=(u')^{-1}\bigl(-e^{\rho t}\xi\bigr)
$$
so that in the case of logarithmic utility, the optimal consumption rate is given by
$$
\hat c_t=-e^{-\rho t}/\xi_t
$$
where the adjoint process $(\xi_t)_{t\ge 0}$ solves the adjoint equation:
$$
d\xi_t=-\partial_nH dt-\xi_t\zeta_t dW^0_t.
$$
Notice that the FOC $\partial_\theta H=0$  gives:
\begin{equation}
\label{fo:first_rt}
r_t= \frac{a-\iota}{q_t} +\phi(\iota)-\delta+\mu^q_t+\sigma\sigma^q_t -\zeta_t
(\sigma+\sigma^q_t)
\end{equation}
and we shall see below that this formula will help us identify the individual expert optimal investment $\hat\theta^e_t$ in terms of the processes $(r_t)_{t\ge 0}$, $(q_t)_{t\ge 0}$, $(\mu^q_t)_{t\ge 0}$, $(\sigma^q_t)_{t\ge 0}$.
Computing $\partial_nH$ from \eqref{fo:H^e} we get
$$
\partial_nH(t,n,\xi,\zeta,c,\iota,\theta)
= \Bigl[\theta \;r_t   +(1-\theta) \; \Bigl( \frac{a-\iota}{q_t} +\phi(\iota)-\delta+\mu^q_t+\sigma\sigma^q_t \Bigr)\Bigr]\xi
-(1-\theta)(\sigma+\sigma^q_t)\zeta\xi
$$
and using \eqref{fo:first_rt} we get
$$
\partial_nH(t,n,\xi,\zeta,c,\iota,\theta)=\xi r_t 
$$
and the adjoint equation rewrites:
\begin{equation}
\label{fo:dxi_t}
\frac{d\xi_t}{\xi_t}= -r_t\;dt -\zeta_t dW^0_t 
\end{equation}
which justifies our choice of the form of the second adjoint variable.
Applying It\^o formula to \eqref{fo:dn^e_t} and \eqref{fo:dxi_t} we get:
$$
\frac{d (\xi_t n_t)}{\xi_t n_t}= -\frac{c_t}{n_t} dt+ \bigl[-\zeta_t+(1-\theta_t)(\sigma+\sigma^q_t)\bigr]dW^0_t
$$
so that, choosing 
\begin{equation}
\label{fo:zeta_t}
\zeta_t=(1-\theta_t)(\sigma+\sigma^q_t),
\end{equation} 
we find that as in the case of the computation of the optimal consumption rate of the households, $\xi_t n_t=-e^{-\rho t}/\rho$ and consequently:
\begin{equation}
\label{fo:expert_c}
\hat c^e_t=\rho n^e_t.
\end{equation}
So the fact that the optimal rate of consumption is proportional to the wealth was not affected by the presence of the random shocks. It is typical in the case of logarithmic utility. Plugging our choice \eqref{fo:zeta_t} for $\zeta_t$ in \eqref{fo:first_rt} we find:
\begin{equation}
\label{fo:r_t}
r_t= \frac{a-\hat\iota_t}{q_t} +\phi(\hat\iota_t)-\delta+\mu^q_t+\sigma\sigma^q_t -(1-\theta^e_t)
(\sigma+\sigma^q_t)^2
\end{equation}
from which we can easily extract $\hat\theta^e_t$ as desired.

%%%%%%%%%%%%%%%%%%%%%%%
\vskip 12pt\noindent
\subsubsection*{\textbf{Clearing constraints and equilibrium. }}

The next step in the search for a general equilibrium for this macro-economic model is to articulate the constraints imposed by the need to have all the markets clear, and to show that one can identify processes $(r_t)_{t\ge 0}$, $(q_t)_{t\ge 0}$, $(\mu^q_t)_{t\ge 0}$, $(\sigma^q_t)_{t\ge 0}$ satisfying these constraints and allowing the simultaneous optimizations of all the agents. Clearing is best expressed in terms of aggregate quantities. For each $i\in\{h,e\}$, we denote by $C^i_t$ the aggregate consumption for the agents of type $i$. Formally we write:
$$
C^h_t=\int_{I^h}\hat c^h_t \;\lambda^h(dh),
\quad\text{and}\quad
C^e_t=\int_{I^e}\hat c^e_t \;\lambda^e(de).
$$
Given \eqref{fo:household_c} and \eqref{fo:expert_c} we see that $C^h_t=\rho N^h_t$ and $C^e_t=\rho N^e_t$ where 
$N^h_t$ and $N^e_t$ are the aggregate worths of the populations of households and experts respectively, i.e.
$$
N^h_t=\int_{I^h}\hat n^h_t \;\lambda^h(dh),
\quad\text{and}\quad
N^e_t=\int_{I^e}\hat n^e_t \;\lambda^e(de).
$$
If we denote by $K_t$ the aggregate physical capital in the economy at time $t$, i.e.
$$
K_t=\int_{I^e}k^e_t \;\lambda^e(de),
$$
the aggregate wealth in the economy is equal to $q_tK_t$

\vskip 2pt
Clearing of the loan market requires that at each time $t$, the aggregate debt of the experts, say $D^e_t$, be equal to the aggregate loans of the households, so that:
$$
D^e_t=\int_{I^h}n^h_t\lambda^h(dh)=N^h_t.
$$
So
$$
N^e_t=q_tK_t-D^e_t=q_tK_t-N^h_t
$$
which implies that 
\begin{equation}
\label{fo:N+N}
q_tK_t=N^h_t+N^e_t. 
\end{equation}
It will be convenient to use the quantity:
\begin{equation}
\label{fo:eta_t}
\eta_t=\frac{N^e_t}{N^h_t+N^e_t}=\frac{N^e_t}{q_tK_t}
\end{equation}
representing the wealth share of the experts. Notice that all these aggregate quantities are random since they depend upon the common noise $\bW^0$ which does not average out in the computation of the aggregates because it is common to all the agents.
\vskip 4pt
Clearing of consumption on the market for goods requires
$$
C_t=(a-\iota^e_t)K_t
$$
in other words $\rho q_tK_t=(a-\iota^e_t(q_t))K_t$ which implies $\rho q_t=a-\iota^e_t(q_t)$ which in turn implies that the process $(q_t)_{t\ge 0}$ is in fact a positive constant, say $q$. As a consequence, $\mu^q_t\equiv 0$ and $\sigma^q_t\equiv 0$, and if we use the function $\phi(\iota)=(1/\kappa)\log(1+\kappa\iota)$ proposed in \cite{BrunnermeierSannikov2} we get
\begin{equation}
\label{fo:q}
q=\frac{1+\kappa a}{1+\kappa\rho}
\qquad\text{and}\qquad
\iota^e=\frac{a-\rho}{1+\kappa\rho}. 
\end{equation}

\vskip 4pt
Capital market clearing yields:
\begin{equation}
\label{fo:capital_clearing}
1-\theta^e_t=\frac{q_tK_t}{N^e_t}=\frac{1}{\eta_t}
\end{equation}

\vskip 4pt
Knowing that $q_t$ has to be a deterministic constant, we can use the facts that 
\begin{equation}
\label{fo:dN^e_t}
\begin{split}
\frac{dN^e_t}{N^e_t}&= \frac{dn^e_t}{n^e_t}\\
&= \Bigl[r_t   +(1-\theta^e_t)^2\sigma^2-\frac{c^e_t}{n^e_t}\Bigr]dt + (1-\theta^e_t)\sigma dW^0_t
\end{split}
\end{equation}
and
\begin{equation}
\label{fo:dK^e_t}
\begin{split}
\frac{dK^e_t}{K^e_t}&= \frac{dk^e_t}{k^e_t}\\
&= \Bigl[ r_t  +(1-\theta^e_t) \sigma^2-\rho\Bigr]dt + \sigma dW^0_t
\end{split}
\end{equation}
to derive from It\^o's formula that:
\begin{equation}
\label{fo:deta^e_t}
\begin{split}
\frac{d\eta_t}{\eta_t}&=\frac{d\bigl(N^e_t/K_t\bigr)}{N^e_t/K_t}\\
&=\Bigl[ -\frac{c^e_t}{n^e_t}+\rho+(\theta^e_t)^2\sigma^2\Bigr] dt -\theta^e_t\sigma dW^0_t\\
&=(\theta^e_t)^2\sigma^2 dt -\theta^e_t\sigma dW^0_t
\end{split}
\end{equation}
which we can rewrite as
\begin{equation}
\label{fo:deta_t_final}
d\eta_t=\sigma^2\frac{(1-\eta_t)^2}{\eta_t} dt +\sigma (1-\eta_t)dW^0_t
\end{equation}
if we use the capital market clearing condition \eqref{fo:capital_clearing}. 

\vskip 12pt\noindent
\subsubsection*{\textbf{Interpretation. }}
\begin{figure}[H]
\centerline{
\includegraphics[width=7cm,height=5cm]{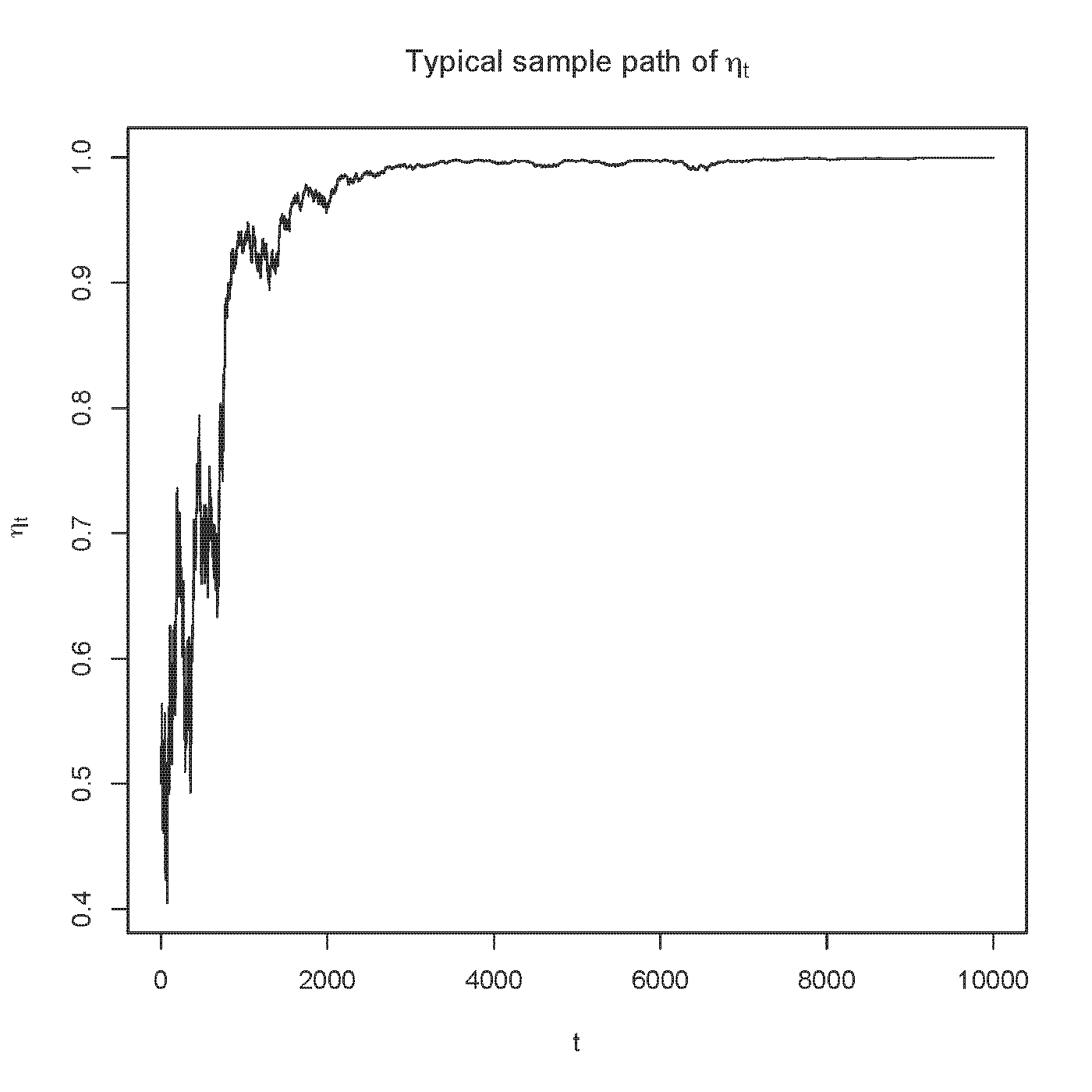}
}
\caption{Typical sample path of $\eta_t$.}
\label{fi:eta_t}
\end{figure}

This is a stochastic differential equation on the open interval $(0,1)$. According to Feller's theory of one dimensional diffusions, the scale function $p(x)$ and the speed measure $m(dx)$ are given by:
$$
p(x)=\frac12\bigl(1-\frac{1}{2x}\bigr)
\qquad\text{and}\qquad
m(dx)=\frac{8}{\sigma^2}\frac{x^2}{(1-x)^2}dx.
$$
Feller's explosion test can be computed and it says that if started inside the interval $(0,1)$, the diffusion remains inside the interval for ever and in fact $\lim_{t\to\infty}\eta_t=1$ almost surely. Note also that the drift is always positive, and very large when $\eta_t$ is small, so up to the fluctuations due to the random shocks (whose sizes $\sigma(1-\eta_t)$ decrease as $\eta_t$ get closer to $1$), one should expect that $\eta_t$ would grow quickly toward $1$ and become mostly flat when it gets close to $1$. This is illustrated in Figure \ref{fi:eta_t}.

From an economic point of view, this means that the proportion of the wealth held by the experts grows quickly toward a high value close to $100\%$, and eventually converges to this $100\%$ level, leaving the households 
helpless.

\vskip 6pt
Finally, revisiting the constraint \eqref{fo:r_t}, we see that in equilibrium we must have:
\begin{equation}
\label{fo:r_t_final}
\begin{split}
r_t&= \frac{a-\iota^e_t}{q_t} +\phi(\iota^e_t)-\delta+\mu^q_t+\sigma\sigma^q_t -(1-\theta^e_t)\sigma^2\\
&= \frac{a-\iota^e_t}{q_t} +\phi(\iota^e_t)-\delta -(1-\theta^e_t)\sigma^2\\
&= \frac{a-\iota^e_t}{q_t} +\phi(\iota^e_t)-\delta-\frac{\sigma^2}{\eta_t}\\
&=\rho+\frac1\kappa\log\Bigl( \frac{1+\kappa a}{1+\kappa\rho}\Bigr)-\delta-\frac{\sigma^2}{\eta_t}
\end{split}
\end{equation}
if we use the function $\phi(\iota)=(1/\kappa)\log(1+\kappa\iota)$ proposed in \cite{BrunnermeierSannikov2}.

\vskip 2pt
\emph{NB :}  This interest rate is negative for small values of $\eta_t$.

\vskip 12pt\noindent
\subsubsection*{\textbf{Conclusion}}
Given the common random shock process $\bW^0$, we solve the stochastic differential equation \eqref{fo:deta_t_final} to find a process $(\eta_t)_{t\ge 0}$ which stays in $(0,1)$. Next we define the short interest rate process $(r_t)_{t\ge 0}$ by \eqref{fo:r_t_final} and with the constant price of capital $q$ given by \eqref{fo:q} all the agents can maximize their expected long run discounted utility of consumption simultaneously and all the markets clear. These are the elements of the desired equilibrium.

%%%%%%%%%%%%%%%%%%%%%%%%%%%
\vskip 12pt\noindent
\subsubsection*{\textbf{Two-Population, Infinite-Horizon Mean Field Game Formulation}}

The model described in this section is the epitome of an \emph{infinite-horizon, two-population Mean Field Game (MFG) with a common noise and no idiosyncratic noise}. We make it explicit directly in the limiting \emph{mean field limit} without motivating it with the description of the finite player analogue. Because we do not know of examples of this type treated in the existing literature, we formulate a rigorous definition in the spirit and with the notations of \cite{CarmonaDelarue_book_I} and \cite{CarmonaDelarue_book_II}, and we accommodate the possibility of idiosyncratic random shocks.

\vskip 4pt
The sources of random shocks are three independent $\RR^d$-valued Wiener processes $\bW^1=(W^1_t)_{t\ge 0}$, standing for the idiosyncratic noise for the players of the first population,  $\bW^2=(W^2_t)_{t\ge 0}$ standing for the idiosyncratic noise for the players of the second population, and $\bW^0=(W^0_t)_{t\ge 0}$ standing for the noise common to all the players. For $i=0,1,2$, we denote by 
$\FF^i=(\cF^i_t)_{t\ge 0}$ the filtration generated by $\bW^i$. 
The MFG problem can be formulated as the conjunction of the following two bullet points:

\begin{enumerate}
\item For any two probability measure $\mu^1_0$ and $\mu^2_0$ on $\RR^d$ and two stochastic flows of (random) probability measures 
$\bmu^1=(\mu_{t}^1)_{t>0}$ and 
$\bmu^2=(\mu_{t}^2)_{t>0}$ on 
$\RR^d$, both adapted to the filtration $\FF^0$ of the common noise,
solve the two optimal control problems:
\begin{equation*}
\sup_{\balpha^1}J^{1,\bmu^1,\bmu^2}(\balpha^1)  
\qquad
\textrm{and}
\qquad 
\sup_{\balpha^2} J^{2,\bmu^1,\bmu^2}(\balpha^2) 
\end{equation*}
over $\FF^1$-progressively measurable $\RR^{k_1}$-valued processes $\balpha^1=(\alpha^1_t)_{t\ge 0}$
and $\FF^2$-progressively measurable $\RR^{k_2}$-valued processes $\balpha^2=(\alpha^2_t)_{t\ge 0}$, where
\begin{equation*}
\begin{split}
&J^{1,\bmu^1,\bmu^2}(\balpha^1) 
= \EE \biggl[ \int_{0}^\infty e^{-\rho t} f_{1}\bigl(t,X_{t}^1,\mu_{t}^1,\mu_{t}^2,\alpha_{t}^1\bigr) dt  \biggr],
\\
&J^{2,\bmu^1,\bmu^2}(\balpha^2)
= \EE \biggl[ \int_{0}^\infty e^{-\rho t} f_{2}\bigl(t,X_{t}^2,\mu_{t}^1,\mu_{t}^2,\alpha_{t}^2\bigr) dt  \biggr],
\end{split}
\end{equation*}
with 
\begin{equation*}
\begin{split}
dX_{t}^1 &= b_{1}\bigl(t,X_{t}^1,\mu_{t}^1,\mu_{t}^2,\alpha_{t}^1\bigr) dt + \sigma_{1}\bigl(t,X_{t}^1,\mu_{t}^1,\mu_{t}^2,\alpha_{t}^1\bigr) dW_{t}^1\\
&\hskip 125pt
+ \sigma_{1,0}\bigl(t,X_{t}^1,\mu_{t}^1,\mu_{t}^2,\alpha_{t}^1\bigr) dW_{t}^0,\\
 dX_{t}^2 &= b_{2}\bigl(t,X_{t}^2,\mu_{t}^1,\mu_{t}^2,\alpha_{t}^2\bigr) dt + \sigma_{2}\bigl(t,X_{t}^2,\mu_{t}^1,\mu_{t}^2,\alpha_{t}^2\bigr) dW_{t}^2\\
 &\hskip 125pt
 + \sigma_{2,0}\bigl(t,X_{t}^2,\mu_{t}^1,\mu_{t}^2,\alpha_{t}^2\bigr) dW_{t}^0,
\end{split}
\end{equation*}
for $t >0$, and $\cL(X_{0}^1)=\mu_{0}^1$
and $\cL(X_{0}^2)=\mu_{0}^2$.
\item Find $\FF^0$-adapted random flows 
$\bmu^1=(\mu_{t}^1)_{t>0}$
and 
$\bmu^2=(\mu_{t}^2)_{t>0}$
such that conditioned on the past of the common noise, almost surely, the marginal distributions of the solutions of the above stochastic control problems coincide with the elements of the probability flows we started from. In other words:  
 \begin{equation*}
\forall t \in [0,T], \quad 
\mu_{t}^1=\cL\bigl(\hat X_{t}^{1,\bmu^1,\bmu^2}|\cF^0_t\bigr),
\quad 
\mu_{t}^2=\cL\bigl(\hat X_{t}^{2,\bmu^1,\bmu^2}|\cF^0_t\bigr),
\end{equation*} 
if  we denote by $\hat \bX^{1,\bmu^1,\bmu^2}$ and 
$\hat \bX^{2,\bmu^1,\bmu^2}$
the solutions of the above optimal control problems.
\end{enumerate} 
In practice, specific assumptions are required on the coefficients $b_1$, $\sigma_1$, $\sigma_{1,0}$, $b_2$, $\sigma_2$, $\sigma_{2,0}$ and the running reward functions $f_1$ and $f_2$ for the stochastic differential equations determining the generic states $X^1_t$ and $X^2_t$ of the two populations to have solutions, and the expected costs to make sense. Moreover, much more restrictive assumptions are required for the existence of a couple of measure flows satisfying the fixed point conditions (2).

\vskip 4pt
Our goal is to explain how the macro-economic model presented in the previous subsection is an instance of such a Mean Field Game. 
\begin{itemize}
\item The individuals in the first population are the households and the individuals of the second population are the  experts.
\item In the present situation, the idiosyncratic noises $\bW^1$ and $\bW^2$ are not present so we can take $\sigma_1=\sigma_2=0$.
\item The generic states $X^1_t$ and $X^2_t$ are the wealths $n^h_t$ and $n^e_t$.
\item As a result, the interpretation of the fixed point condition in item (2) of the above definition of a solution to the MFG is that in equilibrium, the (random) measures $\mu^1_t$ and $\mu^2_t$ should be the conditional distributions of the states $n^h_t$ and $n^e_t$ given the past $\{W^0_s;\;0\le s\le t\}$ of the common noise.
\item In fact as we are about to see, the forms of the coefficients of the state equations as well as of the reward functions are such that the means (first moments) of the probability measures $\mu^1_t$ and $\mu^2_t$ are sufficient statistics. So instead of working with the full random measures $\mu^1_t$ and $\mu^2_t$, we can restrict ourselves to their means $\bar\mu^1_t=\int x\mu^1_t(dx)$ and $\bar\mu^2_t=\int x\mu^2_t(dx)$ which are still functions of the past of the common noise.
\item According to item (1) of the above definition of a solution to the MFG, for each couple of stochastic flows of (random) probability measures 
$\bmu^1=(\mu_{t}^1)_{t>0}$ and 
$\bmu^2=(\mu_{t}^2)_{t>0}$ adapted to the filtration $\FF^0$ of the common noise $\bW^0$, we need to be able to solve the two optimal control problems before we tackle the fixed point problem stated in item (2) of this definition.
Let me argue that this is exactly what we did in the optimization steps of the construction of a general equilibrium for the macro-economic model studied earlier in this section. For the following discussion to be more transparent, we should think of the means $\bar\mu^1_t$ and $\bar\mu^2_t$ as the stochastic processes $(N^h_t)_{t\ge 0}$ and $(N^e_t)_{t\ge 0}$ of the aggregate wealths in the households and experts populations.
\begin{itemize}
\item At each time $t\ge 0$, knowing the averages $\bar\mu^1_t$ and $\bar\mu^2_t$ we can compute $\eta_t=\bar\mu^2_t/(\bar\mu^1_t+\bar\mu^2_t)$ and then the quatity $r_t$ from formula \eqref{fo:r_t_final}.
\item Choosing the control $\alpha^1_t=c^h_t$, the drift function 

\centerline{$b_1(t,n^h,\mu^1_t,\mu^2_t,\alpha^1_t)=r_t n^h_t-c^h_t$,} 
the volatility  $\sigma_{1,0}(t,n^h,\mu^1_t,\mu^2_t,\alpha^1_t)=0$, and the reward function 

\centerline{$f_1(t,n^h,\mu^1_t,\mu^2_t,\alpha^1_t)=\log c^h_t$,}
for the utility of the household, we see that the first optimal control problem in item (1) is exactly what we called the optimization problem of the households which we solved using the Pontryagin maximum principle. 
Notice that in the dynamics of the state, namely in the function $b_1$, the dependence upon the random probability measures $\mu^1_t$ and $\mu^2_t$ appears implicitly through the process $r_t$.
\item Choosing the control $\alpha^2_t=(\iota_t,\theta_t,c^e_t)$, the drift function 

\centerline{\hskip 55pt $b_2(t,n^e,\mu^1_t,\mu^2_t,\alpha^2_t)=\theta_t r_t n^e_t-c^e_t+ (1-\theta_t)n^e_t\bigl(\frac{a-\iota}{q}+\phi(\iota_t)-\delta\bigr)$,}
the volatility 

\centerline{$\sigma_{2,0}(t,n^e,\mu^1_t,\mu^2_t,\alpha^2_t)=(1-\theta_t)n^e_t\sigma$,}
and the reward function

\centerline{$f_2(t,n^e,\mu^1_t,\mu^2_t,\alpha^2_t)=\log c^e_t$}
we see that the second optimal control problem in item (1) is exactly what we called the optimization problem of the experts which we solved using the Pontryagin maximum principle. 
As before, the dependence upon the random probability measures $\mu^1_t$ and $\mu^2_t$ appears implicitly through the process $r_t$.
\end{itemize}
\item The fixed point condition in item (2) of the definition of the MFG guarantees that $\bar\mu^1_t=N^h_t$ and $\bar\mu^2_t=N^e_t$ so that the process $(\eta_t)_{t\ge 0}$  is indeed the wealth share of the experts, and the process $(r_t)_{t\ge 0}$ is indeed the short interest rate process and all the clearing conditions are satisfied.
\end{itemize}

%%%%%%%%%%%%%%%%%%%%%%%%%%%%%%%%%%
\subsection{\textbf{Conclusion}}
So the Nash equilibrium of the two-population Mean Field Game coincides with the general equilibrium constructed in the previous subsections. But what have we gained? Aren't we making matters worse? \emph{Johann Wolfgang von Goethe} once said
\begin{center}
\begin{quote}
\textbf{\emph{\color{blue}
"Mathematicians are like Frenchmen: whatever you say to them they translate into their own language and forthwith it is something entirely different." 
}}
\end{quote}
\end{center}
and I would add that it has to be even worse when the mathematician is French ! Still, once the macro-economic general equilibrium model is reformulated as a Mean Field Game, the technology developed during the last $15$ years to analyze these models can be brought to bear to gain insight in these macro-economic models. Among them, 1) numerical methods to compute solutions and statics, 2) convergence of finite populations models and quantification of the finite size effects, 3) analysis of the uniqueness of the equilibria, or lack thereof, 4) analysis of the centralized optimization counterpart and comparisons of the global welfares (e.g. computation of the price of anarchy), $\ldots$ .

%%%%%%%%%%%%%%%%%%%%%%%%%%%%%%%%%%%%%%%%%%%%%
\section[Moral Hazard \& Contract Theory]{\textbf{Moral Hazard \& Contract Theory}}
%%%%%%%%%%%%%%%%%%%%%%%%%%%%%%%%%%%%%%%%%%%%%
\label{se:contract}
After reviewing some of the major historical developments in contract theory, and explaining the basics of the models, we concentrate on two recent papers using mean field games for the purpose of extending the reach of possible applications of the theory to large populations of agents. Because this topic was not discussed as a possible application of Mean Field Games in the books \cite{CarmonaDelarue_book_I,CarmonaDelarue_book_II}, we spend a significant amount of time reviewing the underpinnings of the economic theory as well as some of the recent applications involving mean field models.

\vskip 6pt
Put in layman's terms, the purpose of economic contract theory is to address questions of the type: a) How should a government control a flu outbreak by encouraging citizens to vaccinate? 
b) How should taxes be levied to influence people's  consumption, saving and investment decisions? 
c) How should an employer incentivize and compensate their employees in order to boost  productivity? 
While these are mundane objectives, they should shed some initial light on the type of problems contract theory can encompass. The economic lingo we shall use and try to elucidate includes terms like
\begin{center}
\emph{Agency Problem, Contract Theory,  Moral Hazard}, and  \emph{Information Asymmetry}. 
\end{center}

\vskip 4pt
\begin{itemize}
\item The \emph{agency problem} refers to a conflict of interest between two parties when one of them is expected to behave in the other party's best interests. 

\item The purpose of \emph{contract theory} is to study how an economic agent can design and structure a contractual agreement to incentivize another agent to behave in his or her best interest. This problem is most often set up when the informations available to the two agents are not the same. 

\item In economics and in finance, \emph{moral hazard} refers to a situation when an agent has an incentive to take excess risk because they do not bear the full consequences of that risk. 

\item In contract theory and economics, \emph{information asymmetry} deals with the study of decisions in transactions where one party has more or better information than the other. 

\item The \emph{principal - agent problem} occurs when one agent makes decisions on behalf of another person or entity called the principal.
\end{itemize}

Earliest contributions to this important field of economics concentrated on 
static one period models. They are attached to the names of Mirrlees  \cite{Mirrlees} and Holmstr\"om \cite{Holmstrom79,Holmstrom82}. The first dynamic models were introduced by Holmstr\"om and Milgrom \cite{HolmstromMilgrom87,HolmstromMilgrom91} and it had to wait almost two decades for Sannikov's breakthrough \cite{Sannikov08,Sannikov12}. By focusing on Markov diffusion models in continuous time, and using the weak formulation of stochastic control problems to capture moral hazard, Sannilov created a new wave of interest, especially among financial mathematicians. The book of 
Cvitanic and Zhang focusing on the use of Backward Stochastic Differential Equations (BSDEs) is a case in point, and the more recent work of Cvitanic, Possama\"{i} and Touzi \cite{CvitanicPossamaiTouzi_contract}
highlighting the real nature of Sannikov's trick, will be instrumental in the developments we present in this section.

\vskip 4pt\noindent
\textbf{NB: } Bengt Holmstr\"om and Oliver Hart were awarded the Nobel Memorial Prize in Economic Sciences in 2016 for their work on Contract Theory.

\subsection{\textbf{Standard Data for a Contract Theory Problem}}
In a classical contract theory model, two parties are present: 
\begin{enumerate}
\item the \textbf{\emph{principal}} who devises a \emph{contract}, according to which \textbf{incentives} are given to, and/or \textbf{penalties} are imposed on, 
\item the  \textbf{\emph{agent}} who may accept the contract and work for the principal. 
\end{enumerate}
The following are the major assumptions which are usually made implicitly or explicitly in a contract theory model.

\begin{itemize}
\item It is assumed that all the agents are \emph{rational} in the sense that  
they behave optimally to maximize their own utilities,
controlling the tradeoff between the rewards/penalties they received and the efforts they put in. 

\item The principal designs the \textbf{contract} 

\item After reviewing the terms of the contract, the agent may walk away. We assume that the agent has a threshold level (e.g. minimum reward, ..... ) below which they think the contract is not worth the effort. We call this threshold the \emph{reservation utility} of the agent.
\item The principal observes the agent actions only partially. This is the source of  information asymmetry and moral hazard in the problem.
\end{itemize}

From a mathematical point of view, the fact that the principal de facto does not see all of the agent behavior forces a special formulation of the optimization problem faced by the principal. As far as we know, Sannikov was the first one to realize that this situation was not accommodated properly by the usual mathematical strong formulation of stochastic games and stochastic control. He proposed to use the weak formulation of stochastic control (also called the martingale approach to stochastic control) to set the principal optimization problem. We now explain what we mean by weak formulation to accommodate moral hazard.

%%%%%%%%%%%%%%%%%%%%%%%%%%%%%%%%
\subsection{\textbf{Weak Formulation for Control \& Two-Player Games}}

We denote by $X^0_t$ the state of the system at time $t$. We shall give detailed explanations of what the state is in each of the applications considered below.
We assume that the system is controlled by one single controller who takes \emph{actions} to implement their control. The actions of the controller affect the state of the system and the level of the \emph{cost / reward}. This describes the set-up of a classical control problem. Games and stochastic games differ in the sense that several controllers can act on the same system, i.e.
several controllers (players, agents) take actions. In this case,
each player has a cost / reward to worry about, and the whole system is affected by the individual actions and their \emph{interactions}. The weak formulation of these optimization problems, also called the
\emph{martingale approach}, is perfectly suited to  information asymmetry and moral hazard.
In this set-up the trajectories $t\hookrightarrow X^0_t(\omega)$ are not affected by the actions taken by the controllers.
Only the likelihood of the scenarios given by the trajectories can be changed by the controls. In other words, only the
\emph{distribution} of the process $\bX^0=(X^0_t)_{0\le t\le T}$ is affected by controls. In words, one can surmise that
the choice of a control is equivalent to the choice of a law for the state process.

\vskip 2pt
Still, it may not be completely clear why the presence of moral hazard and the lack of symmetry of information suggest the use of the weak formulation. Given the terms of the contract, the Agent chooses controls which influence their own reward, as well as the rewards of the Principal.
Let us denote by  $\alpha_t$ the control (effort level) of the agent at time $t$. The agent sees the state $X_t$ as it is impacted by the control, and the agent optimizes their \emph{cost / reward} according to the terms of the contract.

\vskip 2pt
On the other end, through the terms of the contract, the Principal chooses the remuneration conditions of the agent. But he or she does so without observing directly the effort level $\alpha_t$ of the agent, observing only partially the state $X_t$,
and seeing the impact of the agent's effort only through the expected returns he or she is getting,
de facto through the values of expected quantities

\vskip 2pt
Roughly speaking, the principal guesses the expected value of his or her returns through the distribution of the output of the agent.
This is exactly what the weak formulation approach is trying to capture. See below for the mathematical details of this approach.

\vskip 4pt
After this quick and informal review of moral hazard and classical contract theory, we now introduce the application we have in mind in this chapter.

\subsection{\textbf{Contract between One Principal and MANY Agents}}
The main features of the new framework can be summarized as follows: 
\begin{enumerate}
\item \textbf{\emph{ONE principal}} who devises \emph{one single contract}, according to which \textbf{incentives} are given to, and/or \textbf{penalties} are imposed on, 
\item \textbf{\emph{MULTIPLE agents}} who see the same contract and may accept it and work for the principal. 
\end{enumerate}
As before, the major assumptions can be captured in a small number of bulllet points:
\begin{itemize}
\item The principal designs a contract hoping to maximize their own utility;
\item The agents are \emph{rational} so they will also try to maximize their own utilities;
\item The agents have their reservation utilities and so they may decide to walk away;
\item The agents are statistically identical in that their contracts are the same;
\item They behave \textbf{selfishly} and maximize their utilities; 
\item We assume that they reach a Nash equilibrium;
\item The Principal designs the contract anticipating that the agents will reach a Nash equilibrium. 
\end{itemize}
In some sense, we can say that we are considering the problem of one principal contracting a field of agents. As before, the principal does not see (and cannot control) the individual actions taken by the agents. The principal only feels the overall expected value of the reward he or she gets from the actions of the agents. This information asymmetry creates the moral hazard which the model captures through the weak formulation of the optimization problem.

\vskip 4pt
Below, we give mathematical details when the state space is the Euclidean space $\RR^d$ (originally treated by Elie, Mastrolia and Possama\"{i} in \cite{ElieMastroliaPossamai})
and the case of finite states treated by Carmona and Wang in \cite{CarmonaWang_discrete,CarmonaWang_pa}.

%%%%%%%%%%%%%%%%%%%%%%%%%
\subsection{\textbf{Continuous State Case. }}

The weak formulation is best set-up using the canonical representation of the state process by assuming that
$\Omega$ is the space of continuous functions from $[0,T]$ to $E$ (typically $E=\RR$ or $E=\RR^d$), $W_t(\omega)=\omega(t)$ for $t\ge 0$ gives the coordinate process,
$\mathbb{F} := (\mathcal{F}_t)_{t \in [0,T]}$ is the natural filtration generated by the process $\bW=(W_t)_{t\ge 0}$,
$\mu_0$ is a fixed probability measure on $E$ which serves as the initial distribution of the state, i.e. $X_0\sim \mu_0$,
$\mathcal{F} := \mathcal{F}_T$ if we work on a finite time horizon $[0,T]$.
$\mathbb{P}$ is the Wiener measure on $(\Omega, \mathbb{F}, \mathcal{F})$ so that  $\bW$ is a Wiener process, and $X^0_t=\xi_0+\int_0^t\sigma(s,X^0_\cdot)dW_s$ for some Lipschitz function $(s,x)\mapsto \sigma(s,x)$ which is assumed to be bounded from above and below away from $0$ uniformly in $s$ and $x$. In order to be consistent with the existing literature on the subject, we allow the coefficients to depend upon the past history of the states. We use the notation $X_\cdot$ and $x_\cdot$ to denote the whole trajectories of the state. Note also that:
\begin{equation}
\label{fo:dX^0_t}
dX^0_t=\sigma(t,X^0_\cdot)dW_t,\qquad \text{ under \;} \PP
\end{equation}
irrespective of which control is chosen by the agent.

Next we introduce $\AA$, the space of admissible control strategies (representing the agents effort levels). The elements of $\AA$ are adapted processes $\balpha= (\alpha_t)_{0\le t\le T}$ which may satisfy further properties to be specified later on. 
Next we introduce the drift function $b$ (the only part of the dynamics of the state controlled by the agent). We assume that the drift $(t,x,\alpha)\mapsto b(t,x_\cdot,\alpha)\in \RR^d$ is bounded and progressively measurable. 
For each admissible control strategy $\balpha$, we denote by $\PP^{\balpha}$ the state distribution when the effort level of the agent is $\balpha$. It is defined by its density with respect to the measure $\PP$ given by: 
$$
\frac{d\PP^{\balpha}}{d\PP}=\cE\Bigl[\int_0^T\sigma(t,X^0_\cdot)^{-1}b(t,X^0_\cdot,\alpha_t)dW_t\Bigr]
$$
where $\cE(\bM)=\exp[M_t-\frac12<M,M>_t]$ denotes the Doleans exponential of the continuous square integrable martingale $\bM=(M_t)_{0\le t\le T}$. Girsanov theorem implies that:
$$
dX^0_t= b(t,X^0_\cdot,\alpha_t)dt +\sigma(t,X^0_\cdot)dW^{\balpha}_t,\qquad \text{ under } \PP^{\balpha}
$$
where 
$$
W^{\balpha}_t=W_t-\int_0^t\sigma(s,X_\cdot)b(s,X_\cdot,\alpha_s)ds
$$
is a Brownian motion under the measure $\PP^{\balpha}$.
So the same state process $\bX^0$, constructed in \eqref{fo:dX^0_t} independently of the controls $\balpha$, now appears as the state of the process controlled by $\balpha$ if one looks at its evolution under the probability measure $\PP^{\balpha}$. More on that remark below.

\vskip 2pt
We now finalize the weak formulation of the problem by describing the behaviors of the agents in this set-up.
\vskip 1pt
The principal offers a contract $(\br,\xi)$ where
\begin{itemize} 
\item $\br=(r_t)_{0\le t\le T}$ is an adapted process representing the  payment stream;
\item $\xi$ is a random variable representing a terminal payment.
\end{itemize} 
The agent decides whether or not to accept the contract and work for the principal, and if he or she does accept,
chooses an effort level $\balpha=(\alpha_t)_{0\le t\le T}$ to maximize their expected overall reward:
$$
J^{\br,\xi}(\balpha,\mu_0)=\EE^{\PP^{\balpha}}\Bigl[U_A(\xi) +\int_0^T[u_A(r_t)-c(t,X_\cdot,\alpha_t)]dt\Bigr]
$$
where  $u_A$ is the agent running utility, $U_A$ is the agent terminal utility,
and $c(t,x_\cdot,\alpha)$ is the cost for applying the effort level $\alpha$ at time $t$ when the history of the state is $x_{[0,t]}$.

\vskip 2pt
Given this rational expected behavior of the agent, the optimization problem of the principal can be formulated in the following way:

For each contract $(\br,\xi)$, assuming knowledge of the utility and cost functions of the agent
and assuming that the agent is rational, the Principal computes an \emph{optimal} effort level $\balpha^*=(\alpha^*_t)_{0\le t\le T}$ 
$$
\balpha^*\in\arg \inf_{\balpha\in\AA}J^{\br,\xi}(\balpha,\mu_0)
$$
which the agent should choose, and then, search for an optimal contract $(\br^*,\xi^*)$
$$
(\br^*,\xi^*)\in\arg \inf_{(\br,\xi)}\EE^{\PP^{\balpha^*}}\Bigl[U_P\Bigl(X_T-\xi-\int_0^Tr_tdt
\Bigr)\Bigr]
$$
where $U_P$ is the (terminal) utility of the principal.

\vskip 2pt
This is a typical instance of a \emph{Stackelberg game} between the principal \emph{going first} and the agent.

\vskip 12pt\noindent
\subsubsection*{\textbf{What Changes with a Large Number of Agents}}
We assume that the agents, while \emph{competing} with each other, behave similarly (this is the form of symmetry assumption in force in mean field game models), and
because of their large number, their individual influences on the aggregate quantities are \emph{negligeable}.

\vskip 2pt
In these conditions, the optimization problem of the Principal can be formulated as before.
Knowing the utility and cost functions of the agents, the Principal assumes that for each contract $(\br,\xi)$, the agents settle in a Mean Field Nash Equilibrium, so for each $(\br,\xi)$,  the Principal 
\begin{itemize}
\item solves the MFG of the agents
\item computes the effort level $\balpha^*=(\alpha^*_t)_{0\le t\le T}$ of the Nash Equilibria he or she can compute 
\item then search for an optimal contract $(\br^*,\xi^*)$
$$
(\br^*,\xi^*)\in\arg \inf_{(\br,\xi)}\EE^{\PP^{\balpha^*}}\Bigl[U_P\Bigl(X_T-\xi-\int_0^Tr_tdt
\Bigr)\Bigr]
$$
where $U_P$ is the (terminal) utility of the principal.
\end{itemize}

As before, this is a form of Stackelberg game between the principal \emph{going first}, and the field of agents \emph{going next}. But now, the dynamics of the state and the cost/reward functions depend upon the distribution of the state in the sense that:
$$
b(t,X_t,\alpha_t,\mu_t)
\qquad\text{and}\qquad 
c(t,X_t,\alpha_t,\mu_t)
$$
where $\mu_t$ is the distribution at time $t$ of the state $X_t$ under $\PP^{\balpha}$.

\vskip 2pt
Details about the formulation of the problem and an example of solvable model (essentially from the linear-quadratic family) can be found in \cite{ElieMastroliaPossamai}. 

\begin{remark}
One of the major shortcomings of the approach described above is the fact that the agents can only control the drifts of their states. This is due to the reliance on Girsanov's change of measure. Allowing the volatility to be controlled requires the representation of the value functions of the optimization problems by so-called 2BSDEs instead of regular Backward Stochastic Differential Equations (BSDEs). The analysis becomes significantly more technical. The interested reader may want to look at the recent work \cite{Elie_et_al} of Elie, Hubert, Mastrolia, and Possama\"i for an attempt in this direction.
\end{remark}

\vskip 4pt
Next we consider the same contract theory model when the state space is finite. We go over a numerical application in detail to illustrate with a few statics, the informational content of the equilibrium when we can actually compute it.

\subsection{\textbf{The Discrete State Case}}

The framework of the above discussion is based on the theory of diffusion processes in continuous time and its application to problems of stochastic control.
The states are living in Euclidean spaces, their dynamics are modeled by stochastic differential equations, and sophisticated tools from stochastic analysis, starting with Girsanov's theory of changes of measure, are brought to bear in order to formalize the asymmetry of information and the weak formulation appropriate for the optimization of the Principal.
Stochastic dynamical systems taking values with finite state spaces are often used in applications for which numerical implementations are of crucial importance. Strangely enough, what seems like a simplification at first, after all finite state spaces should be easier to handle than continuous spaces, may not always make the theoretical analysis easier. Here, we review recent works attempting to port the strategy outlined in the previous section to this case. In particular, we explain how to set up the weak formulation for Mean Field Games with finitely many states, and we implement the steps previously outlined in the diffusion case in the framework of finitely valued state processes.

\begin{remark}
Mean Field Games with finitely many states have caught the attention of the MFG crowd throughout the past decade.
All the works dated before 2017 are reviewed thoroughly and commented from an historical perspective in 
\cite[Section 7.2]{CarmonaDelarue_book_I}. See also \cite[Section 7.1.9]{CarmonaDelarue_book_II} for a discussion of models with major and minor players, the analysis of which bears much resemblence to some of the steps taken to state and solve contract theory problems with a large number of agents. For the sake of completeness, we mention some of the works on finite state space mean field games which appeared since then, and which we know of. The probabilistic approach to finite state mean field games is advocated by Cecchin and Fischer in \cite{CecchinFischer}, Bayraktar and Cohen derived the equivalent of the master equation  in \cite{BayraktarCohen}, and the convergence problem is studied in \cite{DoncelGastGaujal3} by Doncel, Gast and Gaujal, and in \cite{CecchinPelino} by Cecchin and Pelino. Finally, we note that these models can exhibit all sorts of behavior as shown for example in \cite{CecchinDaiPraFischerPelino} where Cecchin, Dai Pra, Fischer and Pelino identify a two-state model without uniqueness.
\end{remark}

The following is a review of  results of Carmona and Wang borrowed from \cite{CarmonaWang_discrete,CarmonaWang_pa}.

\vskip 12pt\noindent
\subsubsection*{\textbf{The Canonical Process for the Discrete Case}}
In this section, we assume that the state space is the finite set $E=\{e_1,\dots,e_m\}$,  where for the sake of mathematical convenience we shall assume that the $e_i$'s are the unit vectors of the canonical basis of $\mathbb{R}^m$.
The state process $\bX^0 = (X^0_t)_{0\le t\le T}$ will be a continuous-time Markov chain with $m$ states whose sample paths
$t\rightarrow X_t$ are \emph{c\`adl\`ag}, i.e. right continuous with left limits, and continuous at $T$ (i.e. $X_{T-}=X_T$).
In analogy with the Euclidean state space case, we introduce the following
\emph{canonical representation:}
\begin{itemize}
\item $\Omega$ is the space of c\`adl\`ag functions from $[0,T]$ to $E$, continuous at $T$;
\item $X^0_t(\omega) := \omega_t$ is the coordinate process;
\item $\mathbb{F} := (\mathcal{F}_t)_{t \in [0,T]}$ is the natural filtration generated by $\bX^0$;
\item $\mathbf{p}^{\circ}$ is a fixed probability on $E$;
\item $\mathcal{F} := \mathcal{F}_T$;
\item $\mathbb{P}$ is the unique probability on $(\Omega, \mathbb{F}, \mathcal{F})$ for which $\bX^0$ is a continuous-time Markov chain with initial distribution $\mathbf{p}^{\circ}$
and transition rates between any two different states equal to $1$. So if $i\neq j$ and $\Delta t >0$,
$$
\mathbb{P}[X_{t+\Delta t} = e_j | \mathcal{F}_t ] = \mathbb{P}[X_{t+\Delta t} = e_j| X_t]\quad \text{and}\quad
\mathbb{P}[X_{t+\Delta t} = e_j | X_t = e_i] = \Delta t + o(\Delta t)
$$
\end{itemize}
Using the result of \cite{Cohen2008,Cohen2010}
we see that the process $\bX^0$ has the  representation:
$$
X^0_t = X^0_0 + \int_{(0,t]} Q^0\cdot X^0_{t-} dt + \mathcal{M}_t,
$$
where $Q^0$ is the square matrix  whose entries are given by:
\begin{itemize}
\item $Q^0_{i,i}=-(m-1), \qquad i=1,\ldots,m$
\item $Q^0_{i,j}=1$ if $i\ne j$
\end{itemize}
and $\bcM=(\mathcal{M}_t)_{t\ge 0}$ is a $\mathbb{R}^m$-valued $\mathbb{P}$-martingale.
We sometime use the symbol $\cdot$ to emphasize matrix multiplication.
The predictable quadratic variation of the martingale $\bcM$ under $\mathbb{P}$ is given by the formula:
\begin{equation}\label{eq:quad_var}
\langle \bcM, \bcM \rangle_t = \int_0^t \psi_t dt,
\end{equation}
where $\psi_t$ is given by:
\begin{equation}\label{eq:matrix_psi}
\psi_t := diag(Q^0 \cdot X^0_{t-}) - Q^0 \cdot diag( X^0_{t-}) - diag( X^0_{t-}) \cdot Q^0.
\end{equation}

\vskip 12pt\noindent
\subsubsection*{\textbf{Players' Controls. }}
We assume that all the agents can take actions which are elements $\alpha$ of a closed convex subset $A$ of a Euclidean space $\RR^k$. For any agent, the set $\AA$ of admissible (control) strategies is the set of $A$-valued, $\FF$-predictable process $\balpha=(\alpha_t)_{0\le t\le T}$.
The space of probability measures on the state space being the simplex
$$
\cP(E)=\mathcal{S} := \{ p \in \mathbb{R}^m;\; \sum_{i=1}^m p_i = 1, p_i \ge 0\},
$$
the controlled state processes will have dynamics determined by $Q$-matrices
$$
Q(t,\alpha,p,\nu)=[q(t, i, j, \alpha, p,\nu)]_{1\le i,j\le m}
$$
where $q$ is a function
\[
[0,T] \times \{1,\dots,m\}^2 \times A \times \mathcal{S}\times\mathcal{P}(A) \rightarrow q(t, i, j, \alpha, p,\nu)\in\RR.
\] 
We shall assume that:
\begin{itemize}
\item[(i)]  $Q(t, \alpha, p,\nu)$ is a Q-matrix.

\item[(ii)]  $0 < C_1< q(t,i,j,\alpha,p,\nu) < C_2$.

\item[(iii)] For all $(t,i,j)\in[0,T]\times E^2$, $\alpha,\alpha' \in A$, $p,p' \in \mathcal{S}$ and $\nu,\nu'\in\mathcal{P}(A)$, we have:
$$
|q(t,i,j,\alpha,p,\nu) - q(t,i,j,\alpha', p', \nu')| \le C( \|\alpha - \alpha'\| + \|p - p'\| + \mathcal{W}_1(\nu,\nu'))
$$
where $\mathcal{W}_1$ is the $1$-Wasserstein distance on $\cP(A)$.
\end{itemize}

\vskip 4pt
Assumption (i) is natural given that we start from a canonical process $\bX^0$ which is already a continuous time Markov chain. The strictly positive lower bound of assumption (ii) may appear to be restrictive at first, but if we understand that in fact, it is sufficient that it is satisfied for a given power of the matrix, this assumption guarantees that all states are attainable through appropriate actions, and this is a desirable feature for control problems to be solvable. Finally, assumption (iii) is to be expected if one thinks of the mathematical analysis needed to study these models.

\vskip 6pt
Now, given
\begin{itemize}
\item $\balpha=(\alpha_t)_{0\le t\le T}\in\AA$
\item $\bp=(p_t)_{0\le t\le T}$ a flow of probability measures on $E$
\item $\bnu=(\nu_t)_{0\le t\le T}$ a flow of probability measures on $A$
\end{itemize}
we define the martingale  $\bL^{(\balpha,\bp,\bnu)}=(L^{(\balpha,\bp,\bnu)}_t)_{0\le t\le T}$ by
$$
L^{(\balpha,\bp,\bnu)}_t := \int_0^t X_{s^-}^* \cdot(Q(s, \alpha_s, p_s, \nu_s) - Q^0)\cdot\psi_s^+ \cdot d\mathcal{M}_s.
$$
Simple calculations show that
\begin{align*}
\Delta L^{(\balpha,\bp,\bnu)}_t  =&\;\; X_{t-}^*\cdot(Q(t, \alpha_t, p_t, \nu_t) - Q^0)\cdot\psi_t^+\cdot\Delta X_t,
\end{align*}
which is either $0$ when there is no jump at time $t$, or $q(t,i,j,\alpha_t,p_t,\nu_t) - 1$ if the state jumps from state $i$ to state $j$ at time $t$. In any case, $\Delta L^{(\balpha,\bp,\bnu)}_t \ge -1$. Also, the Doleans exponential $\mathcal{E}(\bL^{(\balpha,\bp,\bnu)})$ is uniformly integrable so we can apply the extension of Girsanov's theorem to processes with jumps, and define the probability measure $\mathbb{Q}^{(\balpha,\bp,\bnu)}$ by its density with respect to $\PP$:
$$
\frac{d\mathbb{Q}^{(\balpha,\bp,\bnu)}}{d\mathbb{P}} := \mathcal{E}(\bL^{(\balpha,\bp,\bnu)})_T,
$$
which guarantees that the process $\bcM^{(\balpha,\bp,\bnu)}=(\bcM^{(\balpha,\bp,\bnu)}_t)_{0\le t\le T}$ defined as:
\begin{equation}\label{eq:martingale_M}
\mathcal{M}^{(\balpha,\bp,\bnu)}_t := \mathcal{M}_t - \int_{0}^t ( Q^*(s, \alpha_s, p_s, \nu_s) - Q^0)\cdot  X_{s-} ds,
\end{equation}
is a $\mathbb{Q}^{(\balpha,\bp,\bnu)}$-martingale, and the  canonical decomposition of $\bX^0$ under $\mathbb{Q}^{(\balpha,\bp,\bnu)}$ reads:
\begin{equation}\label{eq:X_decomp}
X^0_t = X^0_0 + \int_0^t Q^*(s, \alpha_s, p_s, \nu_s)\cdot X^0_{s-} dt + \mathcal{M}^{(\balpha,\bp,\bnu)}_t,
\end{equation}
showing that under $\mathbb{Q}^{(\balpha,\bp,\bnu)}$, the stochastic intensity rate of $\bX^0$ is $Q(t, \alpha_t, p_t, \nu_t)$. Notice that $X^0_0$ has still distribution $\mathbf{p}^{\circ}$
and if $\alpha_t = \phi(t, X^0_{t-})$ for some measurable function $\phi$, $\bX^0$ is a continuous-time Markov chain with jump rate intensity  $q(t,i,j,\phi(t,i),p_t,\nu_t)$ under the measure $\mathbb{Q}^{(\balpha,\bp,\bnu)}$.

So as explained earlier in our first mention of the weak formulation, the choice of the control of the agents does not affect the trajectories of the state process, but it does influence the probability distribution, $\mathbb{Q}^{(\balpha,\bp,\bnu)}$ in the present case, which determines the expected costs and rewards of the principal.

\vskip 12pt\noindent
\subsubsection*{\textbf{Principal's Optimization Problem}}
The reward of the Principal depends on  the distribution of the agents' states
and the payments made to the agents. We use the notation
\begin{itemize}
\item $c_0:[0,T]\times\mathcal{S} \rightarrow \mathbb{R}$  for the running cost function 
\item $C_0:\mathcal{S} \rightarrow \mathbb{R}$ for the terminal cost function
\end{itemize}
defining the costs of the Principal.
Now, assuming that all the agents choose $\balpha=(\alpha_t)_{0\le t\le T}$ as their control strategy,  
that the resulting flow of marginal distribution of the agents' states is $\bp=(p(t))_{t\in[0, T]}$,
and the  contract offered by the principal is $(\br, \xi)$,
the \textbf{principal's expected total cost} is given by:
$$
J_0^{\balpha,\bp}(\br,\xi):=\mathbb{E}^{\mathbb{Q}^{(\balpha,\bp)}}\left[\int_0^T [c_0(t, p(t)) + r_t] dt +C_0(p(T)) + \xi\right].
$$

\vskip 12pt\noindent
\subsubsection*{\textbf{Agents' Mean Field Equilibria}}

We assume that, for a given contract $(\br,\xi)$ proposed by the principal, the agents reach a Nash equilibrium as defined rigorously in the following statement. 

\begin{definition}
\label{def:equilibrium}
The couple $(\hat \balpha,\hat \bp)$ is a Nash equilibrium for the contract $(\br,\xi)$,  $(\hat \balpha,\hat \bp) \in \mathcal{N}(\br,\xi)$ in notation, if:

\vspace{2mm}
\noindent (i) $\hat\balpha$  is the \textbf{best response} to the behavior of the other agents, i.e. it minimizes the cost when the agent is committed to the contract $(\br,\xi)$ and the flow of marginal distributions of all the agents is given by the flow $\hat \bp$:
$$
\hat\balpha = \arg\inf_{\balpha \in \mathbb{A}}\mathbb{E}^{\mathbb{Q}^{(\balpha,\hat \bp)}}\left[\int_0^T [c(t, X_t, \alpha_t, \hat p(t)) - u(r_t)] dt - U(\xi)\right].
$$
\noindent (ii) $(\hat\balpha,\hat \bp)$ satisfies the \textbf{fixed point} condition:
\begin{equation}\label{eq:def_nasheq_cons1}
\forall t \in [0,T] \qquad \hat p(t) = \mathbb{E}^{\mathbb{Q}^{(\hat\balpha,\hat \bp)}}[X_t].
\end{equation}
Notice that this equation is equivalent to 
$
\hat p_i(t) = \mathbb{Q}^{(\hat\balpha,\hat \bp)}[X_t = e_i]
$
for all $t \in [0,T]$ and  $i\in\{1,\dots,m\}$.
\end{definition}

\vskip 12pt\noindent
\subsubsection*{\textbf{Principal's Optimal Contracting Problem}}

As we already explained, the Principal minimizes his or her total expected cost assuming the agents reach a Nash equilibrium.
So we only consider contracts $(\br,\xi)$ that result in \emph{at least one} Nash equilibrium. We denote by $\mathcal{C}$ the set of all admissible contracts.
To implement the participation constraint, we disregard the equilibria in which the agent's expected total cost is above a given threshold $\kappa$, i.e. 
\emph{ \emph{take-it-or-leave-it} behavior of the agents in contract theory: if the agents' expected total costs exceed a certain threshold, they should be able to turn down the contract. 
}
In summary, the optimization problem for the principal reads:
{\small
$$
V(\kappa) := \inf_{(\br,\xi)\in\mathcal{C}}\inf_{\substack{(\balpha,\bp) \in \mathcal{N}(\br,\xi)\\ J^{\br,\xi}(\balpha,\bp) \le \kappa}}\mathbb{E}^{\mathbb{Q}^{(\balpha,\bp)}}\left[\int_0^T [c_0(t, p(t)) + r_t] dt +C_0(p(T)) +\xi\right],
$$
}

\vskip12pt\noindent
\subsubsection*{\textbf{Solving the Individual Agent Optimization Problem}}
For the agent's optimization problem, we introduce the Hamiltonian  
$
H: [0,T] \times E \times \mathbb{R}^m\times A \times \mathcal{S} \times \mathbb{R} \rightarrow \mathbb{R} 
$ 
defined by:
$$
H(t,x,z,\alpha,p, r) := c(t,x,\alpha,p) - r + x^*(Q(t, \alpha, p) - Q^0)z.
$$
and $H_i(t,z,\alpha,p, r) = H(t,e_i,z,\alpha,p, r)$.
We assume that there exists a \textbf{unique} minimizer $\hat \alpha_i(t,z,p)$ of $\alpha \rightarrow H_i(t,z,\alpha,p,r)$ and that  it is \emph{uniformly Lipschitz} in $z$, and we use the notations:
$$
\hat H_i(t,z,p, r) = H_i(t,z,\hat\alpha_i(t,z,p),p, r)
\;\text{and}\;
\hat H(t,x,z,p, r) = \sum_{i=1}^m \hat H_i(t,z,p, r) \bf{1}_{x=e_i}
$$
for the maximized Hamiltonians.

\vskip 12pt\noindent
\subsubsection*{\textbf{BSDEs driven by Markov Chains}}
Following the strategy at the root of the weak formulation of stochastic control problems, we introduce the BSDEs :
\begin{equation}\label{eq:bsde_agent_total_cost}
Y_t = -U(\xi) + \int_t^T H(s, X_{s-}, Z_s, \alpha_s, p(s), u(r_t)) ds - \int_t^T Z_s^*d\mathcal{M}_s.
\end{equation}
\begin{equation}\label{eq:bsde_agent_optimization}
Y_t = -U(\xi) + \int_t^T \hat H(s, X_{s-}, Z_s, p(s), u(r_t)) ds - \int_t^T Z_s^*d\mathcal{M}_s.
\end{equation}
and we prove the following representation theorems by inspection. Notice that in the present situation, the BSDEs are driven by continuous time Markov chains.
\begin{lemma}
\label{lem:bsde_characterization}
For each fixed contract $(\br,\xi)$, $\balpha\in\mathbb{A}$ and measurable mapping $\bp: [0,T]\rightarrow\mathcal{S}$,

\vspace{1mm}
\noindent(i) the BSDE (\ref{eq:bsde_agent_total_cost}) admits a unique solution $(\bY, \bZ)$ and we have 
$$
J^{\br,\xi}(\balpha,\bp) = \mathbb{E}^{\mathbb{P}}[Y_0].
$$
\vspace{1mm}
\noindent(ii) The BSDE (\ref{eq:bsde_agent_optimization}) admits a unique solution $(\bY,\bZ)$ and we have 
$$
\inf_{\balpha \in \mathbb{A}}J^{\br,\xi}(\balpha,\bp) = \mathbb{E}^{\mathbb{P}}[Y_0].
$$ 
In addition, the optimal control of the agent is $\hat \alpha(t, X_{t-}, Z_t, p(t))$.
\end{lemma}

\vskip 12pt\noindent
\subsubsection*{\textbf{Nash Equilibria as Solutions of BSDEs}}
Let $(\bY,\bZ,\balpha,\bp,\mathbb{Q})$ be a solution to the \textbf{McKean-Vlasov BSDE} system:
\begin{align}
Y_t =&\;\; -U(\xi) + \int_t^T \hat H(s, X_{s-}, Z_s, p(s), u(r_s)) ds - \int_t^T Z_s^* d\mathcal{M}_s,\label{eq:bsde_mfg_1}\\
\mathcal{E}_t =&\;\;  1 + \int_0^t \mathcal{E}_{s-} X_{s-}^* (Q(s, \alpha_s, p(s)) - Q^0)\psi_s^+ d\mathcal{M}_s,\label{eq:bsde_mfg_2}\\
\alpha_t = &\;\; \hat \alpha(t,X_{t-},Z_t,p(t)),\label{eq:bsde_mfg_3}\\
p(t) = &\;\; \mathbb{E}^{\mathbb{Q}}[X_t],\;\;\frac{d\mathbb{Q}}{d\mathbb{P}} = \mathcal{E}_T.
\label{eq:bsde_mfg_4}
\end{align}
\begin{itemize}
\item $\bY$ is an adapted  c\`adl\`ag process such that $\mathbb{E}^{\mathbb{P}}[\int_0^T Y_t^2] < +\infty$ for all $t \in [0,T]$, 
\item $\bZ$ is an adapted square integrable left-continuous process,
\item $\balpha\in\mathbb{A}$, $\bp:[0,T]\rightarrow\mathcal{S}$ is measurable, $\mathbb{Q}$ is a probability on $\Omega$
\end{itemize}

The following result links the solution of the McKean-Vlasov BSDE (\ref{eq:bsde_mfg_1})-(\ref{eq:bsde_mfg_4}) to the Nash equilibria of the agents.

\begin{theorem}
\label{thm:proba_characterization_ne}
If the BSDE (\ref{eq:bsde_mfg_1})-(\ref{eq:bsde_mfg_4}) admits a solution $(\bY,\bZ,\balpha,\bp,\mathbb{Q})$, then $(\balpha,\bp)$ is a Nash equilibrium. Conversely if $(\hat\balpha,\hat \bp)$ is a Nash equilibrium, then the BSDE (\ref{eq:bsde_mfg_1})-(\ref{eq:bsde_mfg_4}) admits a solution $(\bY,\bZ,\balpha,\bp,\mathbb{Q})$ such that $\balpha = \hat\balpha$, $d\mathbb{P}\otimes dt$-a.e. and $p(t) = \hat p(t)$ $dt$-a.e.
\end{theorem}

\vskip 12pt\noindent
\subsubsection*{\textbf{Principal's Optimal Contracting Problem}}

Recall the optimization problem for the principal:
\[
V(\kappa) := \inf_{(\br,\xi)\in\mathcal{C}}\inf_{\substack{(\balpha,\bp) \in \mathcal{N}(\br,\xi)\\ J^{\br,\xi}(\balpha,\bp) \le \kappa}}\mathbb{E}^{\mathbb{Q}^{(\balpha,\bp)}}\left[\int_0^T [c_0(t, p(t)) + r_t] dt +C_0(p(T)) +\xi\right],
\]

\vspace{2mm}
Unfortunatly, this problem is \emph{totally intractable} !!!!
So we transform it into a more familiar control problem. This is often called the Sannikov trick. Its nature was clearly elucidated by Cvitanic, Possama\"i and Touzi in \cite{CvitanicPossamaiTouzi}.  
We consider the following system of (forward) \textbf{McKean-Vlasov SDEs}:
\begin{align}
Y_t =&\;\;Y_0-\int_0^t\hat H(s, X_{s-}, Z_s, p(s), u(r_s)) ds + \int_0^t Z_s^* d\mathcal{M}_s,\label{eq:forward_y_mkv1}\\
\mathcal{E}_t =&\;\;1 + \int_0^t \mathcal{E}_{s-} X_{s-}^* (Q(s, \alpha_s, p(s)) - Q^0)\psi_s^+ d\mathcal{M}_s,
\label{eq:forward_y_mkv2}\\
\alpha_t = &\;\;\hat \alpha(t,X_{t-},Z_t,p(t)),\label{eq:forward_y_mkv3}\\
p(t) = &\;\;\mathbb{E}^{\mathbb{Q}}[X_t],\;\;\frac{d\mathbb{Q}}{d\mathbb{P}} = \mathcal{E}_T.
\label{eq:forward_y_mkv4}
\end{align}
This is the same type of equations as before, except that we write the dynamic of $\bY$ 
\emph{in the forward direction of time}. That makes the whole difference.
Indeed, if we denote its solution by $(\bY^{\bZ,br, Y_0)},\bZ^{(\bZ,\br,Y_0)},\balpha^{(\bZ,\br,Y_0)}, \bp^{(\bZ,\br,Y_0)}, \mathbb{P}^{(\bZ,\br,Y_0)})$, the expectation under $\mathbb{P}^{(\bZ,\br,Y_0)}$ by $\mathbb{E}^{(\bZ,\br,Y_0)}$, and if we consider the optimal control problem:
\begin{eqnarray*}
&&\tilde V(\kappa) := \inf_{\mathbb{E}^{\mathbb{P}}[Y_0] \le \kappa} \inf_{\substack{\bZ\in\mathcal{H}_{X}^2\\ \br \in \mathcal{R}}}\\&&\hskip 15pt
\mathbb{E}^{(\bZ,\br,Y_0)}\bigg[\int_0^T [c_0(t, p^{(\bZ,\br,Y_0)}(t)) + r_t] dt +C_0(p^{(\bZ,\br,Y_0)}(T))+ U^{-1}(-Y_T^{(\bZ,\br,Y_0)})\bigg],
\end{eqnarray*}
then, as a direct consequence of the previous theorem, we have
$
\tilde V(\kappa) = V(\kappa).
$

\vskip 12pt\noindent
\subsubsection*{\textbf{A Class of Solvable Models}}
While informative at the theoretical level, still, the above results remain of little practical value if they cannot be implemented in the solution of practical problems. In this respect, it is rewarding to discover that under a reasonable set of assumptions, computable solutions can be identified. Here is an example. We fix  $p^{\circ} \in \mathcal{S}$, we assume that the space of actions is a bounded interval, say $A := [\underline \alpha, \overline \alpha]\subset\mathbb{R}^+$, and that the transition rates are linear in the control in the sense that:
\begin{eqnarray*}
q(t,i,j,\alpha,p) &:=& \bar q_{i,j}(t,p) + \lambda_{i,j}(\alpha - \underline\alpha),\;\;\text{for}\;\;i\neq j,\\
q(t,i,i,\alpha,p) &:=& -\sum_{j\neq i}q(t,i,j,\alpha,p),
\end{eqnarray*}
where 
\begin{itemize}
\item $\lambda_{i,j} \in \mathbb{R}^+$ for all $i\neq j$, and $\sum_{j\neq i}\lambda_{i,j} > 0$ for all $i$, 
\item $\bar q_{i,j}:[0,T]\times\mathcal{S}\rightarrow\mathbb{R}^+$ are continuous mappings for all $i\neq j$. 
\end{itemize}
Furthermore, we assume that the agent running costs are of the form:
$$
c(t, e_i, \alpha, p) := c_1(t, e_i, p) + \frac{\gamma_i}{2}\alpha^2,
$$
where $\gamma_i>0$, and the mapping $(t,p)\rightarrow c_1(t, e_i, p)$ is continuous for all $i\in\{1,\dots,m\}$. 
Finally, we assume that the utility function of continuous reward $u$ is continuous, concave and increasing, and that the utility of terminal reward is linear, say $U(\xi) = \xi$.
Under these conditions it is possible to show that the minimizer of the Hamiltonian is given by:
$$
\hat \alpha(t,e_i, z,p) = \hat \alpha(e_i, z) = b\left(-\frac{1}{\gamma_i}\sum_{j\neq i}\lambda_{i,j}(z_j- z_i)\right),
$$
for $i\in \{1,\dots,m\}$, where  $b(z) := \min\{\max\{z,\underline\alpha\},\bar\alpha\}$.
Under these assumptions, one can
\begin{itemize}
\item reduce the problem to the optimal control of a flow of probability measures,
\item and construct an optimal contract!
\end{itemize}
\noindent
See \cite{CarmonaWang_pa} for details. We illustrate this result on a concrete example.

\vskip 12pt\noindent
\subsubsection*{\textbf{A Simple Model of Epidemic Containment}}
I feel compelled to offer a disclaimer before presenting the gory details of the model I propose to use as illustration.
Peiqi Wang and I concocted this model over three years ago for the purpose of illustrating the inner workings of the theory and the analytic computations presented in \cite{CarmonaWang_pa}. In the Spring of 2020, when the paper was accepted for publication in Management Science, the Editor in Chief asked if we could add a discussion to highlight the relevance of this kind of model to the understanding of the COVID-19 pandemic. We obliged, and while doing so, I realized the potential of these new tools to inform policy makers in the control of the spread of epidemics, and the localized re-opening of an economy after shut-down. Given the dire conditions in which we are finding ourselves at this very moment, Aurrell, Dayanikli, Lauri\`ere and I embarked in a systematic investigation of 
what extensions of the model could bring to the understanding of the health and economic consequences of regulations. This effort resulted in \cite{AurrellCarmonaDayanikliLauriere}. The reader interested in applications of similar equilibrium view to epidemic control can also consult the recent work of Elie, Hubert and Turinici  \cite{ElieHubertTurinici}.

\vskip 4pt
Below, we present the model originally introduced in \cite{CarmonaWang_pa}, where plenty numerical illustrations are given illustrating the influence of the various parameters of the model, and in particular, how the contract proposed by the regulator can influence the propensity of the agents to move from one city to another. 
\vskip 2pt
A regulator tries to control the spread of a virus over a time period $[0,T]$.
The jurisdiction of the regulator consists of two cities, say $A$ and $B$.
Each individual is either infected ($I$) or healthy ($H$), lives in city $A$ or $B$. 
So the state space of the model is $E = \{AI, AH, BI, BH\}$
and we denote by $\pi_{AI}, \pi_{AH}, \pi_{BI}, \pi_{BH}$  the proportions of individuals in each of these states.

\vskip 2pt
To describe the time evolution of the state of each individual we introduce the following assumptions:
\begin{itemize}
\item[(1)] the rate of contracting the virus depends on the proportion of infected individuals in the city
so the
\begin{itemize}
\item transition rate from state $AH$ to state $AI$ is $\theta_A^- (\frac{\pi_{AI}}{\pi_{AI} + \pi_{AH}}) $
\item transition rate from state $BH$ to state $BI$ is $\theta_B^- (\frac{\pi_{BI}}{\pi_{BI} + \pi_{BH}}) $. 
\end{itemize}

\item[(2)] the rate of recovery  is a function of the proportion of healthy individuals in the city, so the
\begin{itemize}
\item  transition rate from state $AI$ to state $AH$ is $\theta_A^+(\frac{\pi_{AH}}{\pi_{AI} + \pi_{AH}})$
\item  transition rate from state $BH$ to state $BI$ is $\theta_B^+(\frac{\pi_{BH}}{\pi_{BI} + \pi_{BH}})$. 
\end{itemize}

\item[(3)] Each individual can try to move to the other city:  we denote by $\nu_I\alpha$ the transition rates between the states $AI$ and $BI$, and by $\nu_H\alpha$  the transition rates between the states $AH$ and $BH$. 

\item[(4)] Status of infection does not change when individual moves between cities.
\end{itemize}
The non-negative functions $\theta_A^-$, $\theta_B^-$, $\theta_A^+$ and $\theta_B^+$ are  increasing, differentiable on $[0,1]$. They characterize the quality of health care in the cities $A$ and $B$. So we can change their parameters to make it more or less attractive to individuals to move from one city to the other, or to stay put.
In any case, with these simple prescription, the Q-matrix of the system reads:
$$
\hskip 45pt  AI \hskip 45pt
AH \hskip 45pt
BI \hskip 45pt
BH
$$
\begin{equation*}
Q(t,\alpha,\pi):=\left[
\begin{array}{cccc}
\ldots & \theta_A^+(\frac{\pi_{AH}}{\pi_{AI} + \pi_{AH}}) & \nu_I \alpha & 0\\
\theta_A^-(\frac{\pi_{AI}}{\pi_{AI} + \pi_{AH}}) & \ldots & 0 & \nu_H\alpha\\
\nu_I\alpha & 0 & \ldots & \theta_B^+(\frac{\pi_{BH}}{\pi_{BI} + \pi_{BH}}) \\
0 & \nu_H\alpha & \theta_B^-(\frac{\pi_{BI}}{\pi_{BI} + \pi_{BH}}) & \ldots
\end{array}
\right]
\begin{array}{c}
AI\\
AH\\
BI\\
BH
\end{array}
\end{equation*}

\vskip 2pt\noindent
We now introduce the costs, first for the agents:
\begin{align}
c_1(t, AI, \pi) =&\;\; c_1(t, AH, \pi) := \phi_A\left(\frac{\pi_{AI}}{\pi_{AI} + \pi_{AH}}\right),\label{eq:epicontain_cost_A}\\
c_1(t, BI, \pi) =&\;\; c_1(t, BH, \pi) := \phi_B\left(\frac{\pi_{BI}}{\pi_{BI} + \pi_{BH}}\right),\label{eq:epicontain_cost_B}\\
\gamma_{AI} =&\;\; \gamma_{BI} := \gamma_I,\;\;\gamma_{AH}=\gamma_{BH} := \gamma_H,\label{eq:epicontain_cost_move}
\end{align}
where $\phi_A$ and $\phi_B$ are two increasing functions, and next for the regulator (namely the Principal) for whom the running and terminal costs are given in the form:
\begin{align}
c_0(t,\pi) =&\;\; \exp(\sigma_A \pi_{AI} + \sigma_B \pi_{BI}),\label{eq:epicontain_cost_authority1}\\
C_0(\pi) =&\;\; \sigma_P\cdot(\pi_{AI} + \pi_{AH} - \pi_A^0)^2,\label{eq:epicontain_cost_authority2}
\end{align}
where $\pi_A^0$ is the population of city $A$ at time $0$. 
Choosing the values of the parameters $\sigma_A$, $\sigma_B$ and $\sigma_P$ offer 
\begin{itemize}
\item a trade-off between the control of the epidemic  and  population planning;
\item to try to minimize the infection rate of both cities.
\item In fact, $\sigma_A$, $\sigma_B$ and $\sigma_P$ weigh the relative importance the regulator attributes to each of these objectives.
\end{itemize}
The analysis of this model reduces to the solution of an explicit forward-backward system of Ordinary Differential Equations (ODEs) which can easily be solved numerically, allowing for the computation of statics of the model. Numerical illustrations are provided in \cite{CarmonaWang_pa}.

%%%%%%%%%%%%%%%%%%%%%%
\subsection{Comparison with Plain Nash Equilibria}
It is natural and enlightening to compare the equilibrium computed from the solution of the principal-agent problem to the Nash equilibrium of the mean field game reached by the individuals in the absence of the regulator. In its absence,  the states of the individuals are still governed by the same transition rates, but the individuals' rewards or penalties from the authority are not present in the objective functions they optimize. In other words, they minimize selfishly their expected costs
\begin{equation}
\label{fo:expected_cost}
\mathbb{E}^{\mathbb{Q}^{(\balpha,\pi)}}\left[\int_0^T c(t, X_t, \alpha_t, \pi(t)) dt\right],
\end{equation}
as part of a regular mean field game, and it is plain to compute some of the numerical characteristics of its Nash equilibrium.
Note that this formula for the expected costs of the agents does not contain the payment stream $\br=(r_t)_{0\le t\le T}$ and the terminal payment $\xi$ which enter  the costs to the agents as part of the covenants between them and the principal. This comparison it very much in the spirit of the computation of the so-called   \emph{Price of Anarchy} in classical game theory.

\vskip 2pt
Following the analytical approach to finite-state mean field games introduced in \cite{GomesMohrSouza_continuous}, it is straightforward to derive the system of forward-backward ODEs characterizing the Nash equilibrium. See the system of ODEs (12)-(13) in \cite{GomesMohrSouza_continuous} or \cite[Section 7.2]{CarmonaDelarue_book_I}. In the particular case of the model discussed in this section, the exact form of this system of ODEs is given in the appendix of \cite{CarmonaWang_pa}.

%%%%%%%%%%%%%%%%%%%%%%%%%%%%%%%%%%%%%%%%%%%%%%%%%%%%%%
% \section{Bibliography}
%%%%%%%%%%%%%%%%%%%%%%%%%%%%%%%%%%%%%%%%%%%%%%%%%%%%%%

\bibliographystyle{amsplain}
%\bibliography{games}

\begin{thebibliography}{10}

\bibitem{AchdouBueraLLMoll}
Y.~Achdou, F.~Buera, J.M. Lasry, P.L. Lions, and B.~Moll, \emph{Partial
  differential equation models in macroeconomics}, Philosophical Transactions
  of the Royal Society \textbf{372} (2014).

\bibitem{AchdouGiraudLasryLions}
Y.~Achdou, P.N. Giraud, J.M. Lasry, and P.L. Lions, \emph{A long-term
  mathematical model for mining industries}, Applied Mathematics and
  Optimization \textbf{74} (2016), 579--618.

\bibitem{AchdouHanLLMoll}
Y.~Achdou, J.~Han, J.M. Lasry, and B.~Moll P.L.~Lions, \emph{Income and wealth
  distribution in macroeconomics: A continuous-time approach}, Tech. report,
  http://www.nber.org/papers/w23732, 2017.

\bibitem{AidBaseiPham}
R.~A{\"{i}}d, R.~Dumitrescu, and P.~Tankov, \emph{A mckean-vlasov approach to
  distributed electricity generation development.}, Mathematical Methods of
  Operations Research \textbf{91} (2019), 269--310.

\bibitem{AidDumitrescuTankov}
\bysame, \emph{The entry and exit game in the electricity markets: a mean field
  game approach.}, Tech. report, arXiv.org/2004.14057, 2020.

\bibitem{AlasseurBenTaharMatoussi}
C.~Alasseur, I.~Ben Tahar, and A.~Matoussi, \emph{An extended mean field game
  for storage in smart grids}, arXiv:1710.08991, 2017.

\bibitem{AlmgrenChriss}
R.~Almgren and N.~Chriss, \emph{Optimal execution of portfolio transactions},
  Journal of Risk \textbf{3} (2001), 5Ð39.

\bibitem{AurrellCarmonaDayanikliLauriere}
A.~Aurrell, R.~Carmona, G.~Dayanikli, and M.~Lauri\`ere, \emph{Optimal
  incentives to mitigate epidemics: a stackelberg mean field game approach},
  Tech. report, https://arxiv.org/abs/2011.03105, in preparation.

\bibitem{BahnHaurieMalhame}
O.~Bahn, A.~Haurie, and R.~Malham\'e, \emph{Limit game models for climate
  change negotiations}, Advances in Dynamic and Mean Field Games. (J.~Apaloo
  and B.~Viscolani, eds.), Annals of the International Society of Dynamic
  Games, vol 15, Birkh{\"{a}}user, 2017, pp.~27--47.

\bibitem{BayraktarCohen}
E.~Bayraktar and A.~Cohen, \emph{Analysis of a finite state many player game
  using its master equation}, Tech. report, {\tt arXiv arXiv:1707.02648}, 2017.

\bibitem{Bertucci}
C.~Bertucci, \emph{Optimal stopping in mean field games, and obstacle problem
  approach}, Tech. report, arXiv:1704.06553v2, 2017.

\bibitem{BBLL}
C.~Bertucci, L.~Bertucci, J.M. Lasry, and P.L. Lions, \emph{Mean field game
  approach to bitcoin mining}, Tech. report, arXiv:12004.08167v1, 2020.

\bibitem{BrunnermeierPedersen}
M.~Brunnermeier and L.~Pedersen, \emph{Predatory trading}, Journal of Finance
  \textbf{60} (2005), 1825--1863.

\bibitem{BrunnermeierSannikov1}
M.~Brunnermeier and Y.~Sannikov, \emph{On the optimal inflation rate}, vol.
  106, 2016, pp.~484--489.

\bibitem{BrunnermeierSannikov2}
\bysame, \emph{Macro, money and finance: A continuous time approach},  (2017),
  1497--1546.

\bibitem{bueler}
B.~Bueler, \emph{Solving an equilibrium model for trade of ${\rm co_{2}}$
  emission permits}, European Journal of Operational Research \textbf{102}
  (1997), no.~2, 393--403.

\bibitem{BenazzoliCiampiDIPersio}
L.~Ciampi C.~Benazzoli and L.~DiPersio, \emph{Mean field games with controlled
  jump–diffusion dynamics: Existence results and an illiquid interbank market
  model}, Stochastic Processes and their Applications \textbf{30} (2020).

\bibitem{CardaliaguetLehalle}
P.~Cardaliaguet and C.A. Lehalle, \emph{Mean field game of controls and an
  application to trade crowding}, Tech. report, arXiv:1610.09904.

\bibitem{CarlinLoboViswanathan}
B.~Carlin, M.~Lobo, and S.Viswanathan, \emph{Episodic liquidity crises:
  Cooperative and predatory trading}, Journal of Finance \textbf{65} (2007),
  2235-- 2274.

\bibitem{CarmonaDayanikliLauriere}
R.~Carmona, G.~Dayanikli, and M.~Lauri\`ere, \emph{Mean field game models for
  renewable investment in the electricity markets}, Tech. report, Princeton
  University, 2020.

\bibitem{CarmonaDelarue_book_I}
R.~Carmona and F.~Delarue, \emph{Probabilistic theory of mean field games: vol.
  i, mean field fbsdes, control, and games}, Stochastic Analysis and
  Applications, Springer Verlag, 2017.

\bibitem{CarmonaDelarue_book_II}
\bysame, \emph{Probabilistic theory of mean field games: vol. ii, mean field
  games with common noise and master equations}, Stochastic Analysis and
  Applications, Springer Verlag, 2017.

\bibitem{CarmonaDelarueLacker_timing}
R.~Carmona, F.~Delarue, and D.~Lacker, \emph{Mean field games of timing and
  models for bank runs}, Applied Mathematics and Optimization \textbf{76}
  (2017), 217--260.

\bibitem{CFHP}
R.~Carmona, M.~Fehr, J.~Hinz, and A.~Porchet, \emph{Market design for emissions
  markets trading schemes}, SIAM Review \textbf{52} (2010), 403--452.

\bibitem{CarmonaFouque_delay}
R.~Carmona, J.P. Fouque, M.~Moussavi, and L.H. Sun, \emph{Systemic risk and
  stochastic games with delay}, Tech. report, 2018.

\bibitem{CarmonaFouqueSun}
R.~Carmona, J.P. Fouque, and L.H. Sun, \emph{Mean field games and systemic
  risk: a toy model}, Communications in Mathematical Sciences \textbf{13}
  (2015), 911--933.

\bibitem{CarmonaLacker1}
R.~Carmona and D.~Lacker, \emph{A probabilistic weak formulation of mean field
  games and applications}, Annals of Applied Probability \textbf{25} (2015),
  1189--1231.

\bibitem{CarmonaWang_pa}
R.~Carmona and P.~Wang, \emph{Finite-state contract theory with a principal and
  a field of agents}, Management Science (2020).

\bibitem{CarmonaWang_discrete}
\bysame, \emph{A probabilistic approach to extended finite state mean field
  games}, Mathematics of Operations Research (2020).

\bibitem{CarmonaWebster_fs}
R.~Carmona and K.~Webster, \emph{The self financing condition in high frequency
  markets}, Finance \& Stochastics \textbf{23} (2019), 729 -- 759.

\bibitem{CarmonaYang}
R.~Carmona and J.~Yang, \emph{Predatory trading: a game on volatility and
  liquidity}, Quantitative Finance \textbf{(under revision)} (2011).

\bibitem{CarteaJaimungal}
A.~Cartea, S.~Jaimungal, and J.~Penalva, \emph{Algorithmic and high- frequency
  trading}, Mathematics, Finance and Risk, Cambridge University Press, 2015.

\bibitem{CecchinFischer}
A.~Cecchin and M.~Fischer, \emph{Probabilistic approach to finite state mean
  field games}, Applied Mathematics \& Optimization (2018).

\bibitem{CecchinPelino}
A.~Cecchin and G.~Pelino, \emph{Convergence, fluctuations and large deviations
  for finite state mean field games via the master equation}, Tech. report,
  {\tt arXiv:1707.01819}, 2018.

\bibitem{CecchinDaiPraFischerPelino}
A.~Cecchin, P.~Dai Pra, M.~Fischer, and G.~Pelino, \emph{On the convergence
  problem in mean field games: A two state model without uniqueness}, SIAM
  Journal on Control and Optimization \textbf{57} (2020), 2443--2466.

\bibitem{ChanSircar}
P.~Chan and R.~Sircar, \emph{Fracking, renewables \& mean field games}, SIAM
  Review \textbf{59} (2017), 588--615.

\bibitem{Cohen2010}
S.~Cohen and R.~Elliott, \emph{Comparisons for backward stochastic differential
  equations on markov chains and related no-arbitrage conditions}, The Annals
  of Applied Probability, no.~1, 267--311.

\bibitem{Cohen2008}
\bysame, \emph{Solutions of backward stochastic differential equations on
  {M}arkov chains}, Communications in Stochastic Analysis (2008), no.~2,
  251--262.

\bibitem{CvitanicPossamaiTouzi}
J.~Cvitanic, D.~Possamai, and N.~Touzi, \emph{Dynamic programming approach to
  principal-agent problems}, Tech. report, September 2017.

\bibitem{CvitanicPossamaiTouzi_contract}
J.~Cvitanic, D.~Possama{\"\i}, and N.~Touzi, \emph{Dynamic programming approach
  to principal-agent problems}, Finance and Stochastics \textbf{22} (2018),
  1--37.

\bibitem{DjehicheBarreiro-GomezTembine}
B.~Djehiche, J.~Barreiro-Gomez, and H.~Tembine, \emph{Electricity price
  dynamics in the smart grid: A mean-field-type game perspective}, 23rd
  International Symposium on Mathematical Theory of Networks and Systems Hong
  Kong University of Science and Technology, 2018.

\bibitem{DoncelGastGaujal3}
J.~Doncel, N.~Gast, and B.~Gaujal, \emph{Discrete mean field games: Existence
  of equilibria and convergence}, Tech. report, {\tt arXiv 1909.01209}, 2019.

\bibitem{Elie_et_al}
R.~Elie, E.~Hubert, T.~Mastrolia, and D.~Possama{\"{i}}, \emph{Mean-field moral
  hazard for optimal energy demand response management.}, Tech. report, arXiv
  1902.10405, 2019.

\bibitem{ElieHubertTurinici}
R.~Elie, E.~Hubert, and G.~Turinici, \emph{Contact rate epidemic control of
  covid-19: an equilibrium approach}, Mathematical Modelling of Natural
  Phenomena \textbf{15} (2020).

\bibitem{ElieMastroliaPossamai}
R.~Elie, T.~Mastrolia, and D.~Possama{\"\i}, \emph{A tale of a principal and
  many many agents}, Mathematics of Operations Research \textbf{44} (2019),
  440--467.

\bibitem{FongGossnerHornerSannikov}
K.~Fong, O.~Gossner, J.~H{\"{o}}rner, and Y.~Sannikov, \emph{Coordination under
  private monitoring: from bank runs to the prisoner's dilemma}, Tech. report,
  Princeton University, 2014.

\bibitem{FouqueLangsam}
J.P. Fouque and J.~Langsam (eds.), \emph{Handbook on systemic risk}, Cambridge
  University Press, 2013.

\bibitem{BouveretDumitrescuTankov}
R.~Dumitrescu G.~Bouveret and P.~Tankov, \emph{Mean-field games of optimal
  stopping: a relaxed solution approach},  (2020).

\bibitem{GiraudGueantLasryLions}
P.N. Giraud, O.~Gu{\'{e}}ant, J.M. Lasry, and P.L. Lions, \emph{A mean field
  game model of oil production in presence of alternative energy producers},
  Tech. report, to appear.

\bibitem{GolosovHasslerKrusellTsyvinski}
M.~Golosov, J.~Hassler, P.~Krusell, and A.~Tsyvinski, \emph{Optimal taxes on
  fossil fuel in general equilibrium}, Econometrica \textbf{82} (2014), 41--88.

\bibitem{GomesMohrSouza_continuous}
D.A. Gomes, J.~Mohr, and R.R. Souza, \emph{Continuous time finite state mean
  field games}, Applied Mathematics \& Optimization \textbf{68} (2013),
  99Ð143.

\bibitem{GraberBensoussan}
P.J. Graber and A.~Bensoussan, \emph{Existence and uniqueness of solutions for
  {B}ertrand and {C}ournot mean field games.}, Tech. report, {\tt arXiv
  1508.05408v1}, 2015.

\bibitem{FDD09}
O.~Gu{\'{e}}ant, J.M. Lasry, and P.L. Lions, \emph{Mean field games and oil
  production}, Finance and Sustainable Development : Seminar's lectures.
  (2009).

\bibitem{GueantLasryLions.pplnm}
\bysame, \emph{Mean field games and applications}, Paris Princeton Lectures in
  Mathematical Finance IV (R.~Carmona et~al., ed.), Lecture Notes in
  Mathematics, vol. 2003, Springer Verlag, 2010.

\bibitem{Haurie}
A.~Haurie and L.~Viguier, \emph{A stochastic dynamic game of carbon emissions
  trading}, Environmental Modeling and Assessment \textbf{8} (2003), no.~3,
  239--248.

\bibitem{HeXiong}
Z.~He and W.~Xiong, \emph{Dynamic debt runs}, Review of Financial Studies
  \textbf{25} (2012), 1799 -- 1843.

\bibitem{HinzWarsaw}
J.~Hinz, \emph{An equilibrium model for electricity auctions}, Appl. Math.
  (Warsaw) \textbf{30} (2003), 243--249.

\bibitem{Holmstrom79}
B.~Holmstrom, \emph{Moral hazard and observability}, The Bell Journal of
  Economics \textbf{10} (1979), no.~1, 74--91.

\bibitem{Holmstrom82}
\bysame, \emph{Moral hazard in teams}, The Bell Journal of Economics
  \textbf{13} (1982), no.~2, 324--340.

\bibitem{HolmstromMilgrom87}
B.~Holmstrom and P.~Milgrom, \emph{Aggregation and linearity in the provision
  of inter-temporal incentives}, Econometrica \textbf{55} (1987), 303--328.

\bibitem{HolmstromMilgrom91}
Bengt Holmstrom and Paul Milgrom, \emph{Multitask principal-agent analyses:
  Incentive contracts, asset ownership, and job design}, Journal of Law,
  Economics, \&amp; Organization \textbf{7} (1991), 24--52.

\bibitem{KambhuWeidmanKrishnan}
J.~Kambhu, S.Weidman, and N.~Krishnan (eds.), \emph{New directions for
  understanding systemic risk: A report on a conference cosponsored by the
  federal reserve bank of new york and the national academy of sciences},
  National Research Council, 2007.

\bibitem{KrusellSmith}
P.~Krusell and Jr. A.~Smith, \emph{Income and wealth heterogeneity in the
  macroeconomy}, Journal of Political Economy \textbf{106} (1998), 867--896.

\bibitem{LachapelleLasryLehalleLions}
A.~Lachapelle, J.M. Lasry, C.A. Lehalle, and P.L. Lions, \emph{Efficiency of
  the price formation process in presence of high frequency participants: a
  mean field games analysis}, Mathematics and Financial Economics \textbf{10}
  (2016), 223 -- 262.

\bibitem{LRS}
Z.~Li, M.~Reppen, and R.~Sircar, \emph{A mean field games model for
  cryptocurrency mining}, Tech. report, arXiv:1912.01952v1, 2019.

\bibitem{LudkovskiSircar}
M.~Ludkovski and R.~Sircar, \emph{Game theoretic models for energy production},
  Commodities, Energy and Environmental Finance, Springer, 2015.

\bibitem{Mirrlees}
J.~Mirrlees, \emph{The optimal structure of incentives and authority within an
  organization}, The Bell Journal of Economics \textbf{7} (1976), no.~1,
  105--131.

\bibitem{MorrisShin1998}
S.~Morris and H.S. Shin, \emph{Unique equilibrium in a model of self-fulfilling
  currency attacks}, American Economic Review (1998), 587--597.

\bibitem{NadtochyiShkolnikov}
S.~Nadtochiy and M.~Shkolnikov, \emph{Mean field systems on networks, with
  singular interaction through hitting times}, Annals of Probability
  \textbf{48} (2020), 520--1556.

\bibitem{Nutz}
M.~Nutz, \emph{A mean field game of optimal stopping}, Tech. report, 2018.

\bibitem{RochetVives}
J.C. Rochet and X.~Vives, \emph{Coordination failures and the lender of last
  resort}, Journal of the European Economic Associateion \textbf{2} (2004),
  1116 -- 1148.

\bibitem{Sannikov08}
Yuliy Sannikov, \emph{A continuous-time version of the principal-agent
  problem}, The Review of Economic Studies \textbf{75} (2008), no.~3, 957--984.

\bibitem{Sannikov12}
\bysame, \emph{Contracts: The theory of dynamic principal-agent relationships
  and the continuous-time approach}, 10th World Congress of the Econometric
  Society, 2012.

\bibitem{ShrivatsFirooziJaimungal}
A.~Shrivats, D.~Firoozi, and S.~Jaimungal, \emph{A mean field game approach to
  equilibrium pricing optimal generation, and trading in solar renewable energy
  certificate (srec) markets.}, Tech. report, arXiv 2003.04938, 2020.

\end{thebibliography}
%\printbibliography

\end{document}